\documentclass[acmtog]{acmart}
\usepackage{multirow}
\usepackage{graphicx}
\usepackage{amsmath}
\usepackage[normalem]{ulem}
\usepackage[ruled,vlined]{algorithm2e}
\usepackage{soul}
\AtBeginDocument{%
  \providecommand\BibTeX{{%
    \normalfont B\kern-0.5em{\scshape i\kern-0.25em b}\kern-0.8em\TeX}}}

\setcopyright{rightsretained}
\acmJournal{TOG}
\acmYear{2022}\acmVolume{41}\acmNumber{4}\acmArticle{128}\acmMonth{7} \acmDOI{10.1145/3528223.3530096}

\acmJournal{TOG}
\acmVolume{41}
\acmNumber{4}
\acmArticle{128}
\acmMonth{7}

\acmSubmissionID{304}

\begin{document}

\title{Iterative Poisson Surface Reconstruction (iPSR) for Unoriented Points}

% \affiliation{%

 \author{Fei Hou}
 \email{houfei@ios.ac.cn}
 \orcid{0000-0001-8226-6635}
 \author{Chiyu Wang}
 \email{wangcy@ios.ac.cn}
 \orcid{0000-0002-6427-2211}
 \author{Wencheng Wang}
 \email{whn@ios.ac.cn}
 \orcid{0000-0001-5094-4606}
 \affiliation{%
   \institution{State Key Laboratory of Computer Science, Institute of Software, Chinese Academy of Sciences \& University of Chinese Academy of Sciences}
   \city{Beijing}
   \country{China}
 }

 \author{Hong Qin}
 \email{qin@cs.stonybrook.edu}
 \orcid{0000-0001-7699-1355}
 \affiliation{%
   \institution{Department of Computer Science, Stony Brook University}
   \city{New York}
   \country{USA}
 }

 \author{Chen Qian}
 \email{qianchen@sensetime.com}
 \affiliation{%
   \institution{SenseTime Research \& Tetras.AI}
   \city{Beijing}
   \country{China}
 }

 \author{Ying He}
 \authornote{Corresponding author}
 \email{yhe@ntu.edu.sg}
 \orcid{0000-0002-6749-4485}
 \affiliation{%
   \institution{School of Computer Science and Engineering \& S-Lab, Nanyang Technological University}
   \country{Singapore}
 }

%\renewcommand{\shortauthors}{Authors et al.}

\begin{teaserfigure}
\centering
\includegraphics[width=3.5in]{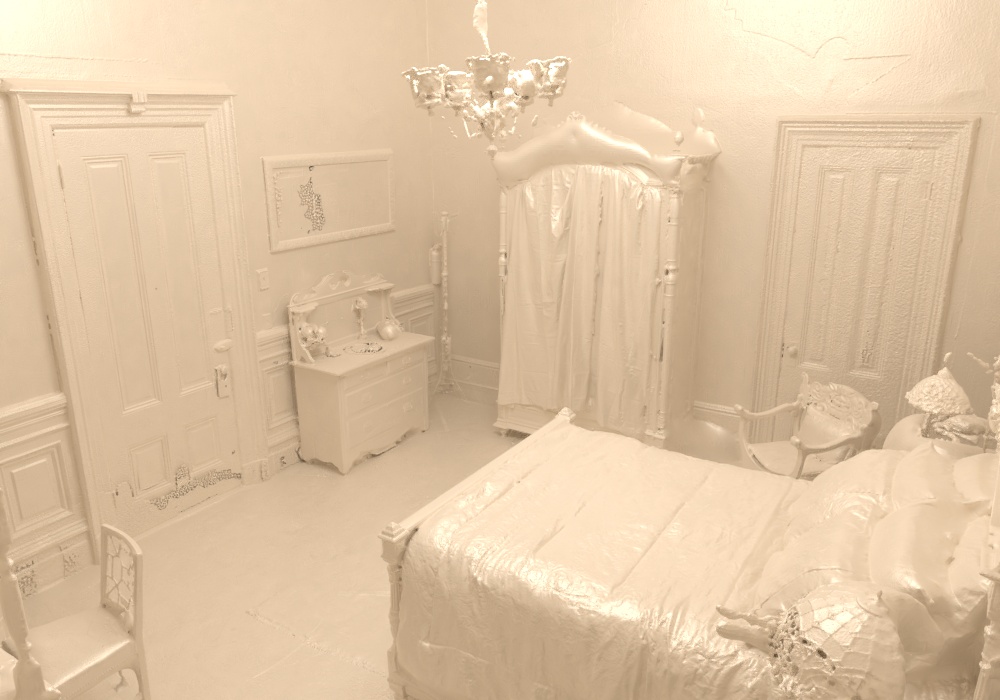}
\includegraphics[width=3.5in]{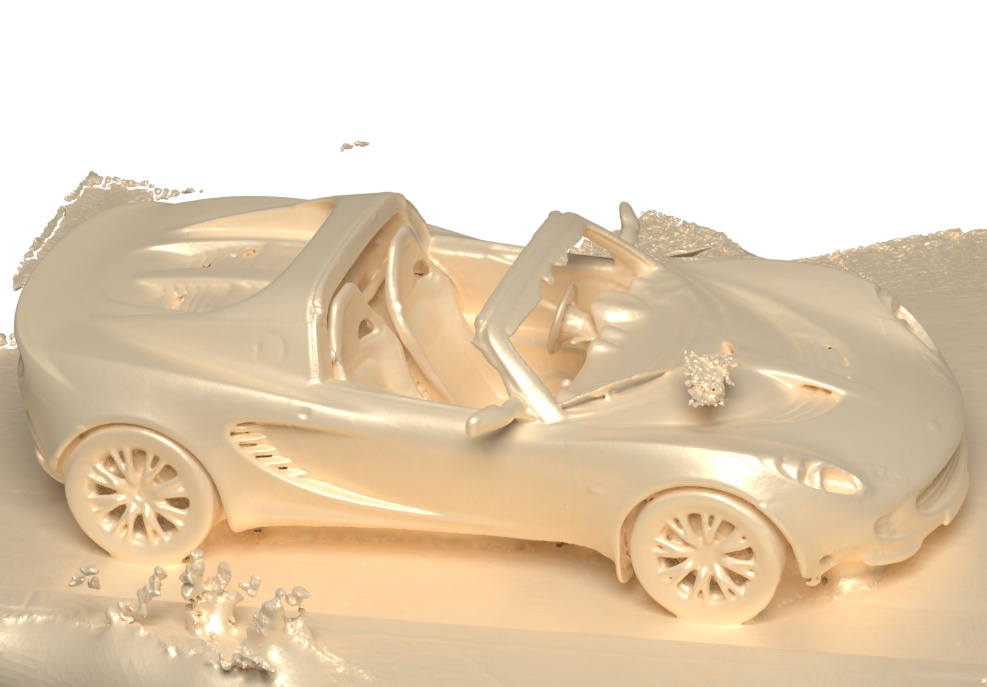}
\caption{Our method extends the popular Poisson surface reconstruction technique by eliminating the requirement of point orientation. As a result, it can directly apply to points without normal information. Left: an indoor scene with 17 million points~\cite{Park2017}. Right: an outdoor scene with 1.89 million points~\cite{Choi2016}. We set the octree depth 12 in both examples.}
\label{fig:teaser}
\end{teaserfigure}
\begin{abstract}
Poisson surface reconstruction (PSR) remains a popular technique for reconstructing watertight surfaces from 3D point samples thanks to its efficiency, simplicity, and robustness. Yet, the existing PSR method and subsequent variants work only for oriented points. This paper intends to validate that an improved PSR, called iPSR, can completely eliminate the requirement of point normals and proceed in an iterative manner. In each iteration, iPSR takes as input point samples with normals directly computed from the surface obtained in the preceding iteration, and then generates a new surface with better quality. Extensive quantitative evaluation confirms that the new iPSR algorithm converges in 5-30 iterations even with randomly initialized normals. If initialized with a simple visibility based heuristic, iPSR can further reduce the number of iterations. We conduct comprehensive comparisons with PSR and other powerful implicit-function based methods. Finally, we confirm iPSR's effectiveness and scalability on the AIM@SHAPE dataset and challenging (indoor and outdoor) scenes. Code and data for this paper are at \url{https://github.com/houfei0801/ipsr}.
\end{abstract}

\if 0
\begin{CCSXML}
<ccs2012>
 <concept>
  <concept_id>10010520.10010553.10010562</concept_id>
  <concept_desc>Computer systems organization~Embedded systems</concept_desc>
  <concept_significance>500</concept_significance>
 </concept>
 <concept>
  <concept_id>10010520.10010575.10010755</concept_id>
  <concept_desc>Computer systems organization~Redundancy</concept_desc>
  <concept_significance>300</concept_significance>
 </concept>
 <concept>
  <concept_id>10010520.10010553.10010554</concept_id>
  <concept_desc>Computer systems organization~Robotics</concept_desc>
  <concept_significance>100</concept_significance>
 </concept>
 <concept>
  <concept_id>10003033.10003083.10003095</concept_id>
  <concept_desc>Networks~Network reliability</concept_desc>
  <concept_significance>100</concept_significance>
 </concept>
</ccs2012>
\end{CCSXML}
\ccsdesc[500]{Computer systems organization~Embedded systems}
\ccsdesc[300]{Computer systems organization~Redundancy}
\ccsdesc{Computer systems organization~Robotics}
\ccsdesc[100]{Networks~Network reliability}
\fi
\keywords{Unoriented points; Poisson surface reconstruction; Iterative algorithm}

\maketitle

\section{Introduction}
For more than a decade, Poisson surface reconstruction
(PSR)~\cite{Kazhdan2006,Kazhdan2013} has been a well-known
technique for producing watertight surfaces from oriented point
samples. Its key idea is to compute a signed distance field by solving
Poisson's equation, resulting in a sparse linear system, hence is
computationally efficient and also works for large-scale inputs. Other
noticeable advantages include being resilient to noisy data and
tolerant to registration artifacts. Nevertheless, its strong
requirement on point orientation severely confines its potentially
widespread applications. Despite significant research progress pertinent to machine learning in recent years, precisely predicting point orientation from raw, noisy points remains an insurmountable challenge.

Some recent works have shown earlier attempts in bridging such gaps,
with an ultimate goal of inferring implicit surfaces from unoriented
points. Huang et al.~\shortcite{Huang2019} formulated an elegant
variational framework using Duchon's energy~\cite{Duchon1977}. Their method,
called variational implicit point set surfaces (VIPSS), does not need
domain discretization and works well for both exact interpolation and
approximation towards linear geometry reproduction. However, it
involves dense matrix formulation, thereby is only limited to
small-scale point clouds. Metzer et al.~\shortcite{Metzer2021}
addressed the problem of orienting point clouds by separating its
global and local components into two sub-problems. In the local phase,
it trains a deep neural network to learn a coherent normal direction
per patch, while in the global phase, it propagates the orientation
across all coherent patches using a dipole propagation. Their method
is able to predict accurate normals for most of the samples, yet there
are frequently a few patches whose normals are flipped, yielding
artifacts in the corresponding reconstructed regions. Deep
learning based surface reconstruction
methods~\cite{Erler2020,Park2019,Groueix2018} were proposed recently.
Although they are excellent to reconstruct data belonging to the same class of the training data with strong prior knowledge, they are not robust to the reconstruction of other classes of data, so they are not ideal for general purpose usage.

This paper showcases our new research effort towards an enhanced PSR without the need of point orientation information (critical to surface reconstruction in prior algorithms). We wish to validate that an improved PSR can completely eliminate the strong requirement of point normals and naturally proceed towards the final reconstruction in an iterative manner. The key insight in our pursuit of the new algorithm is that, when assigning random normals to the input points, PSR generates a surface that is often far from the correct final shape, but this intermediate surface can still afford valuable information for updating point orientations, from which an even better surface can be generated. Specifically in each iteration, our new algorithm takes as input point samples with normals directly computed from the surface obtained in the preceding iteration, and then reconstructs a new surface with better quality. Because of the algorithm's iterative
nature, we call our method iterative Poisson surface reconstruction, or iPSR. Extensive quantitative evaluation on the AIM$@$SHAPE dataset confirms that the new iPSR algorithm converges in 5-30 iterations (with average 10 iterations) even with completely randomized normals at the initialization stage. One observation during
our extensive experiments is that, when initialized with a simple
visibility based heuristic~\cite{Katz2007}, our iPSR can further
reduce the number of iterations by 45\% on models with 100K+ points. We conduct thorough, in-depth comparisons with PSR and other powerful implicit-function based techniques through comprehensive experiments, which all ascertain iPSR's effectiveness and scalability on benchmark dataset and challenging scene data. It is also worth mentioning that other inherent advantages include being
robust to outliers, noisy, non-uniform, and/or sparse point cloud data. 

%Figure~\ref{fig:teaser} illustrates two  examples of large-scale 3D scene reconstruction using iPSR. The indoor scene is scanned by a Lidar scanner~\cite{Park2017}, and we use the merged and resampled points as our input. The outdoor scene is scanned by a RGBD camera~\cite{Choi2016}. As only raw depth images and the reconstructed surface mesh are available, we take the mesh vertices as input to iPSR.

\section{Related Work}
\subsection{Implicit Function Methods}
A large repository of existing techniques result from implicit methods, which essentially generate a distance field and extract iso-surface to reconstruct the surface. In principle, the implicit method can guarantee a watertight surface reconstruction, where the generated surface may not pass through the sample points, but it tends to be more robust for noisy inputs.

MPU~\cite{Ohtake2003} blends local quadratic functions to generate implicit fields. Poisson surface reconstruction~\cite{Kazhdan2006,Bolitho2009} as well as screened Poisson surface reconstruction~\cite{Kazhdan2013,Kazhdan2020} fit a smoothed 0-1 indicator function blurred near the modeled surface. The gradient of the indicator function is derived from surface normals and the indicator function is fitted by a solvable (screened) Poisson's equation. Manson et al.~\shortcite{Manson2008} reconstructed the indicator function by wavelets. Calakli and Taubin~\shortcite{Calakli2011} generated a smooth approximation to the signed distance field of a surface. Taking advantage of both indicator function and signed distance function, Lu et al.~\shortcite{Lu2018} generated the implicit field using a modified Gauss formula with higher accuracy. However, all of them demand oriented normals as input, which may not be easily obtained in advance.

Another group of methods try to infer normal directions automatically
from classic PCA~\cite{Hoppe1992} to Voronoi diagrams~\cite{Alliez2007,Merigot2011} to estimate
normal directions, or specially for
surfaces with sharp features~\cite{Li2010,Boulch2012}. But they fail
to address the normal orientation problem. The consistent orientation
is the key for surface reconstruction, which can be classified into
local or global methods~\cite{Kazhdan2020}. Local methods fit a local
surface first and then blend them together or propagate the normals
greedily. Global methods infer all the normals simultaneously by
optimizing a global function.

\paragraph{Local Methods.} The pioneering work~\cite{Hoppe1992} propagates
normal orientations along neighboring centers whose directions are
nearly parallel, which is a greedy algorithm seeking to orient normal
on a minimum spanning tree. However, the neighboring size is crucial
to the algorithm~\cite{Mitra2003}. More reliable measures are proposed
later for orientation propagation~\cite{Xie2003}\cite{Huang2009}\cite{Huang2013}.
Still, the propagation strategy is greedy in nature. Some methods are
proposed to reconstruct surface of noisy point cloud by least square
fitting~\cite{Mitra2003,Xie2003}, yet they are not suitable for sparse
point clouds. Inspired by dipole, Metzer et al.~\shortcite{Metzer2021}
proposed to orient points in the local and global phases,
respectively.  The method is efficient, but is not robust to complicated data due to its intrinsic propagation nature.

\paragraph{Global Methods.} In contrast to the aforementioned local
approaches, global methods are more reliable to reconstruct 3D models. After estimating normal directions from Voronoi diagram, Alliez et al.~\shortcite{Alliez2007} evaluated the implicit field as well as orienting normals by solving a generalized eigenvalue problem to maximize an anisotropic Dirichlet energy. Mullen et al.~\shortcite{Mullen2010} computed an unsigned distance approximation of the input data first, and then estimated its sign by minimizing a quadratic energy. Their method is robust to noise and outliers.
Schertler et al.~\shortcite{Schertler2017} proposed a global graph-based minimization approach to orienting normals. Most of the approaches decompose the estimation of normal directions and orientations into two steps. Wang et al.~\shortcite{Wang2012} proposed a variational framework, which integrates the two steps together. Recently, Huang et al.~\shortcite{Huang2019} fitted surface and normals by Duchon's energy.
It works well for sparse uniform samples and wireframes, but it fails to reconstruct dense point clouds due to high computational cost.
Peng et al.~\shortcite{Peng2021} exploited fast Fourier transform based Poisson solver to differentiate the indicator function with respect to point positions and normals, so that they optimize point positions and the corresponding normals to minimize reconstruction error. It is much more reliable than previous methods since both point positions and normals are optimized in a unified framework to minimize the reconstruction error. However, it requires thousands of epochs to converge, hence too slow to reconstruct large-scale models. In addition, in our experiments we observe that it is not robust to reconstructing high-genus models.

\subsection{Other Techniques}
Besides implicit methods, there are a plethora of earlier works in surface reconstruction using combinatorial methods, such as ball-pivoting~\cite{Bernardini1999}, power crust~\cite{Amenta2001}, and tight cocone~\cite{Dey2003}. Since these methods rely on either Delaunay triangulation or its dual Voronoi diagram, they do not require point normals and can be directly applied to raw points. However, they are vulnerable to noise and outliers, and fail to function with sparse points. Lazar et al.~\shortcite{Lazar2018} formulated an elegant combinatorial optimization for surface reconstruction subject to topological constraints. The method is robust and can be applied to reconstruction from cross-sectional slices and iso-surfacing an intensity volume. However, it requires topological constraints (i.e., genus) as input, which is hard to obtain as prior knowledge for models with complex geometry and/or arbitrary topology.

Recently, as an emerging method, deep learning has shown promise in point orientation and surface reconstruction.
Using an automatically generated prior shape as the initial mesh, Point2Mesh~\cite{Hanocka2020} continuously deforms it to shrink-wrap the input point cloud and generates a watertight triangle mesh of the same topology. It works well for genus-0 models, but extending it to high-genus models is non-trivial due to lack of techniques for generating initial meshes. Point2Surf~\cite{Erler2020} and Iso-Points~\cite{Wang2021} are appropriate for dense noisy point clouds, but they are not able to reconstruct sparse points such as the wireframe samples used in~\cite{Huang2019}. IGR~\cite{Gropp2020} uses a multilayer perceptron to represent 3D shapes and adopts a simple regularization term to train it. The method works for raw points, but it is sensitive to noisy input.
Implicit occupancy network represents 3D surfaces as a continuous decision boundary of a deep neural network classifier~\cite{Mescheder2019}. It does not require discretization and can represent shape in a continuous manner. However, due to fully-connected network architecture, it cannot reconstruct high frequency surface detail.
Later, Peng et al.~\shortcite{Peng2020} proposed convolutional occupancy network to handle surface details and scenes.
IM-Net~\cite{Chen2019}, as an implicit field decoder, trains a binary classifier to indicate whether a point is outside a shape or not. It can be used for 3D reconstruction with good visual quality.
The local implicit grid representations~\cite{Jiang2020} learn shape priors at a micro scale and leverage them in a macro scale for 3D reconstruction. The method works well for 3D scenes containing man-made objects that are smooth at a ``part'' scale, however, it cannot deal with models with rich geometric detail. DeepSDF~\cite{Park2019} encodes a shape into a feature vector applicable for shape interpolation and completion, but it is not used for surface reconstruction. In essence, the deep learning based methods always rely on the training process of certain datasets, which are specifically tailored for certain types of models and their subsequent reconstruction task. In contrast, our newly-developed iPSR is a general purpose framework well suitable in a wide range of model types, without the need of point orientation information at all.

\section{Preliminaries}
Poisson surface reconstruction~\cite{Kazhdan2006} takes as input a set of oriented points $\{s_i\}_{i=1}^m$ sampling the boundary of a watertight surface. Throughout the paper, we denote by $M$ the solid bounded by the watertight surface. Then $\partial M$ is the 2-manifold closed surface that we aim to reconstruct. Assume the input model is uniformly scaled into a unit box, we denote by $\vec{n}(x)$ the given unit inward normal for a sample point $x=s_i$. Transform the discrete vectors $\vec{n}$ into a continuous vector field $\vec{V}$ using a smoothing filter $F(\cdot)$,
$$\vec{V}(x)=\oint_{\partial M}F_y(x)\vec{n}(y)dy,$$
where $F_p(q)=F(q-p)$ is the translation to point $p$.
In~\cite{Kazhdan2006}, $F_y(x)$ is the Gaussian function centered at $x$. The Poisson surface reconstruction algorithm~\cite{Kazhdan2006} computes an indicator function $\chi:\mathbb{R}^3\rightarrow \mathbb{R}$ whose gradient approximates $V$ by minimizing the energy $$\int_{[0,1]^3}\left\|\nabla\chi(x)-\vec{V}(x)\right\|dx.$$ The indicator function is defined to have value $1$ inside and value $0$ outside the model. Thus, the function value is $\frac{1}{2}$ for points on the surface $\partial M$. Using the Euler-Lagrange formulation, the minimum is obtained by solving Poisson's equation $$\Delta\chi(x)=\nabla\cdot V,$$ with Dirichlet boundary condition $\chi(x)=0$ for $x$ on the boundary of $[0,1]^3$. Using octree for decomposing the domain and locally supported basis functions for domain discretization, Kazhdan et al. showed that the Poisson equation becomes a well-defined sparse linear system, which can be solved easily and efficiently. 

The screened Poisson surface reconstruction algorithm~\cite{Kazhdan2013} adds into the energy an additional term that penalizes the surface from deviating the samples. Minimizing such an energy can be interpreted as a screened Poisson equation, which is also a sparse linear system. 

\begin{figure}
    \centering
    \includegraphics[width=1.50in]{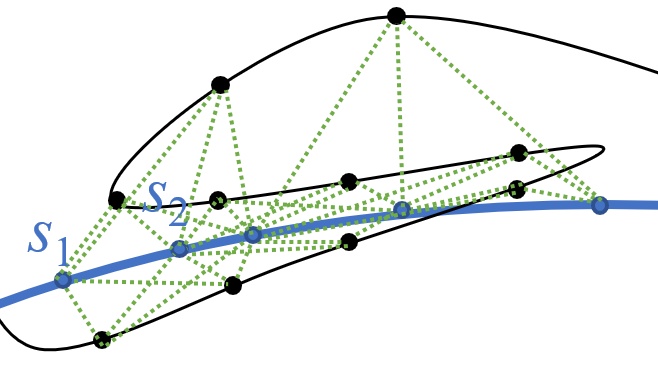}
    \includegraphics[width=1.50in]{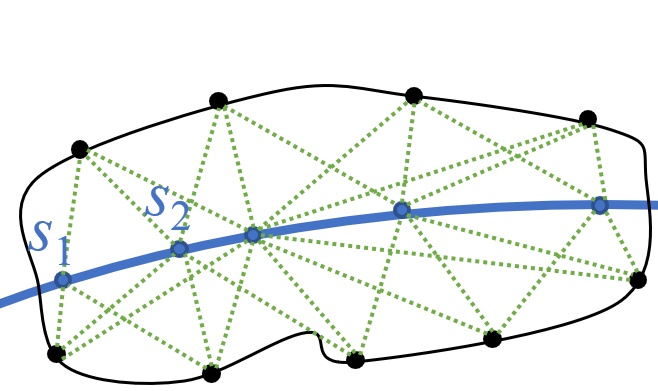}\\
    \makebox[1.50in]{(a) 3 layers}
    \makebox[1.50in]{(b) 2 layers}\\
    \caption{Illustration of the $k$-nearest neighbor searching using 2D examples. The blue curve represents part of the boundary surface of a solid $M$, on which sample points (blue dots) are located. The black curve is part of the iso-surface $\mathcal{F}$ produced by screened PSR. Due to incorrect normals used by PSR, the computed iso-surface $\mathcal{F}$ is twisted and has different topology than $\partial M$. After applying the marching cube algorithm, we obtain a triangular mesh representing $\mathcal{F}$. We draw the center of each triangular face as a black dot. The green dashed lines show the relation between triangle centers and sample points. To avoid visual clutter, we set a small $k=3$ in both examples, i.e., each black dot is linked to 3 closest green dots. Consider two samples $s_i$, $i=1,2$ and denote by $n_i$ the number of centers associated to sample $s_i$. We show two representative layered structures in (a) and (b), respectively. (a) is odd-layered and we have $n_1=4$ and $n_2=7$. (b) is even-layered and $n_1=3$ and $n_2=5$. We then update the normal of sample $s_i$ by \textbf{weighted} averaging the normals of the associated black dots. It is worth noting that the parity of $n_i$ is not important, since we adopt weighted average to update normals, i.e., the normal of each black dot is multiplied by the triangle area. As a result, what matters most is whether these black dots are from an odd-layered structure. We observe that (1) if the iso-surface has an odd-layer structure, it is very likely that the weighted average of normals is inward; and (2) if the iso-surface has an even-layer structure locally, weighted normal averaging will turn the structure into odd-layered in future iterations.}
    \label{fig:2d_illustration}
\end{figure}

\begin{figure*}[!htbp]
\centering
\boxed{
    \includegraphics[width=0.908in]{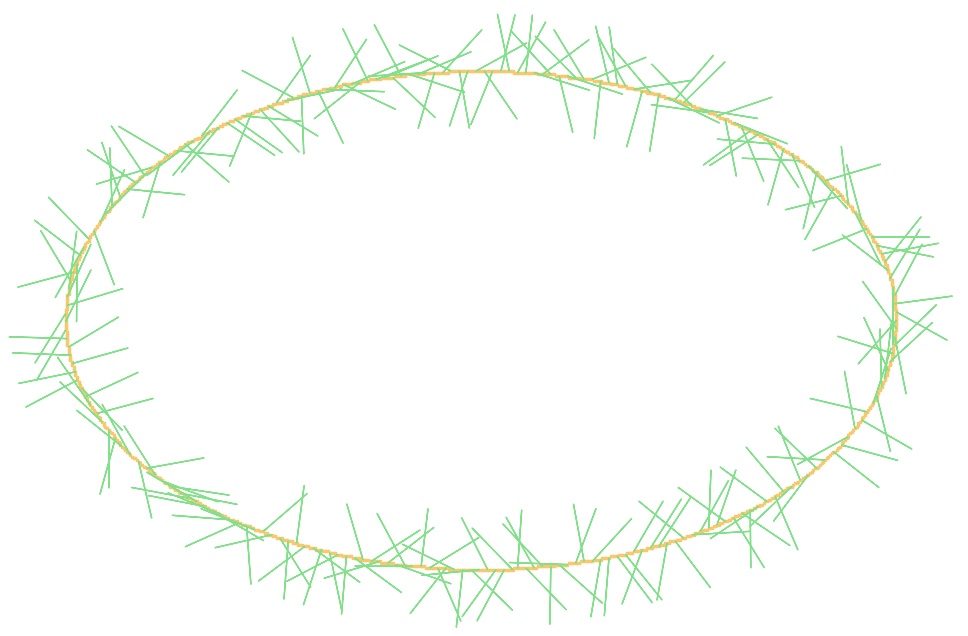}
    }
    \boxed{
    \includegraphics[width=0.908in]{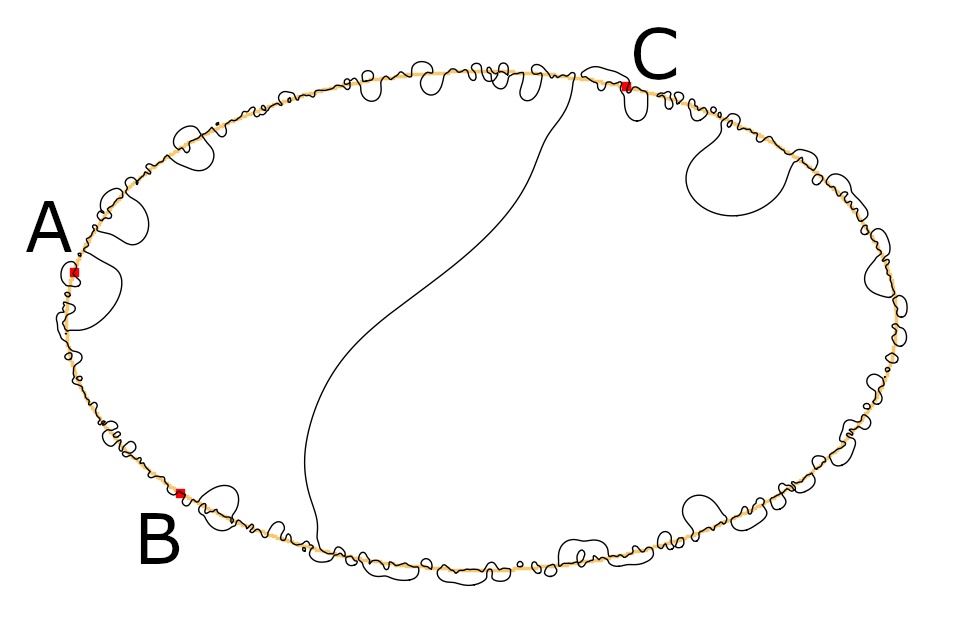}
    \includegraphics[width=0.908in]{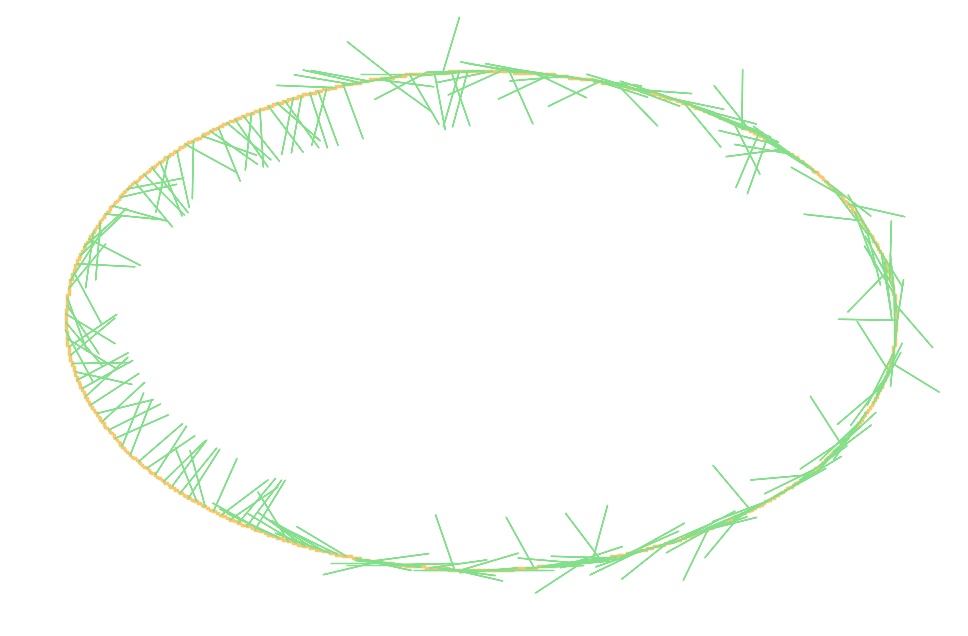}}
    \boxed{
    \includegraphics[width=0.908in]{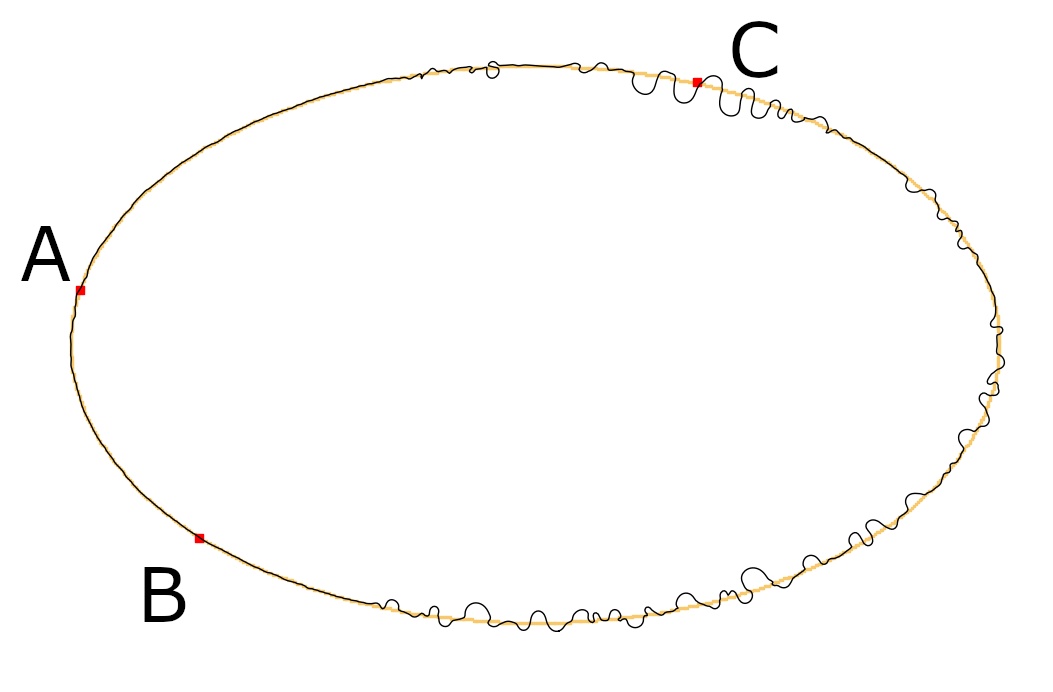}
    \includegraphics[width=0.908in]{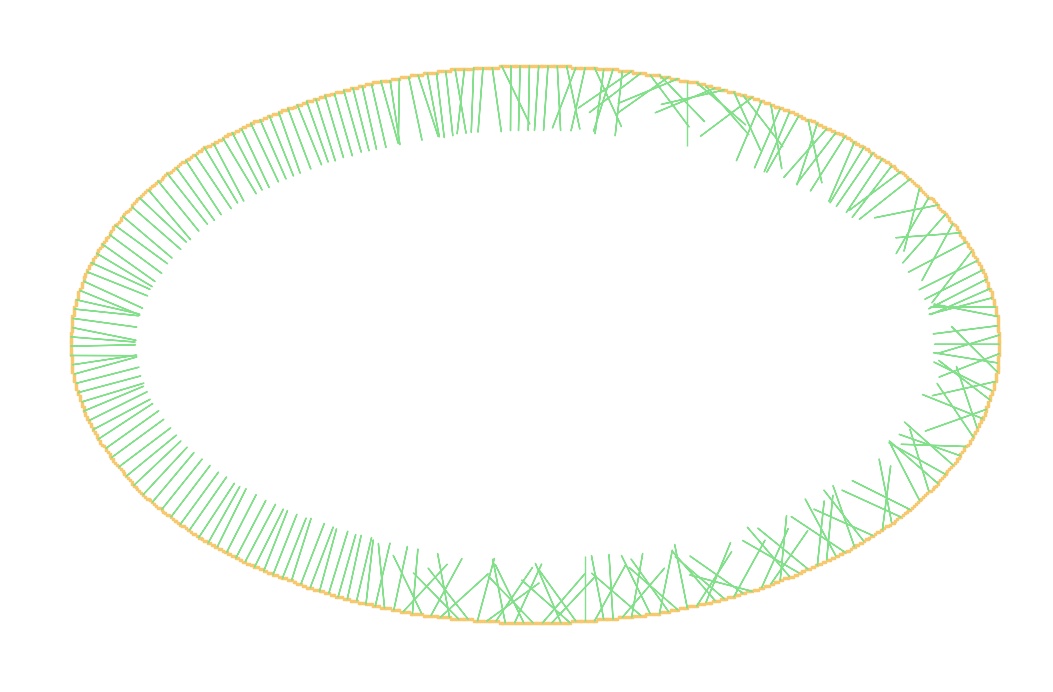}}
    \boxed{
    \includegraphics[width=0.908in]{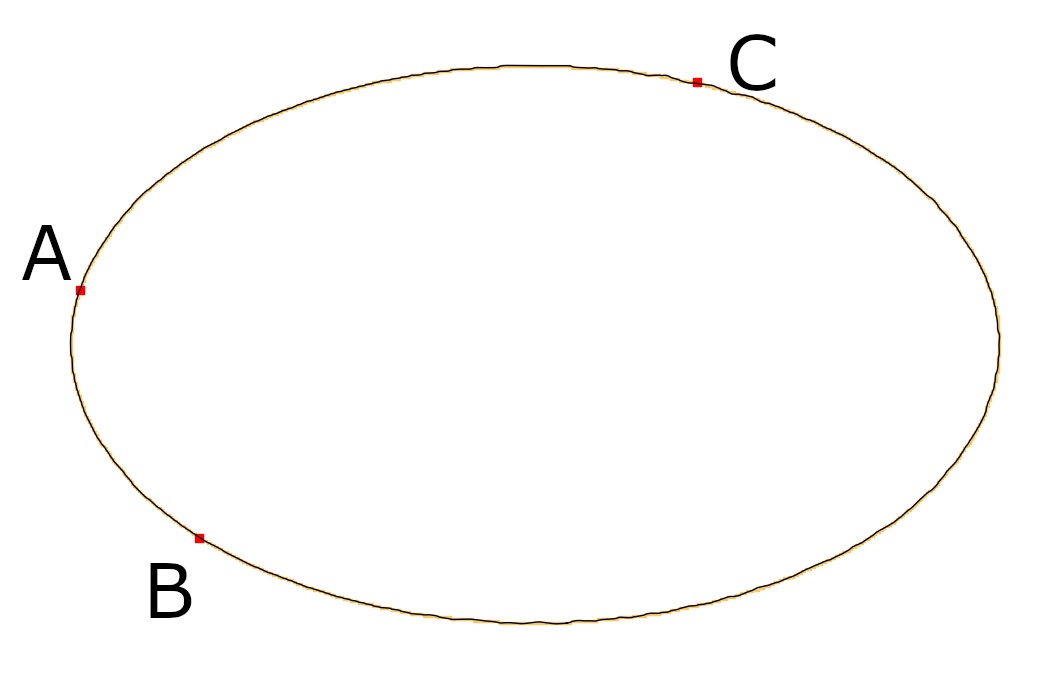}
    \includegraphics[width=0.908in]{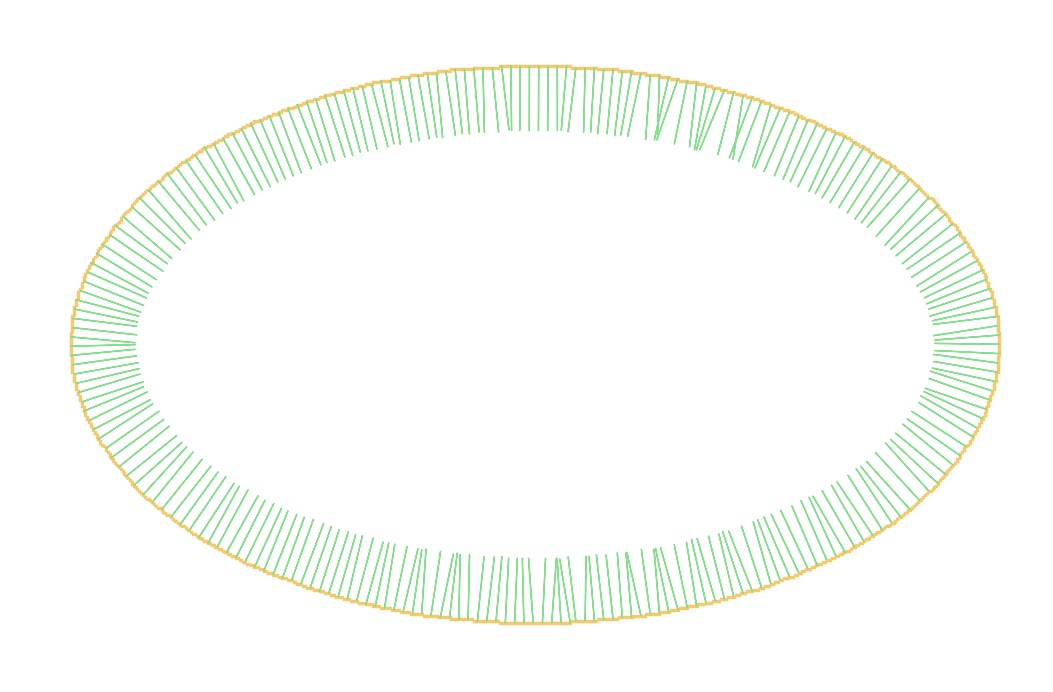}}\\
    \makebox[0.96in]{(a)}
    \makebox[0.93in]{(b)}
    \makebox[0.93in]{(c)}
    \makebox[0.93in]{(d)}
    \makebox[0.93in]{(e)}
    \makebox[0.93in]{(f)}
    \makebox[0.93in]{(g)}\\
    \vspace{0.05in}
    \boxed{
    \includegraphics[width=0.70653in]{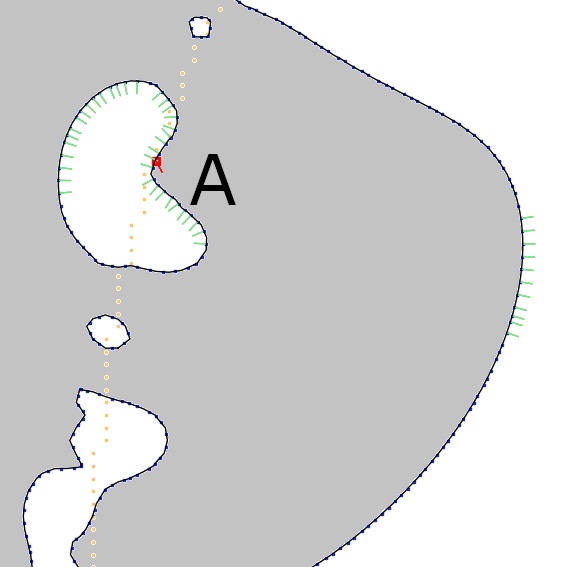}
    \hspace{0.02in}
    \includegraphics[width=0.70653in]{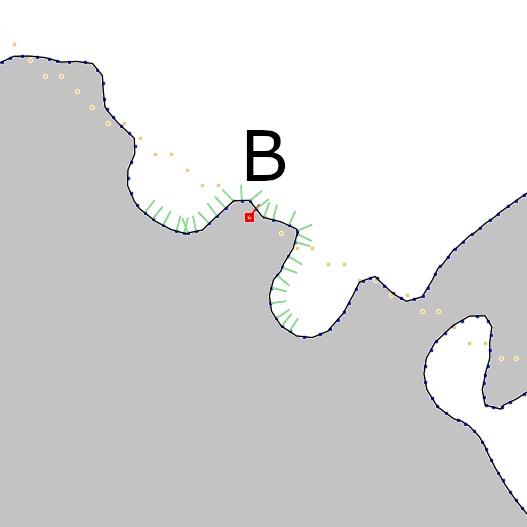}
    \hspace{0.02in}
    \includegraphics[width=0.70653in]{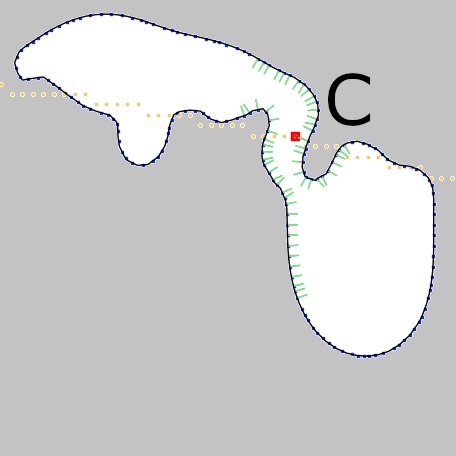}
    }
    \boxed{
    \includegraphics[width=0.70653in]{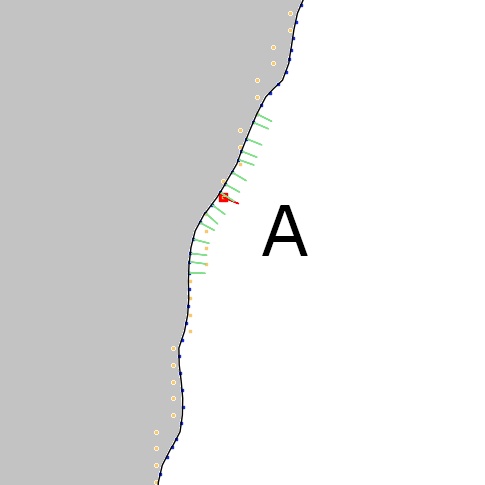}
    \hspace{0.02in}
    \includegraphics[width=0.70653in]{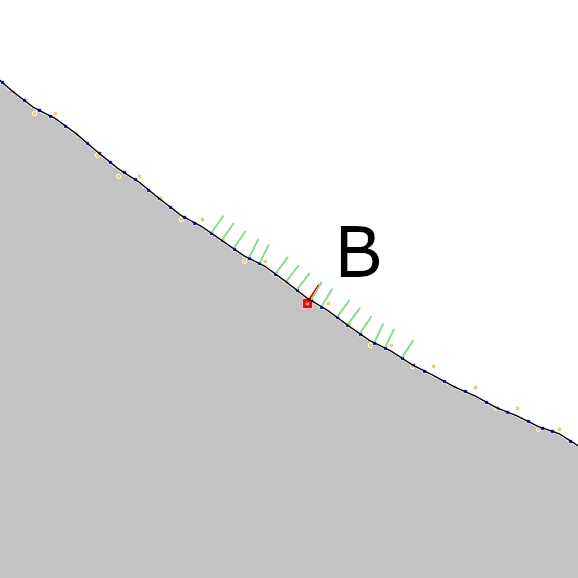}
    \hspace{0.02in}
    \includegraphics[width=0.70653in]{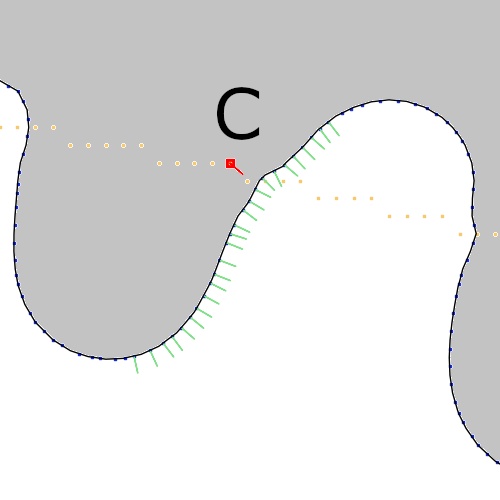}
    }
    \boxed{
    \includegraphics[width=0.70653in]{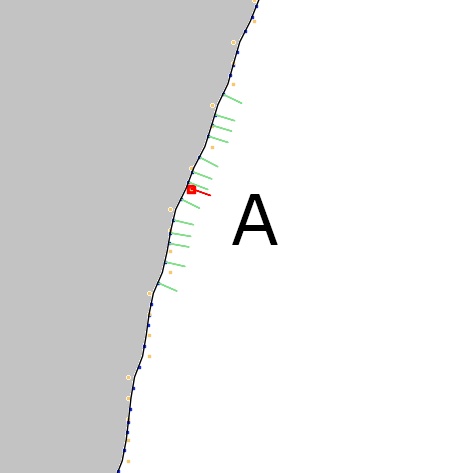}
    \hspace{0.02in}
    \includegraphics[width=0.70653in]{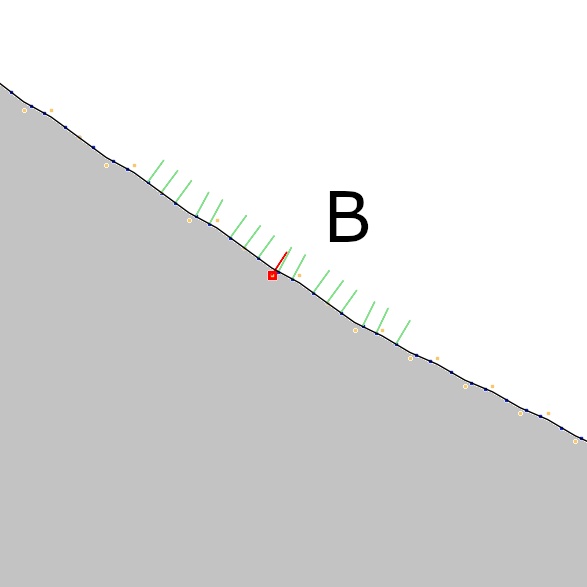}
    \hspace{0.02in}
    \includegraphics[width=0.70653in]{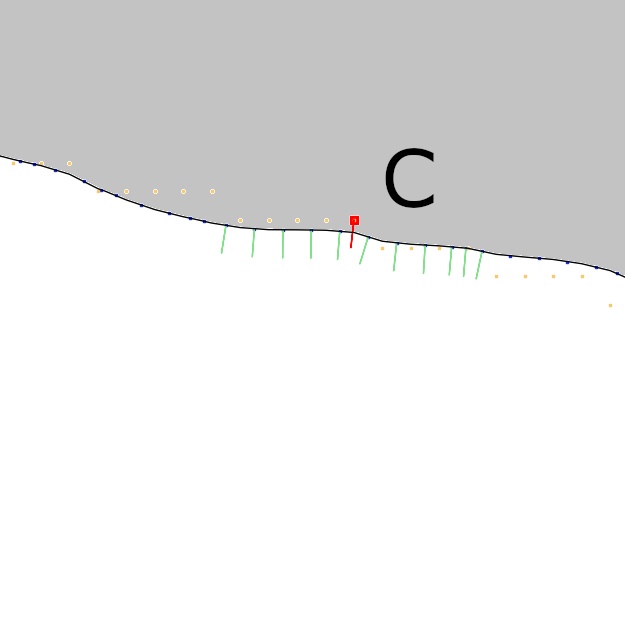}
    }\\
    \makebox[0.7067in]{triple layers}
    \makebox[0.7067in]{single layer}
    \makebox[0.7067in]{double layers}
    \makebox[0.7067in]{single layer}
    \makebox[0.7067in]{single layer}
    \makebox[0.7067in]{single layer}
    \makebox[0.7067in]{single layer}
    \makebox[0.7067in]{single layer}
    \makebox[0.7067in]{single layer}\\
    \makebox[2.13in]{Iteration 1}
    \makebox[2.13in]{Iteration 2}
    \makebox[2.13in]{Iteration 3}
    \caption{Layered structure and normal updating. In the close-up views, we shade the exterior region and draw only part of the normals to avoid visual clutter. See the texts for details.}
    \label{fig:ellipse}
\end{figure*}

\begin{figure*}[!htbp]
    \centering
    \boxed{
    \includegraphics[width=0.9in]{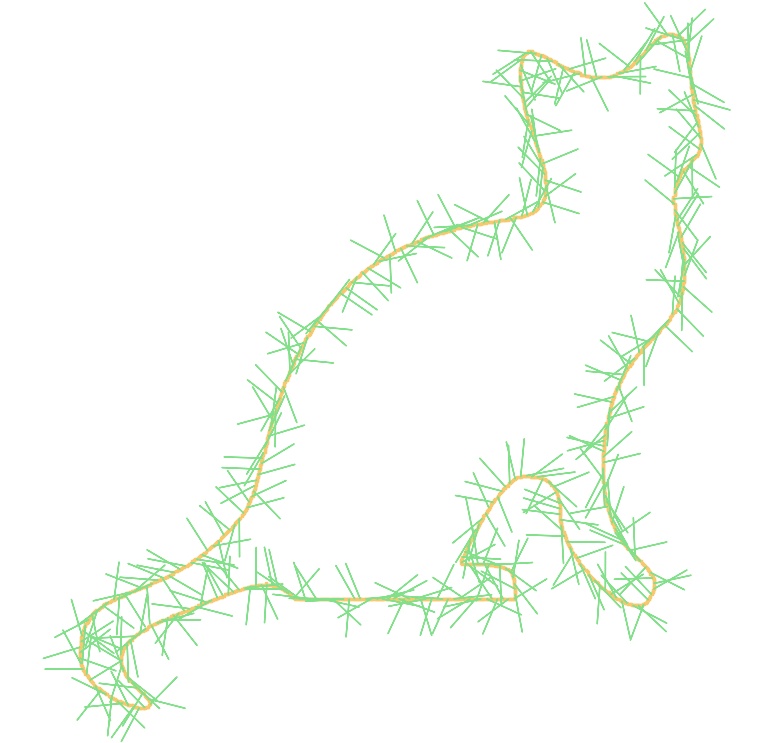}}
    \boxed{
    \includegraphics[width=0.9in]{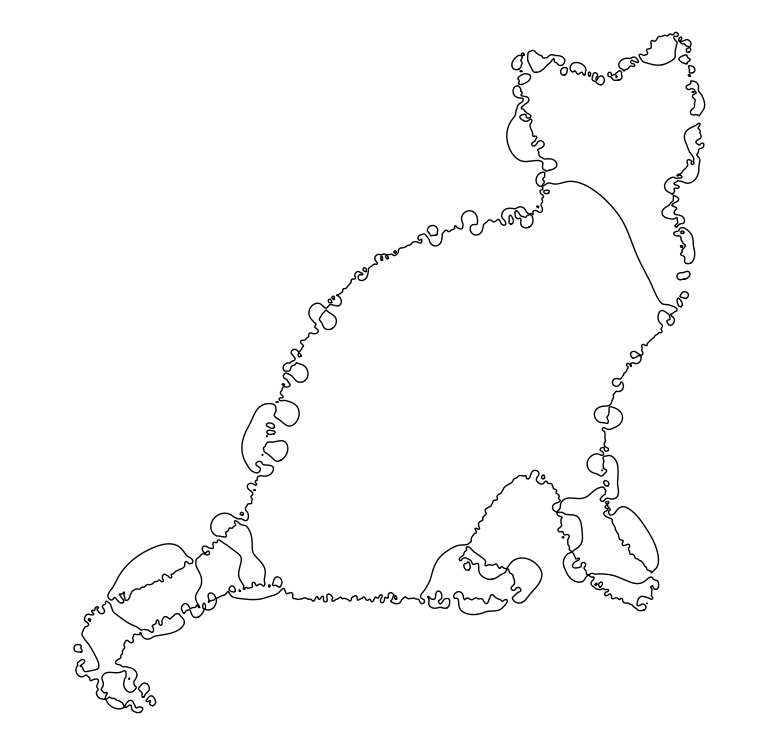}
    \includegraphics[width=0.9in]{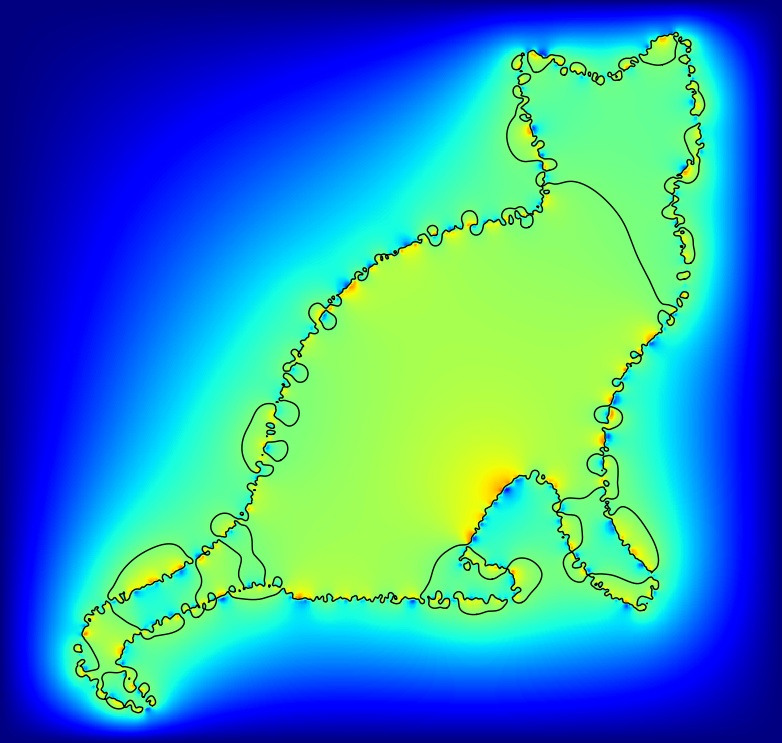}
    \includegraphics[width=0.9in]{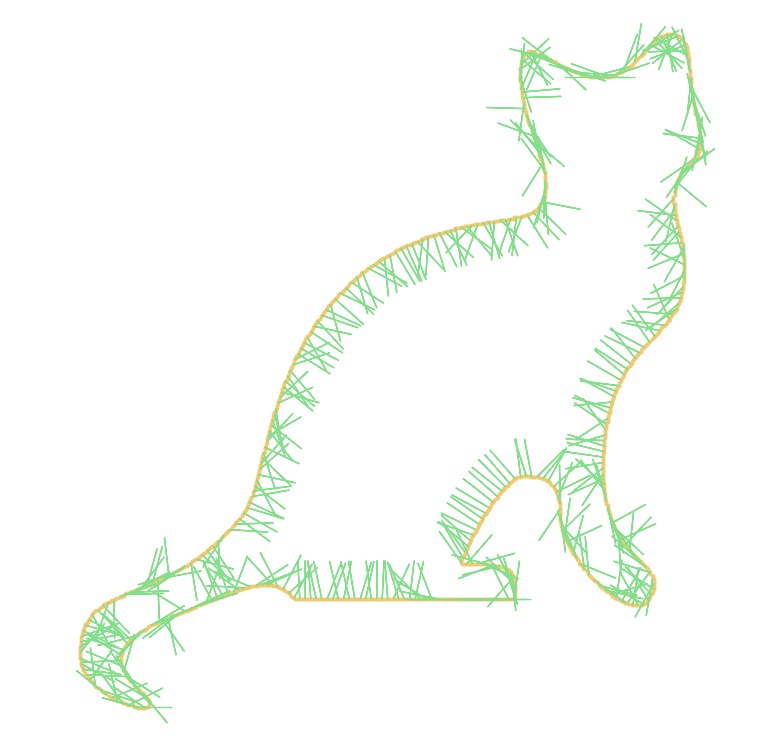}}
    \boxed{
    \includegraphics[width=0.9in]{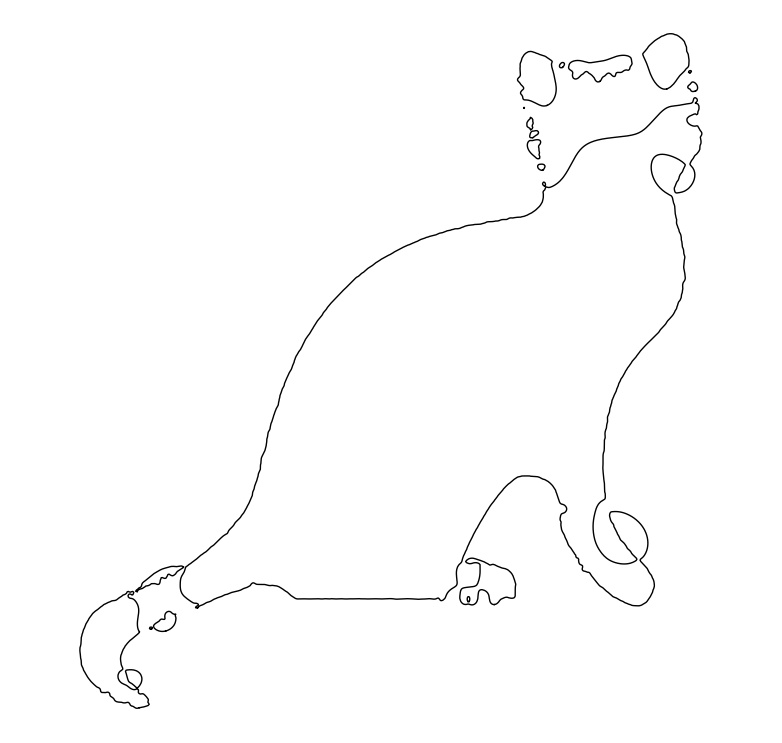}
    \includegraphics[width=0.9in]{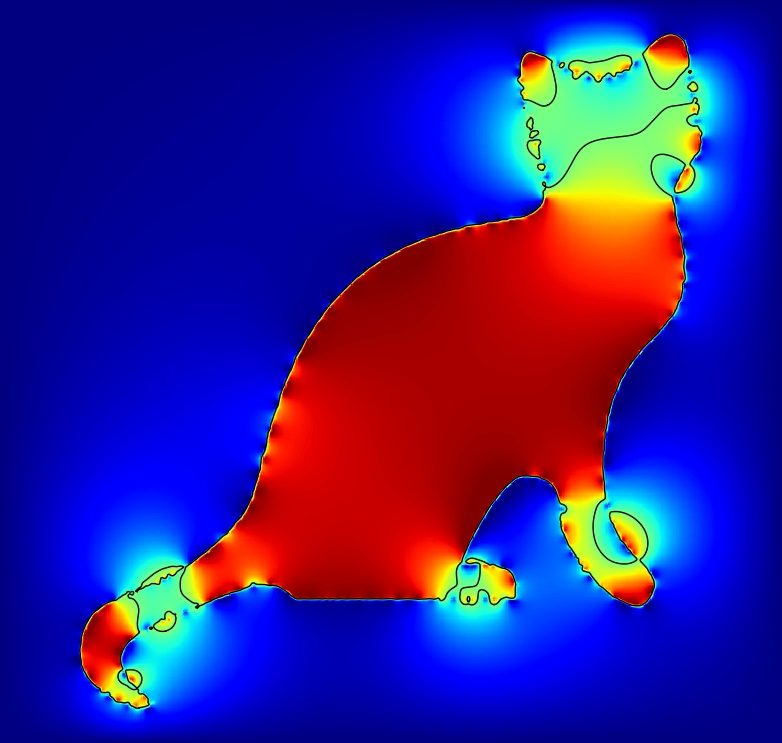}
    \includegraphics[width=0.9in]{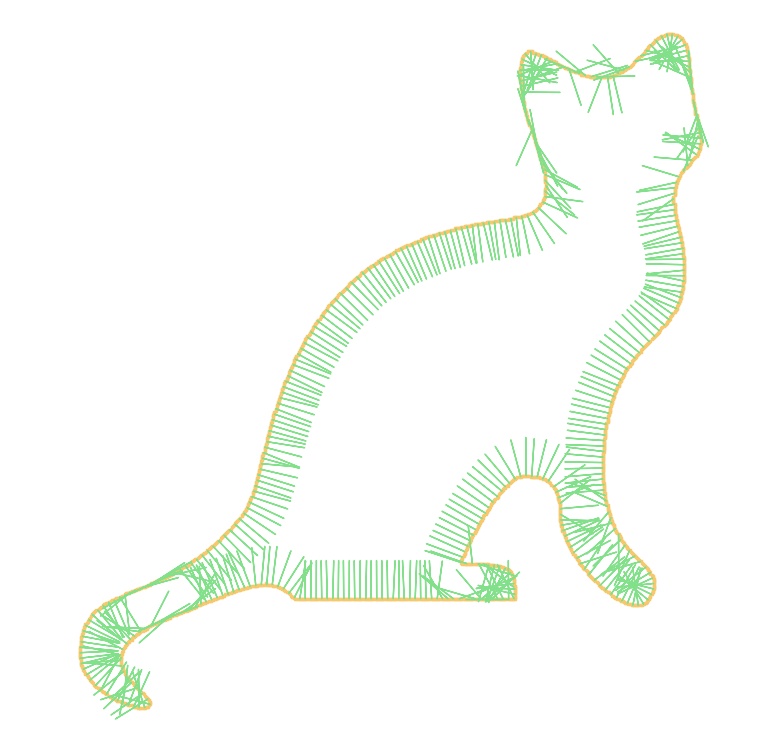}
    }\\
    \makebox[1.0in]{Random normals}
    \makebox[2.8in]{Iteration 1}
    \makebox[2.8in]{Iteration 2}\\
    \boxed{
    \includegraphics[width=0.9in]{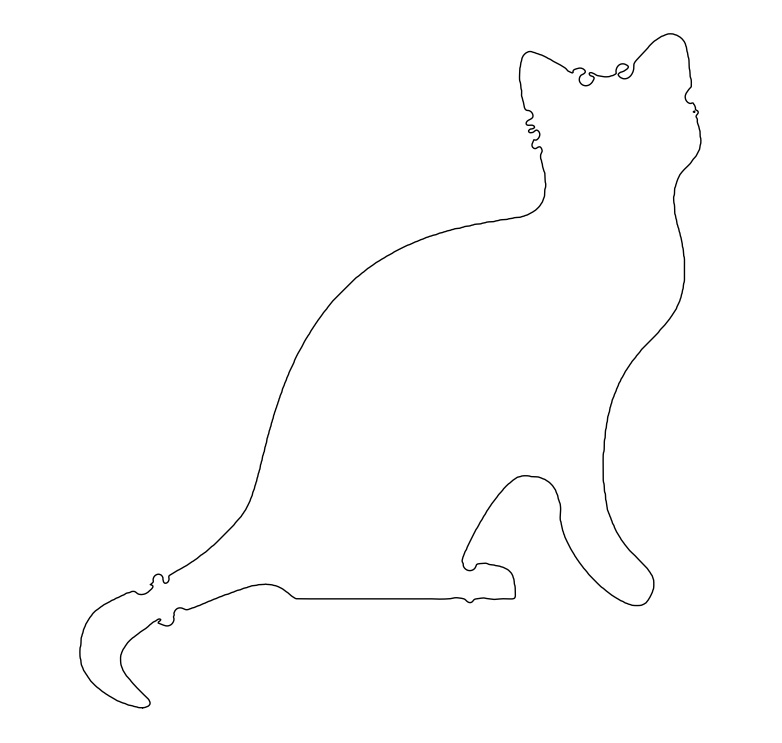}
    \includegraphics[width=0.9in]{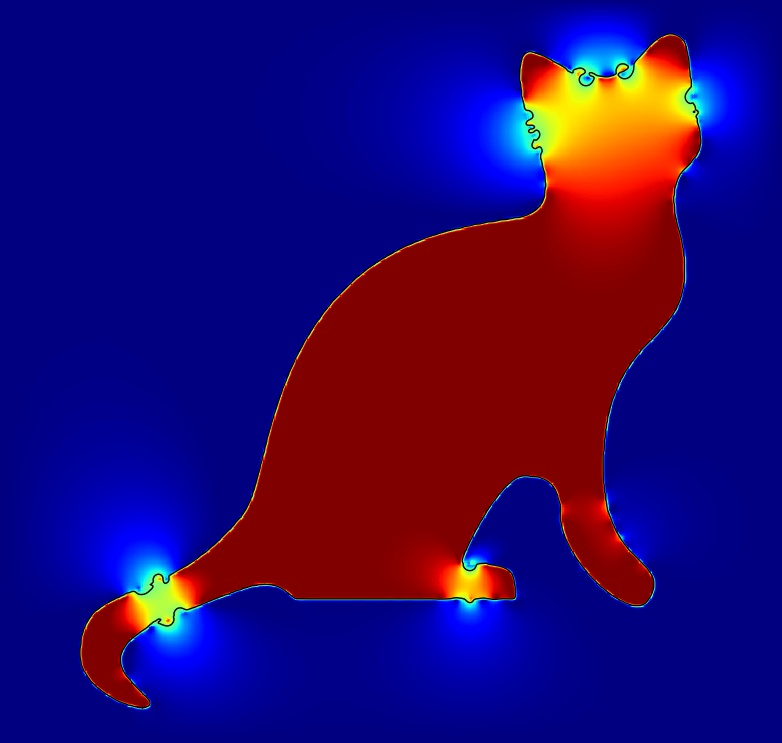}
    \includegraphics[width=0.9in]{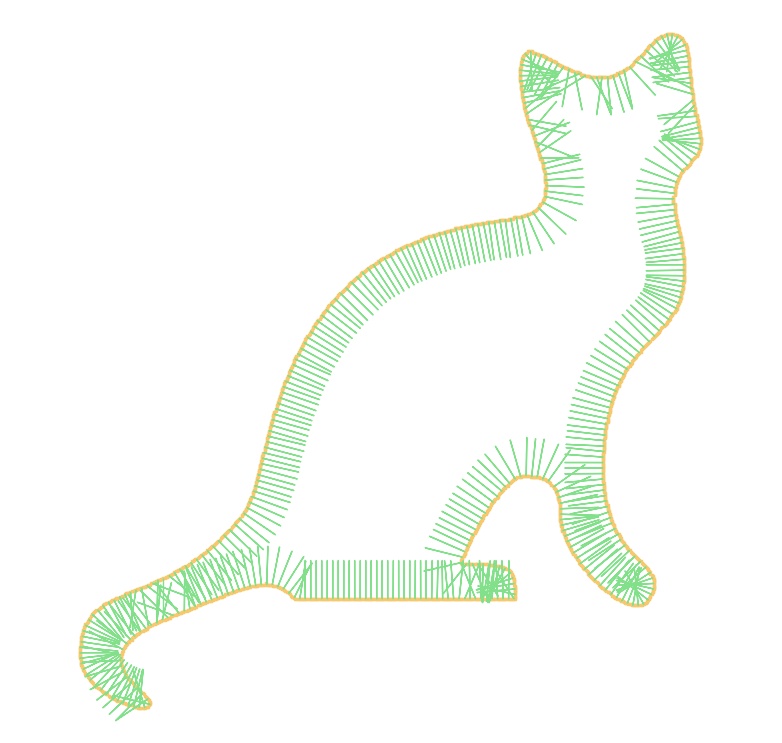}
    }
    \boxed{
    \includegraphics[width=0.9in]{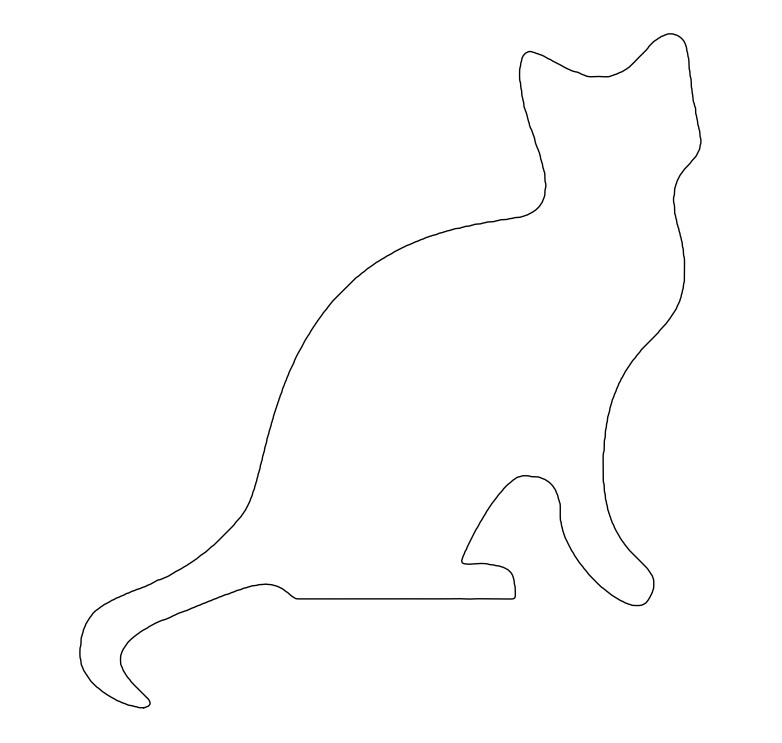}
    \includegraphics[width=0.9in]{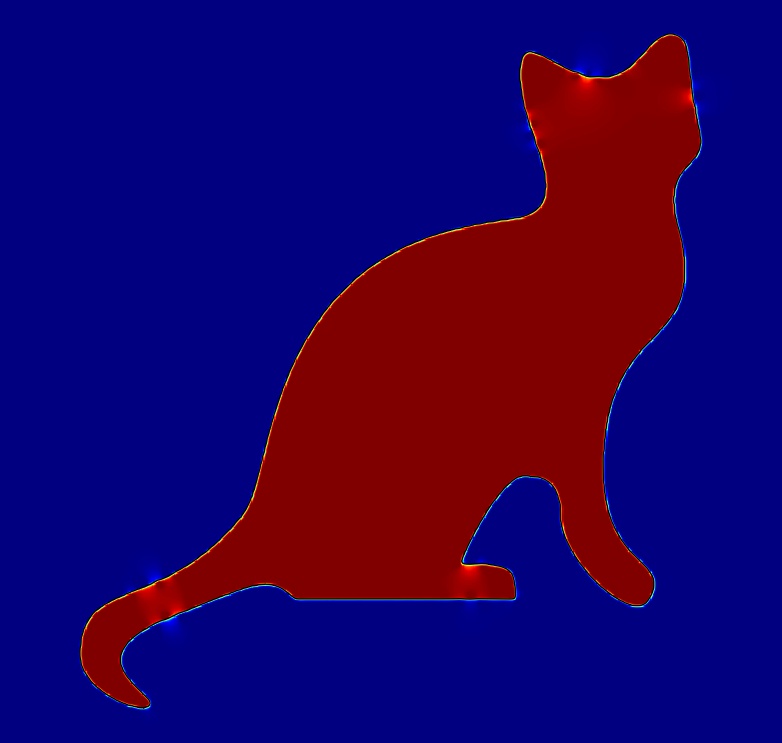}
    \includegraphics[width=0.9in]{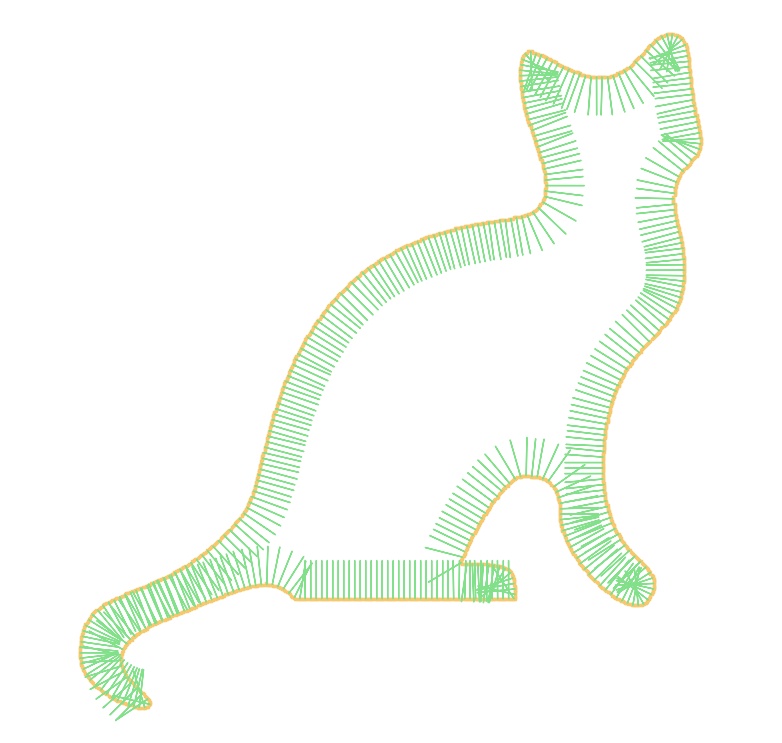}
    }
    \boxed{
    \includegraphics[width=0.9in]{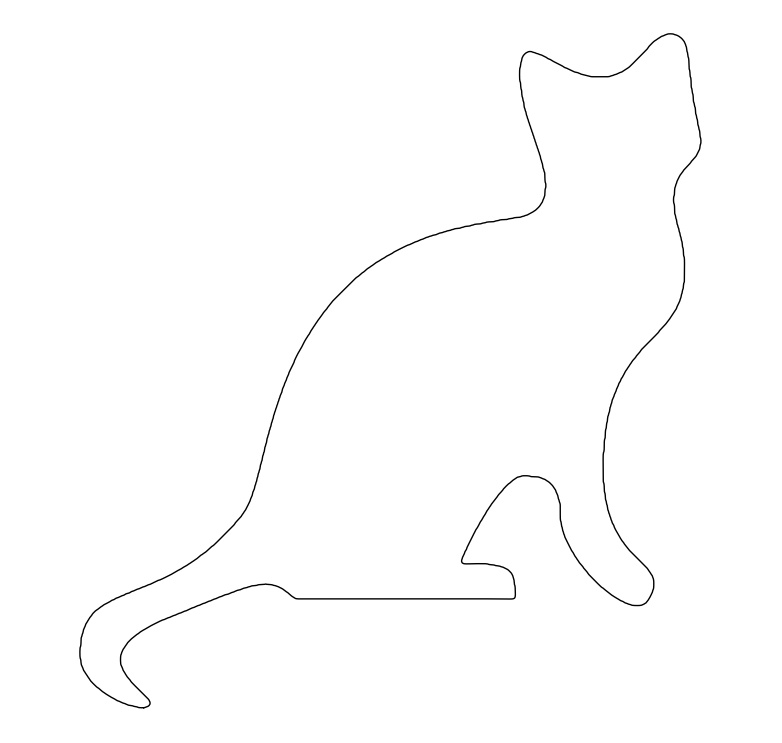}}\\
    \makebox[2.8in]{Iteration 3}
    \makebox[2.8in]{Iteration 4}
    \makebox[1.0in]{Output}
\caption{Illustration of the algorithmic pipeline on a 2D model.
In each iteration, we show the color-coded indicator function, the extracted iso-curve and the updated normals (only part of the normals are shown to avoid visual clutter). }
\label{fig:cat}
\end{figure*}

\begin{algorithm}
\SetAlgoLined
\SetKwInOut{Input}{input}
\SetKwInOut{Output}{output}
\Input{Unoriented points $\mathcal{P}=\{p_1,p_2,\ldots,p_m$\}, the maximum octree depth $D$, the convergence  threshold $\delta$ and the screened PSR weight $\alpha$}
\Output{A watertight surface approximating the points}
\BlankLine
construct an octree of depth $D$ to discretize $\mathcal{P}$\;
$\mathcal{S}\leftarrow$ nodes of the octree\;
$n\leftarrow |\mathcal{S}|$\;
construct a kd-tree for $\mathcal{S}$ for nearest sample searching\;
initialize normal for each sample $s_i\in \mathcal{S}$ randomly\;

\While{not convergence or exceeding maximum number of iterations}
{
    compute indicator function $\chi$ by applying screened PSR with parameter $\alpha$ to $\mathcal{S}$\;
    extract the iso-surface $\mathcal{F}$ with iso-value $\frac{1}{n}\sum_{j=1}^n\chi(s_j)$ by marching cubes\; 
    \For{each sample $s_j\in\mathcal{S}$}
    {
    $s_j\mathrm{.face\_list}\leftarrow \emptyset$
    }
    \For{each triangular face $f_j\in\mathcal{F}$}
    {
        $\vec{n}_j\leftarrow f_j$'s inward normal, i.e., towards interior of the component $f_j$ belonging to\;
        $a_j\leftarrow $ the area of triangle $f_j$\;
        use kd-tree to find top-$k$ samples in $\mathcal{S}$ that are closest to $f_j$\;
        add $f_j$ to the associated face list of each of the $k$ samples\;
    }
    \For{each sample $s_j\in\mathcal{S}$}
    {
        $\vec{n}(s_j)\leftarrow \sum_{i=1}^{m_j}a_{j_i}\vec{n}_{j_i};$~~$/\ast$ $m_j=s_j.\textrm{face\_list.count()}$~~$\ast/$\\
        normalize $\vec{n}(s_j)$\;
    }
    $d\leftarrow $ the average of top $0.1\%$ normal difference of samples between the previous and the current iterations\;
    \If{$d<\delta$}
    {
    $convergence \leftarrow true$
    }
}
apply screened PSR to the sample points $\{s_i\}$ with predicted normals $\vec{n}(s_i)$ and return the iso-surface $\mathcal{F}$ extracted from the computed indicator function\;
\caption{\label{alg:ipsr}iPSR}
\end{algorithm}

\begin{figure*}[!htbp]
    \centering
    \includegraphics[width=0.1625\linewidth]{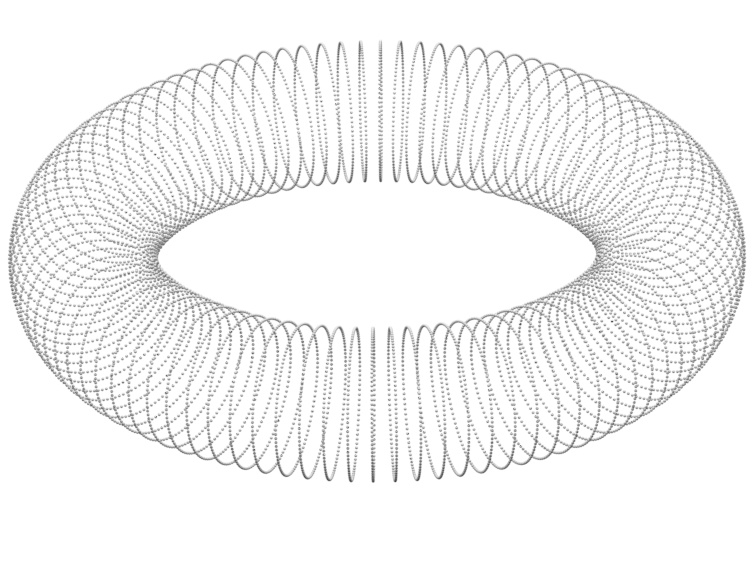}
    \includegraphics[width=0.1625\linewidth]{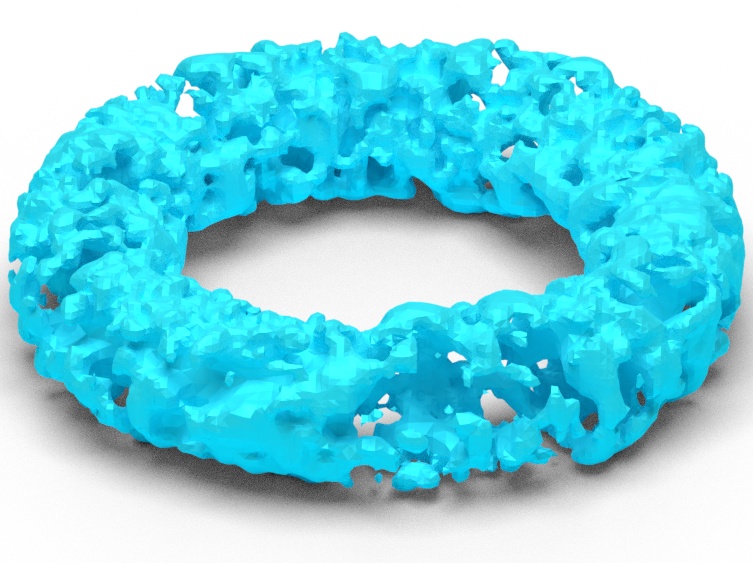}
    \includegraphics[width=0.1625\linewidth]{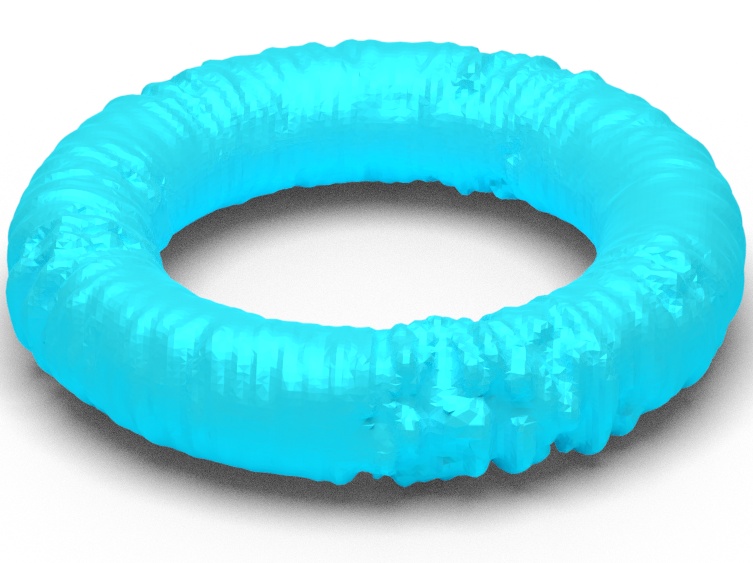}
    \includegraphics[width=0.1625\linewidth]{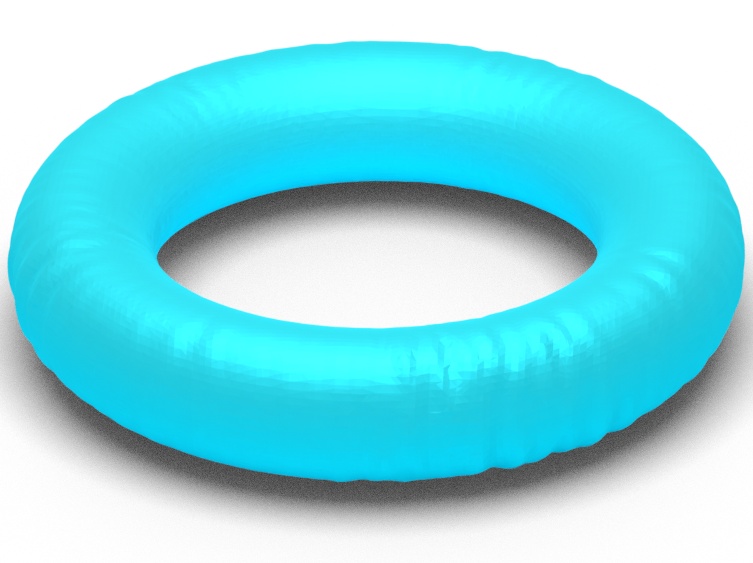}
    \includegraphics[width=0.1625\linewidth]{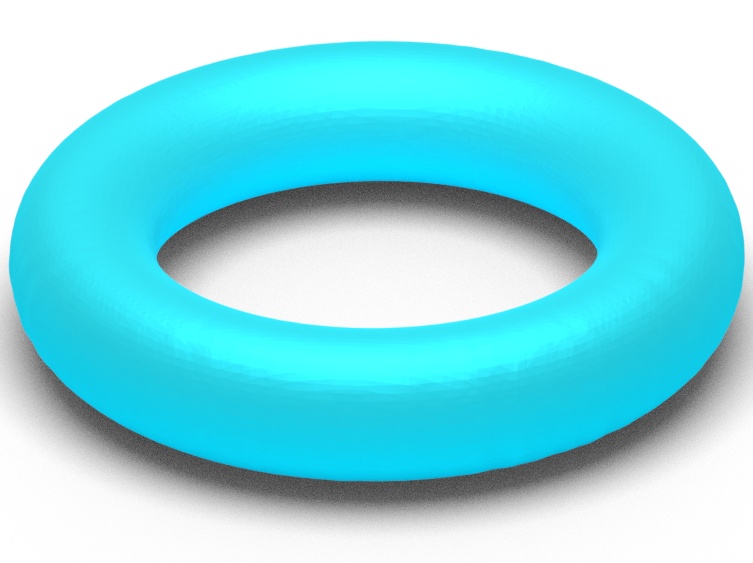}
    \includegraphics[width=0.1625\linewidth]{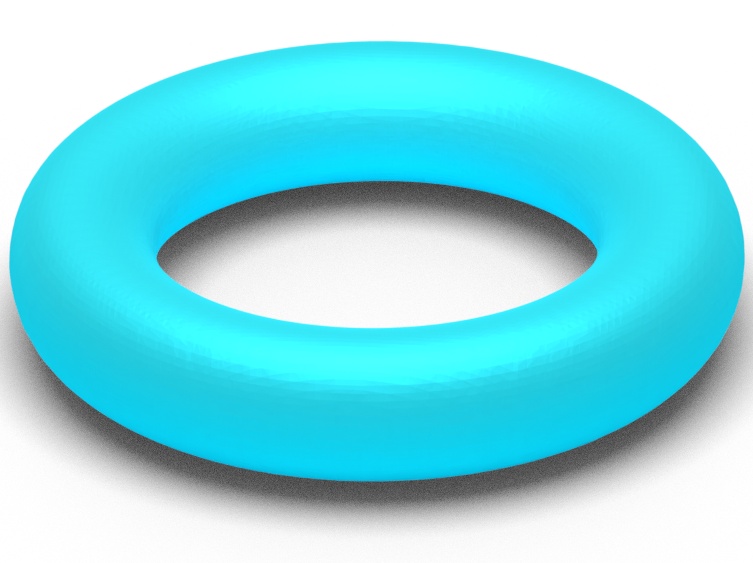}\\
    \includegraphics[width=0.1392\linewidth]{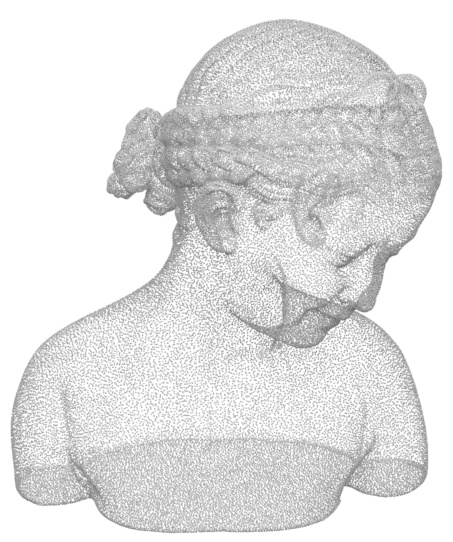}
    \includegraphics[width=0.1392\linewidth]{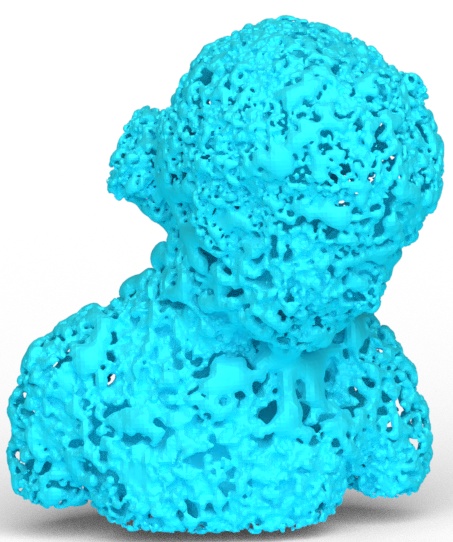}
    \includegraphics[width=0.1392\linewidth]{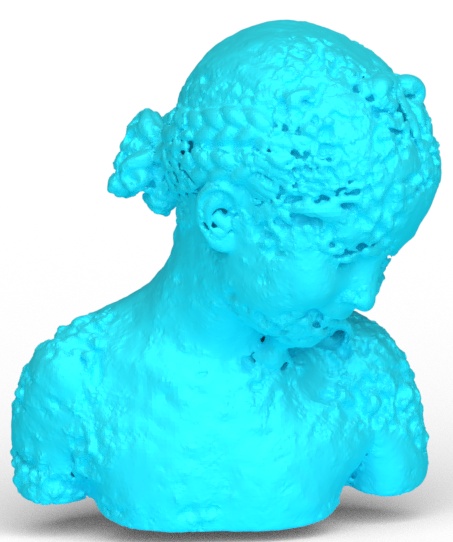}
    \includegraphics[width=0.1392\linewidth]{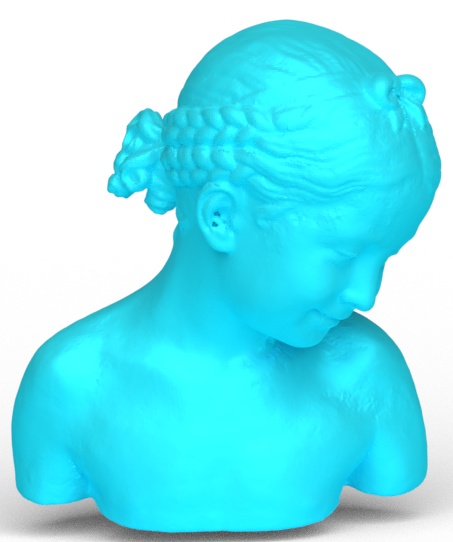}
    \includegraphics[width=0.1392\linewidth]{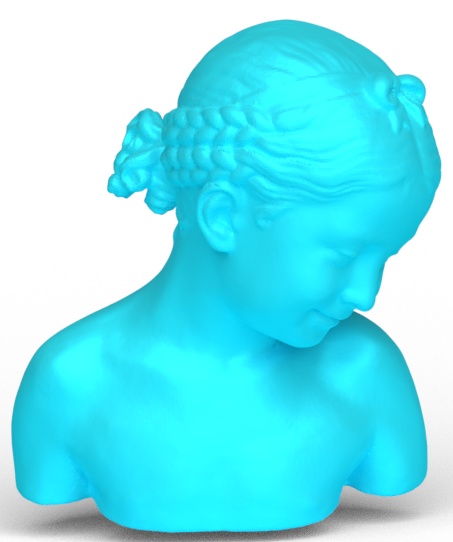}
    \includegraphics[width=0.1392\linewidth]{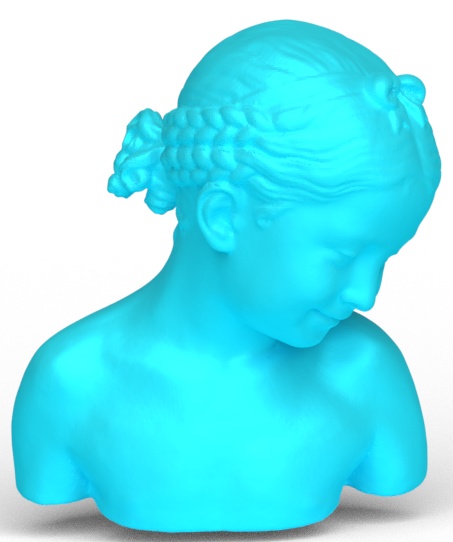}
     \includegraphics[width=0.1392\linewidth]{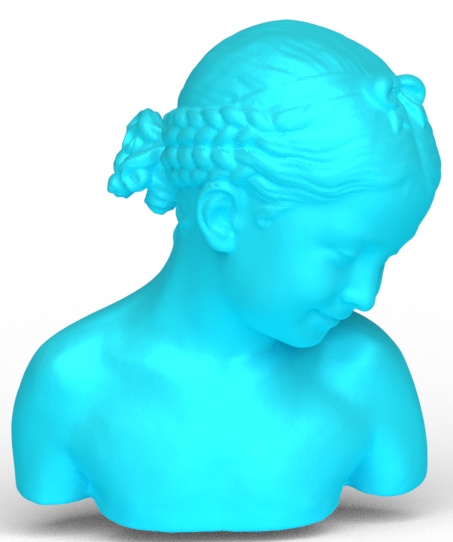}\\
    \includegraphics[width=0.106\linewidth]{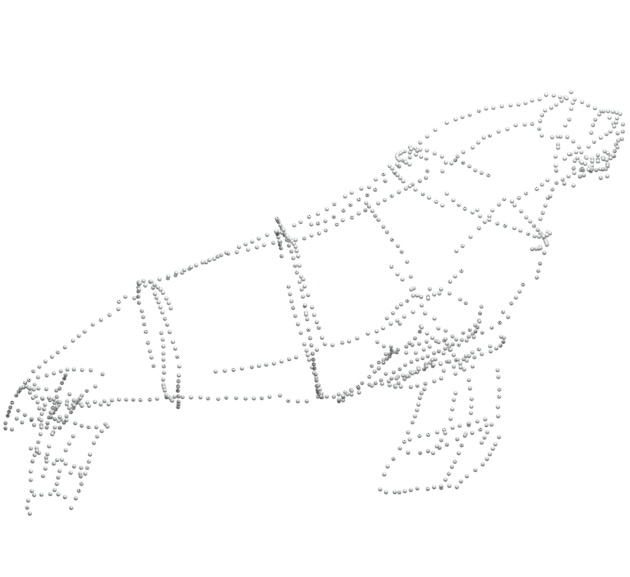}
    \includegraphics[width=0.106\linewidth]{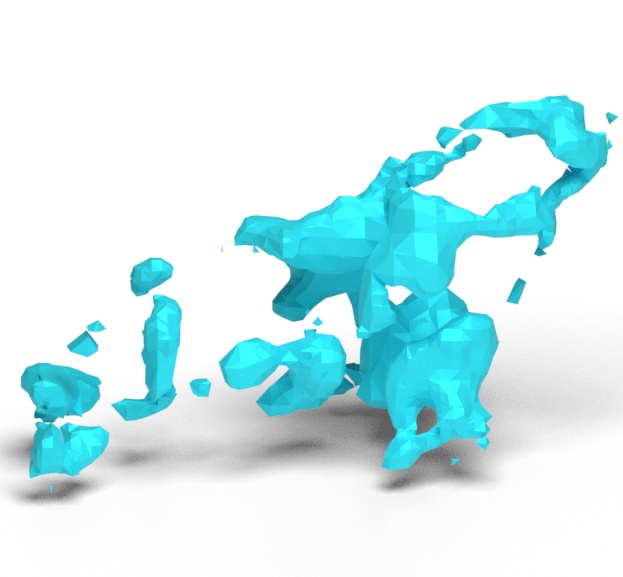}
    \includegraphics[width=0.106\linewidth]{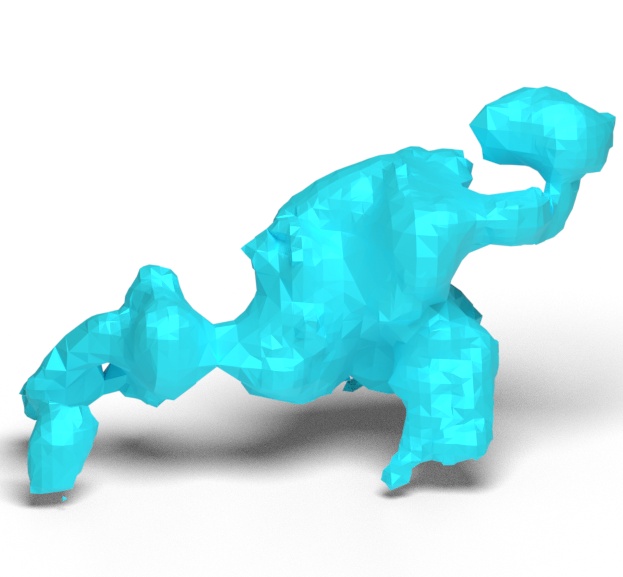}
    \includegraphics[width=0.106\linewidth]{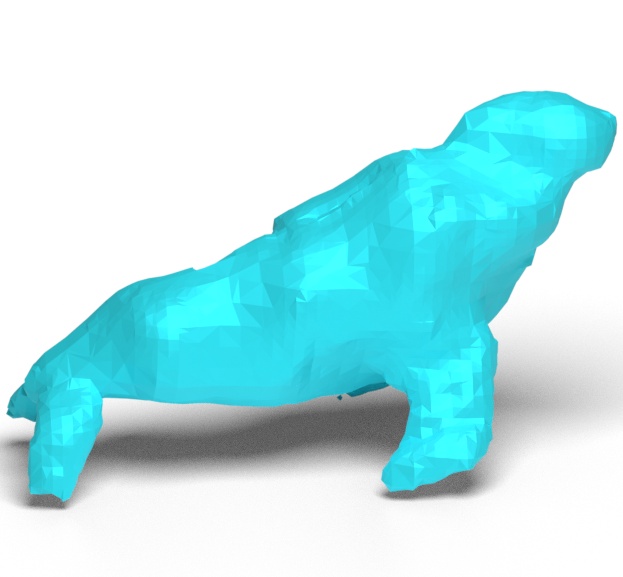}
    \includegraphics[width=0.106\linewidth]{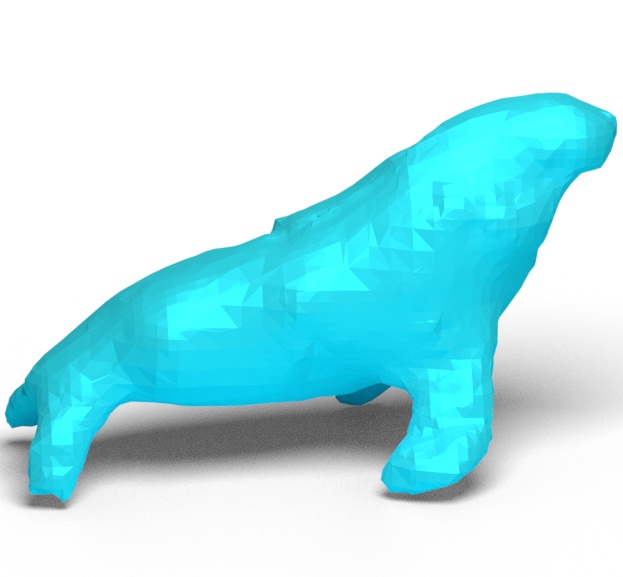}
    \includegraphics[width=0.106\linewidth]{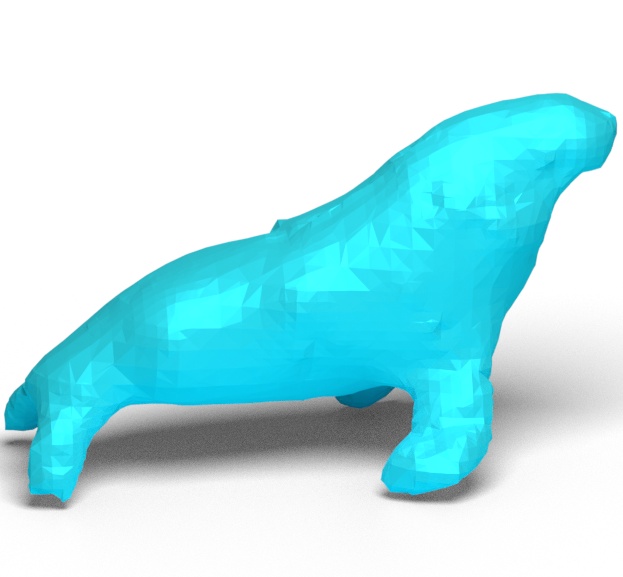}
    \includegraphics[width=0.106\linewidth]{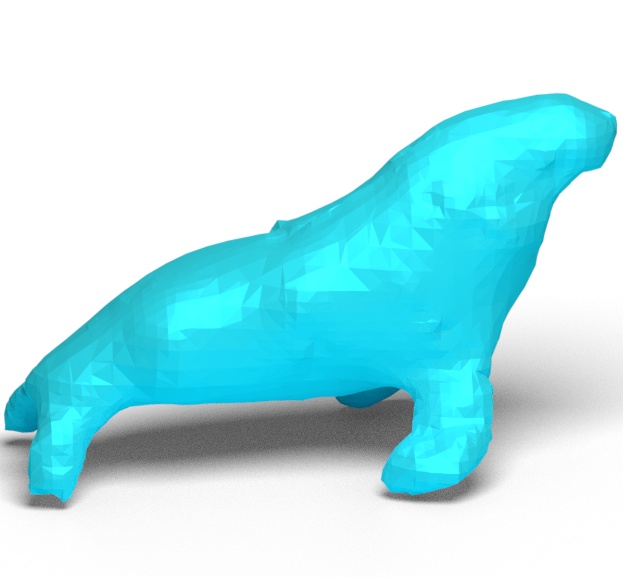}
    \includegraphics[width=0.106\linewidth]{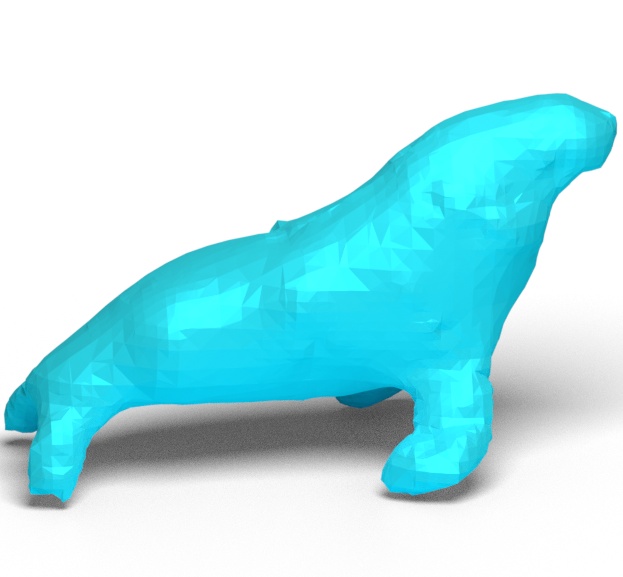}
    \includegraphics[width=0.106\linewidth]{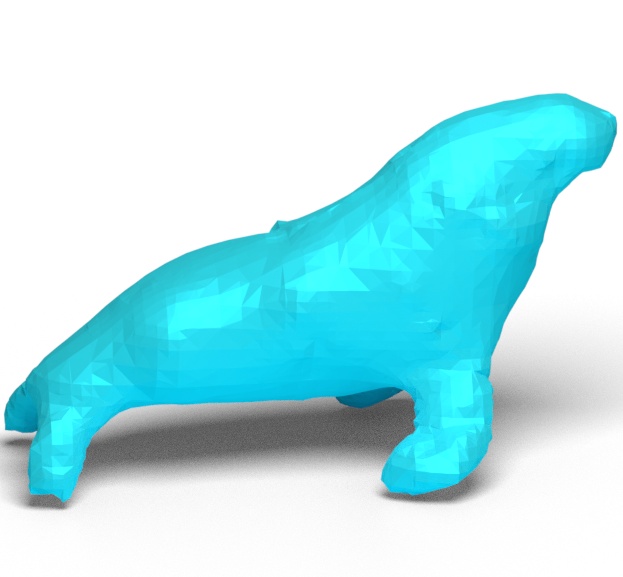}\\
    \includegraphics[width=0.106\linewidth]{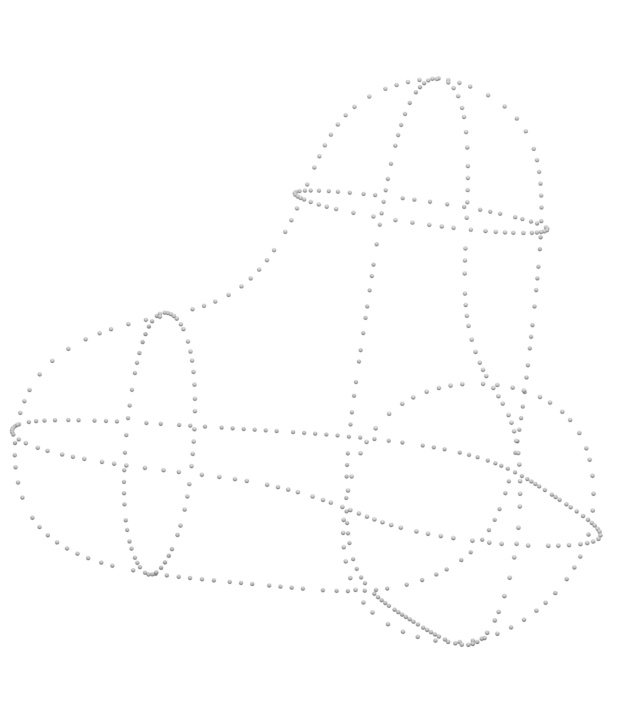}
    \includegraphics[width=0.106\linewidth]{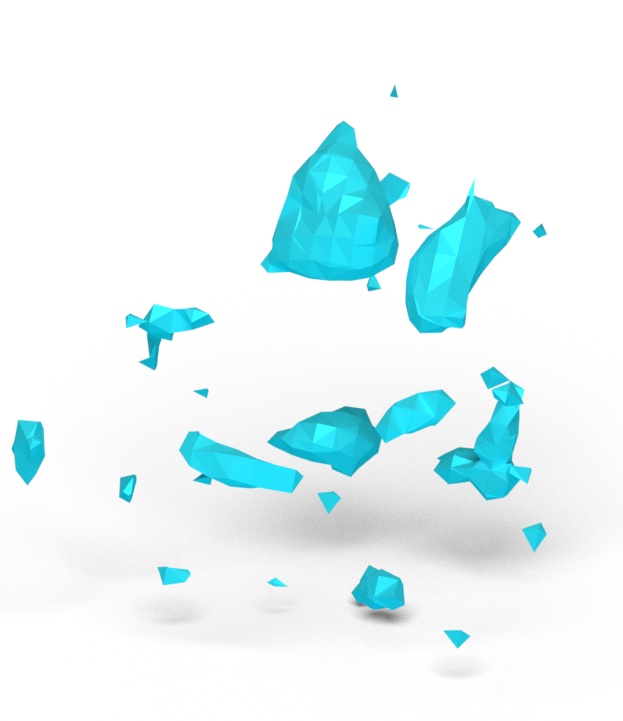}
    \includegraphics[width=0.106\linewidth]{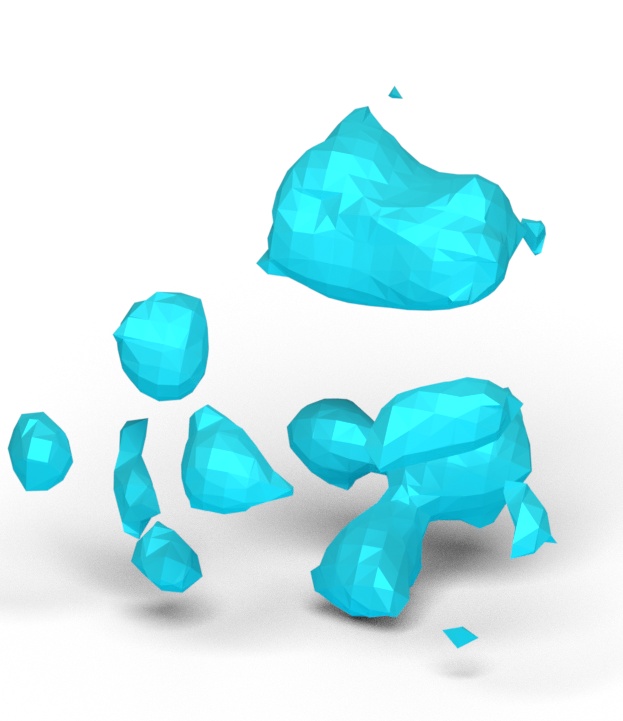}
    \includegraphics[width=0.106\linewidth]{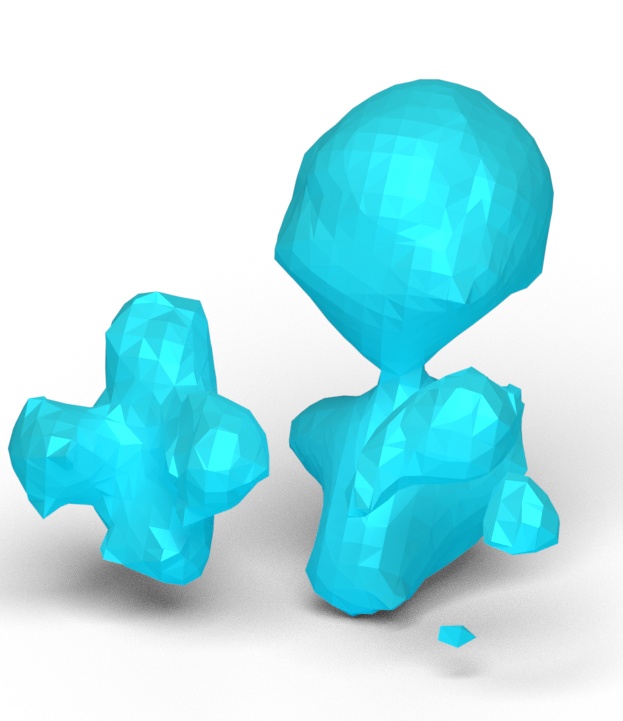}
    \includegraphics[width=0.106\linewidth]{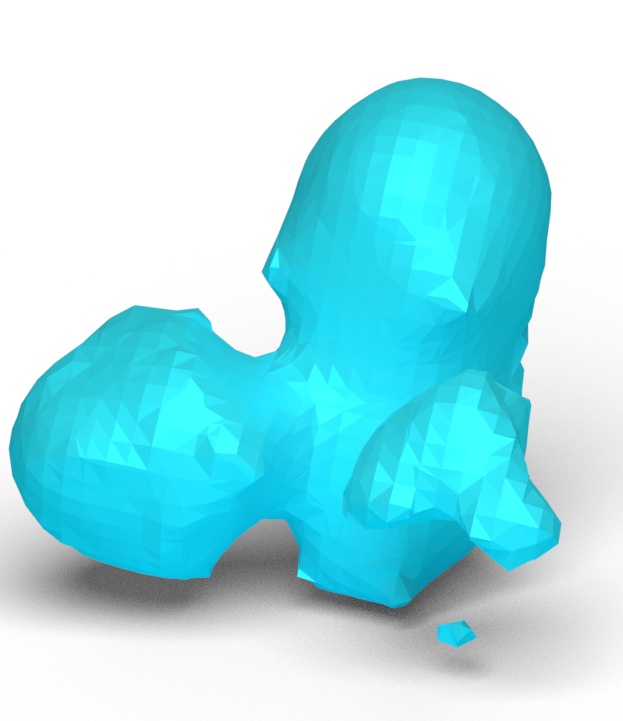}
    \includegraphics[width=0.106\linewidth]{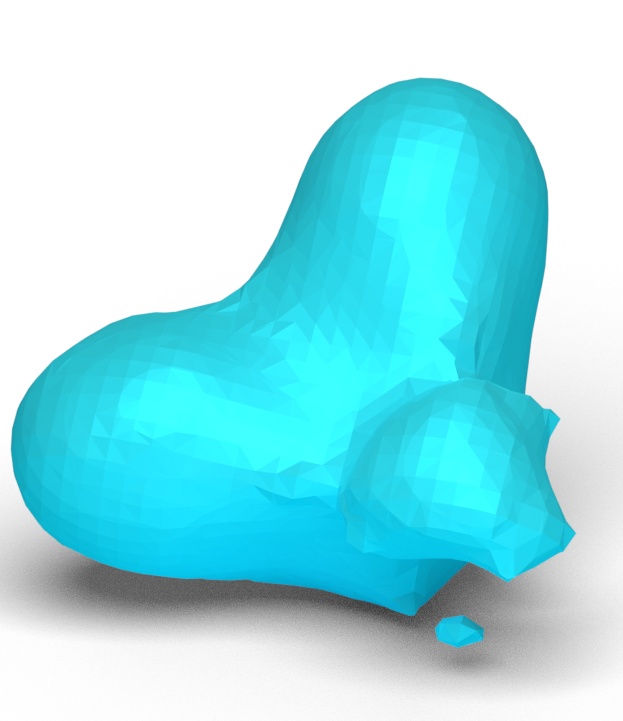}
    \includegraphics[width=0.106\linewidth]{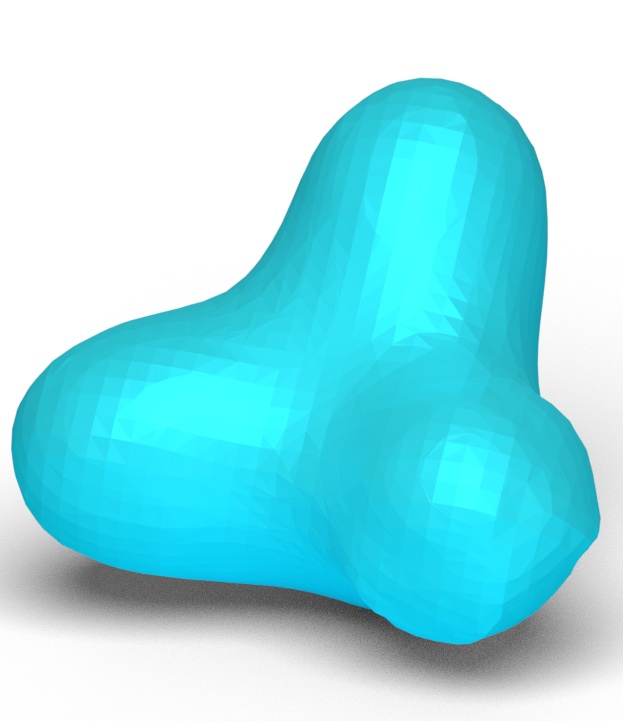}
    \includegraphics[width=0.106\linewidth]{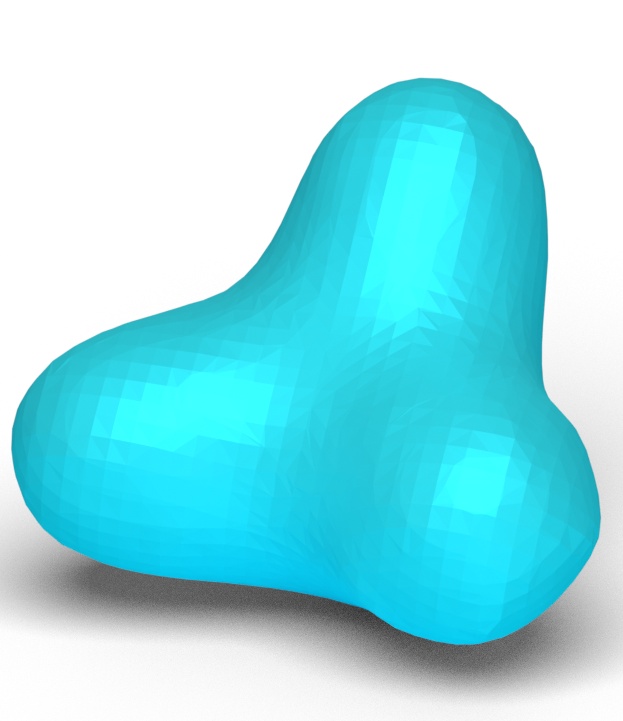}
    \includegraphics[width=0.106\linewidth]{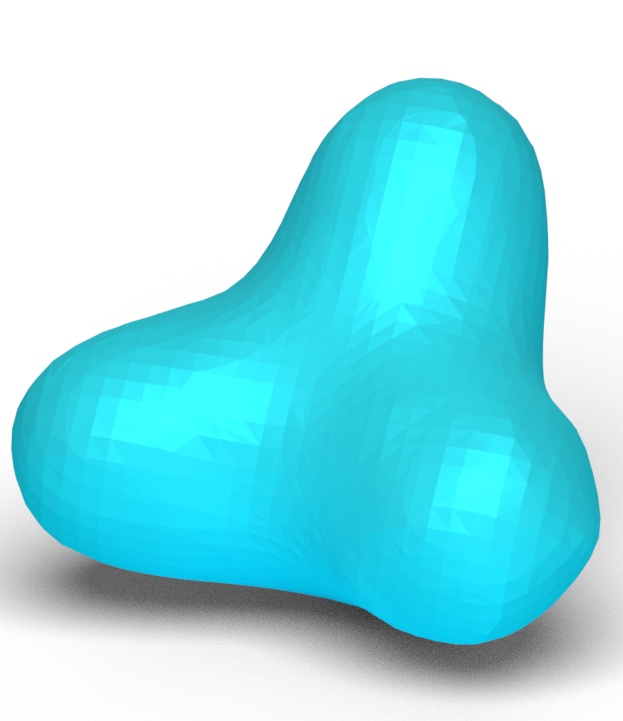}\\
    \caption{Given points with randomly assigned normals, our method iteratively applies screened PSR to reconstruct a watertight surface, from which it updates the normals of all points. The algorithm terminates when  normal changes are smaller than a threshold. It takes iPSR 4, 5, 7, and 7 iterations to converge on Torus, Bimba, Walrus and Trebol, respectively, so there are 5, 6, 8, and 8 reconstructed surfaces shown here. See also the accompanying video for animation. }
    \label{fig:Convergence_procedure}
\end{figure*}

\section{Method}

\paragraph{Motivation.}
Given sample points with incorrect normals, PSR produces an iso-surface  $\mathcal{F}$ which is usually highly twisted and has multiple connected components. Each of the connected components is a closed surface.  Figure~\ref{fig:Convergence_procedure} (column 2) shows a few such examples. Let $s$ be an arbitrary sample point. Consider a ray which is from the interior of solid $M$ to infinity and is perpendicular to $\partial M$ at $s$ so that the ray may intersect the iso-surface $\mathcal{F}$ at locations near $s$. The key observation is that if the starting point of the ray is also inside the iso-surface $\mathcal{F}$, the ray crosses $\mathcal{F}$ an odd \footnote{If the ray is tangential to an iso-surface, we count two intersection points.} number of times, say $2n+1$, no matter how many connected components $\mathcal{F}$ has. Since all connected components of the iso-surface $\mathcal{F}$ are \textit{closed} surfaces, we can compute the \textit{inward} normal for each intersection point, i.e., the normal is towards the interior of the component the point belonging to. Among the intersections, there are $n+1$ points whose normals of the iso-surface $\mathcal{F}$ are towards the interior of solid $M$. For the rest, their normals are pointing to the exterior of $M$. Assume the normals are almost collinear. Then, averaging the normals of the $2n+1$ intersections yields an inward normal of $M$.

Inspired by this observation, we adopt the following strategy to update normals. We call the local neighborhood of an iso-surface around an intersection point a ``layer''. Notice that the intersections are located at an odd number of layers of the iso-surface produced by PSR. Therefore, one could use the representative points on each layer to correct normals instead of explicitly finding the intersection points. 

Let the iso-surface $\mathcal{F}$ be discretized by marching cubes~\cite{Lorensen1987}.
Assume the discretization resolution is sufficiently high so that the resulting triangle mesh preserves the topology of the layered structures, i.e., there are mesh vertices sampled on each layer. For each triangular face of the discretized iso-surface, we find $k$ sample points that are nearest to it. Going through all the faces, we can connect the input points with the triangles of $\mathcal{F}$. As a result, we can use the triangular faces associated to each sample point as the representatives of the layered structure around it. 
In particular, for each sample $s$, we compute the \textbf{area-weighted} sum of the normals of its associated triangular faces. Also, we can view each layer as a plane whose normal represents the layer's normal. Thus, the resulting average normal can be viewed as the average of the normals of all the layers around $s$. If there are an odd number of layers around sample $s$, the average of layer normals is towards the interior of solid $M$. This implies that the parity of the number of associated representative triangular faces to each sample $s$ does not matter. As long as these faces are located on the layers, they can be used to represent the layered structure so that the area-weighted averaging yields a vector approximating the average of the normals of all layers. Therefore, the average of layer normals is conceptually equivalent to the average of the normals of the $2n+1$ intersection points along the ray from interior of $M$ to infinity (which is mentioned at the beginning of Section 4).

We call the local neighborhood of a sample $s$ ``odd-layered'' (resp. ``even-layered'') if there are an odd (resp. even) number of layers of the iso-surface around $s$. When the normals for points around $s$ are perturbed, the indicator function usually has oscillations, leading to small, fragmented iso-surfaces in the local neighborhood of $s$. As a result, a ray starting from an exterior point of the iso-surface $\mathcal{F}$, crossing $s$ and travelling to infinity produces an even-layered structure.
We observe that even with randomly initialized normals, the iso-surface computed in the first iteration has a large number of odd-layered structures. Moreover, an even-layered structure can turn into odd-layered in the future iterations, but not the other way around. As a result, through the iterative procedure, more and more samples exhibit an odd-layered structure around them, and our area-weighted normal averaging strategy can make the normals more and more accurate. See Figure~\ref{fig:2d_illustration} for a 2D illustration of the normal updating strategy.

\paragraph{Overview.}
Our method is a fairly straightforward realization of the aforementioned normal averaging strategy detailed in Algorithm~\ref{alg:ipsr}. Given a set of unoriented points $\mathcal{P}$ as input, we first construct an octree with maximum depth $D$ (specified by the user) and use the octree nodes as the sample set $\mathcal{S}$. To facilitate sample search, we also construct a
kd-tree for the samples. We initially assign each sample a random normal vector, before our method proceeds in an iterative manner. In each iteration, we apply the screened PSR to the sample set $\mathcal{S}$ with the current normals and obtain an indicator function $\chi$. We apply the octree-oriented marching cube~\cite{Kazhdan2006,Kazhdan2013} to extract the iso-surface $\mathcal{F}$ with iso-value $\frac{1}{n}\sum_{i=1}^n\chi(s_i)$, where $n=|\mathcal{S}|$ is the number of samples. Then we update the normal for each sample by averaging the iso-surface's inward normals of triangular faces associated with $s$. The algorithm continues until the point normals do not change any more. Figure~\ref{fig:cat} illustrates the entire pipeline using a 2D example, and Figure~\ref{fig:Convergence_procedure} shows the iterative results on 4 typical 3D models. We document more implementation details next.

\paragraph{Visibility-based initialization.} Although our algorithm converges with random initialization, in practice it can run faster if given a better initialization. Besides random initialization, we adopt a simple visibility based method~\cite{Katz2007} to estimate initial normals. Specifically, we scale the input points into a unit cube and create a concentric cube with edge length three. Then we set 26 viewpoints (camera positions) using the large cube: 8 are at the corners of the cube, 6 at the face centers, and 12 at the edge midpoints. From each viewpoint $v_i$, we apply the hidden point removal operator~\cite{Katz2007} to determine the samples that are visible from $v_i$. Then we initialize the normal for every sample by averaging the directions of all visible rays. If a sample $s_j$ is invisible to any of the viewpoints, we simply assign a fixed normal $[1,0,0]$. Katz et al.'s method is simple and highly efficient thanks to its linear time and space complexities. Experimental results on the Aim$@$Shape dataset show that the visibility initialization can reduce the number of iterations by 45\% on models with more than 100K points.

\paragraph{Normal processing.} We first apply the octree-oriented
marching cube~\cite{Kazhdan2006,Kazhdan2013} to extract the
iso-surface $\chi^0$, which may contain multiple connected components especially in the first few iterations, and obtain a triangulated mesh. Then we compute the normal vector for each face using a cross product. For each triangular face of the iso-surface, we find the top-$k$ samples that are closest to it. In this way, we build relationship between triangular faces and samples. Finally, we update the normal $\vec{n}(s_i)$ by computing the average of the normals of the faces which are linked with $s_i$. In our implementation, we empirically set $k=10$. See Section~\ref{sec:results} for discussions on the choice of $k$. 

\paragraph{Terminating condition.} We compute the normal variance between the samples of the current and the preceding iterations. The algorithm terminates when the average of normal variance is less than a user-specified threshold $\delta$.

\paragraph{A 2D toy example.}
Figure~\ref{fig:ellipse} illustrates the layered structure of the iso-surface using an ellipse. (a) Each sample point is initialized with a random normal. Among them, 49\% normals are inward, and the rest are either outward or tangential to the curve. (b) Applying screened PSR produces many disconnected closed curves in the first iteration. Let us examine 3 samples $A$, $B$ and $C$, whose local neighborhoods exhibit different layered structures. Sample $A$ is triple-layered, $B$ single-layered and $C$ double-layered. Using the average normals of representative points on the layers, we make $A$'s and $B$'s normals inward, while $C$'s normal is still outward. (c) The normal averaging strategy improves the normals effectively, and 78\% of normals become inward after the first iteration.
(d) Taking the updated normals as input, screened PSR produces a much-improved shape, which is already simply connected. All of the three representative samples $A$, $B$ and $C$ exhibit odd-layered structures.
(e) After iteration 2, 100\% normals are inward. (f) Applying screened PSR yields an almost correct ellipse. (g) shows the averaged normals of (f).

\begin{figure*}[!htbp]
    \centering
    \includegraphics[width=1.60in]{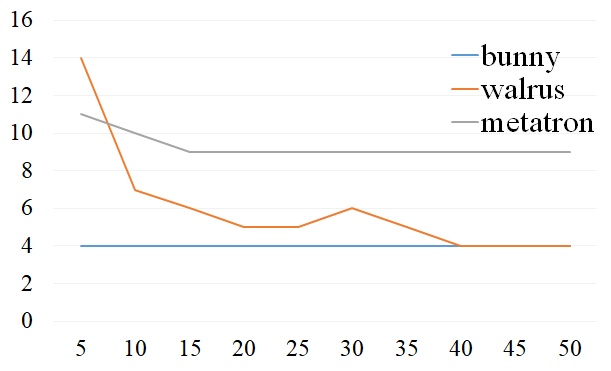}
    \includegraphics[width=1.60in]{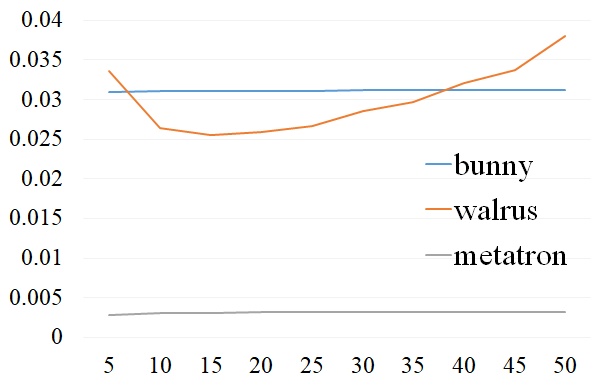}
    \includegraphics[width=1.60in]{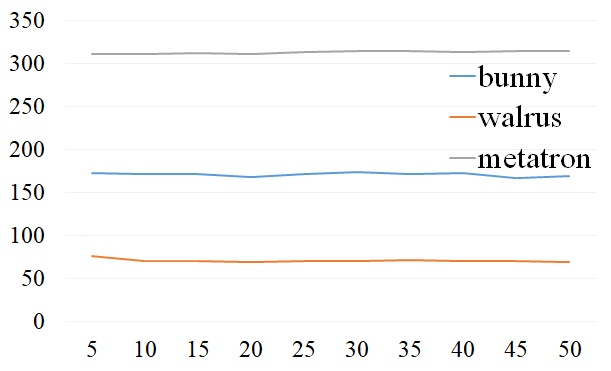}
    \includegraphics[width=1.60in]{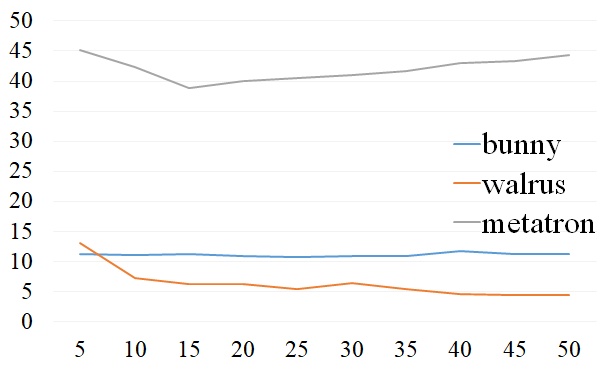}\\
    \includegraphics[width=1.60in]{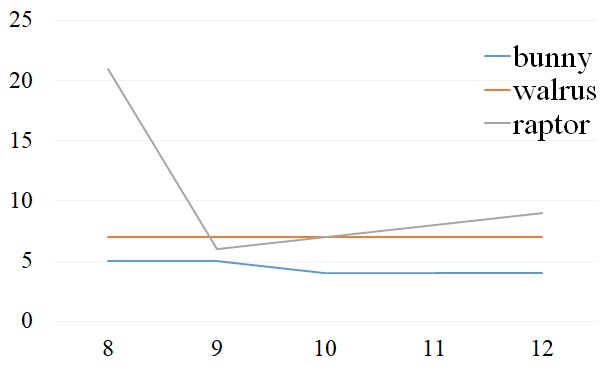}
    \includegraphics[width=1.60in]{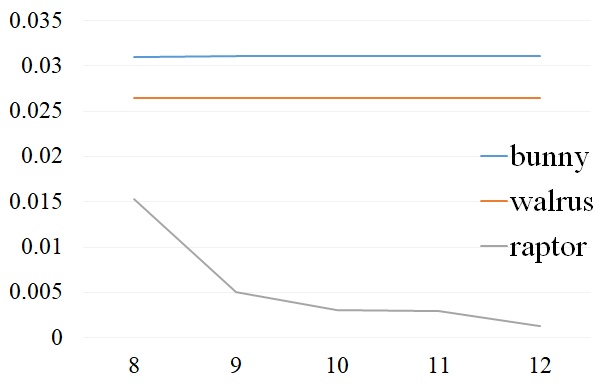}
    \includegraphics[width=1.60in]{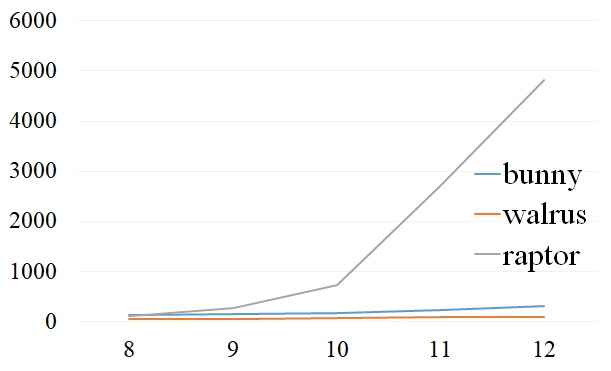}
    \includegraphics[width=1.60in]{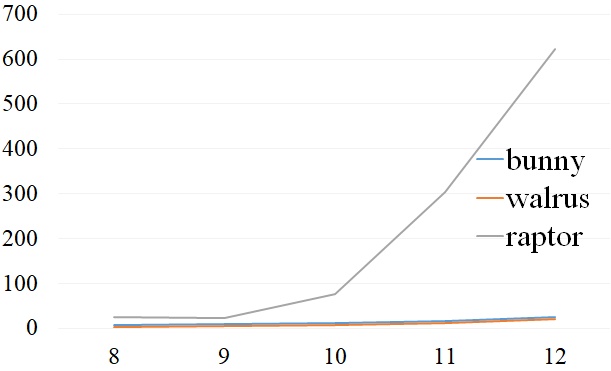}\\
    \makebox[1.60in]{(a) Number of iterations}
    \makebox[1.60in]{(b) Reconstruction error}
    \makebox[1.60in]{(c) Peak memory (MB)}
    \makebox[1.60in]{(d) Time (s)}\\
    \caption{Stress tests of neighborhood parameter $k$ and the maximum octree depth $D$. We examine $k\in[5,50]$ and $D\in[8,12]$ on 3 representative models. The vertical axes are the number of iterations, reconstruction error, peak memory and running time, and the horizontal axes are the values of $k$ (top) and $D$ (bottom).}
    \label{fig:various_neighbor}
\end{figure*}

\begin{figure}[!htbp]
    \centering
    \includegraphics[width=1.6in]{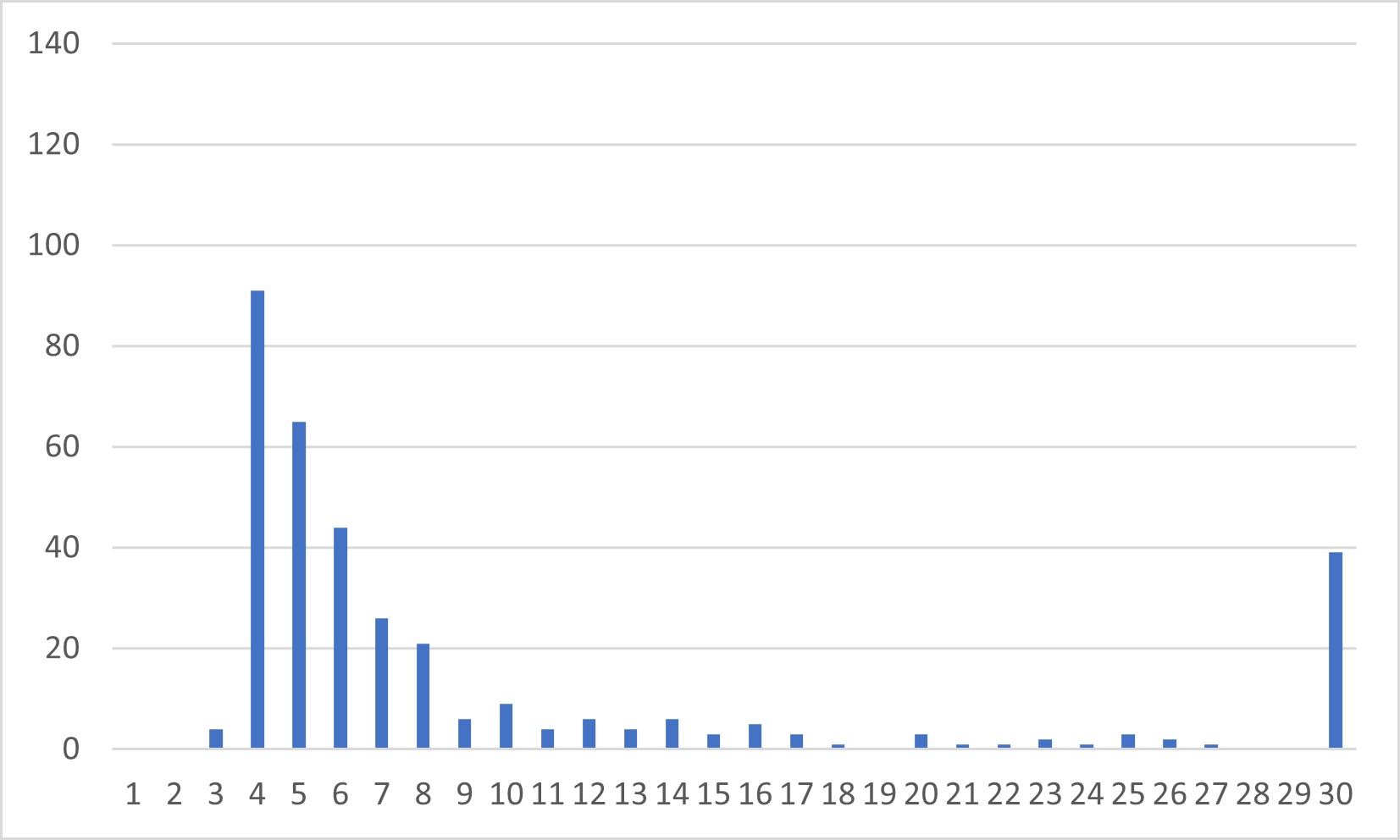}
    \includegraphics[width=1.6in]{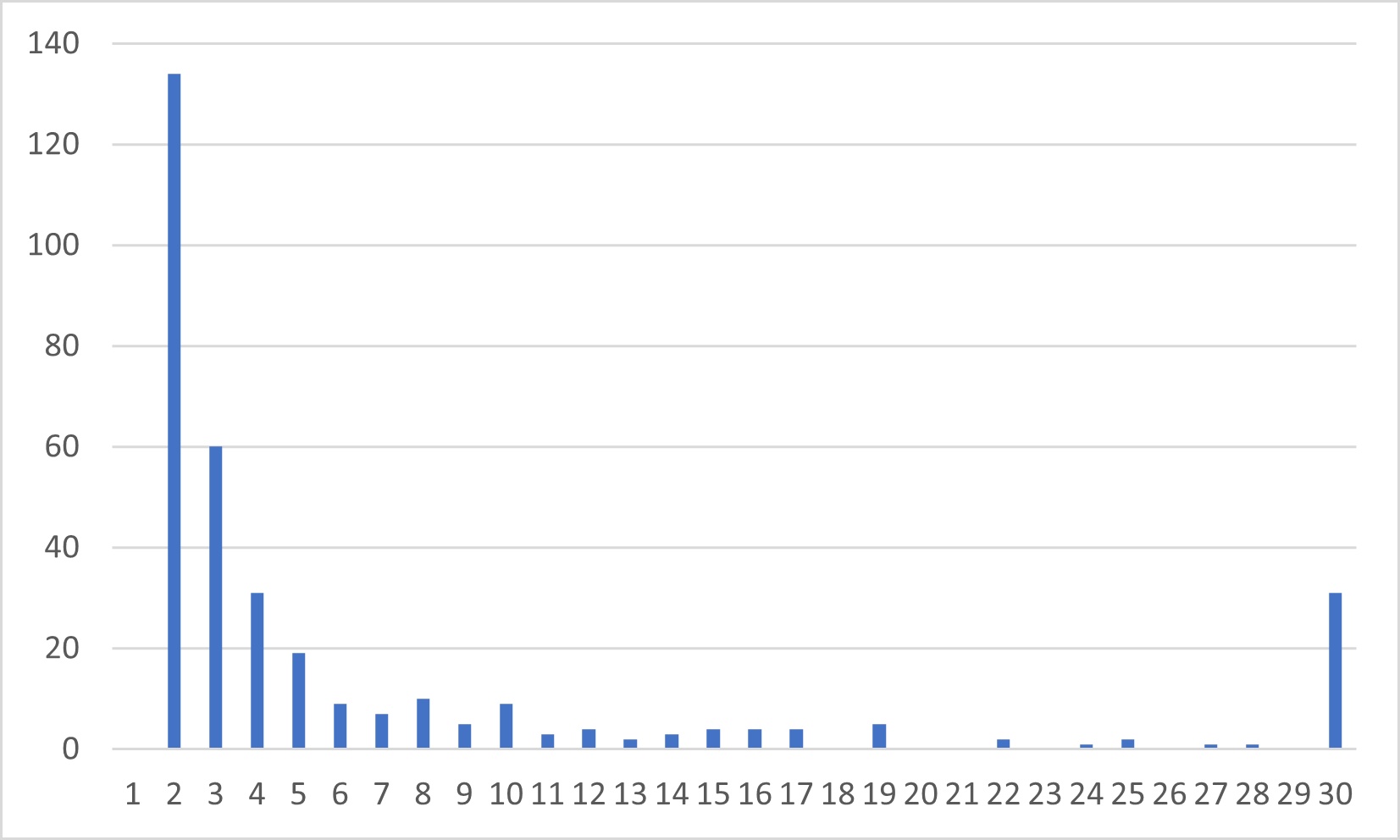}\\
    \makebox[1.6in]{Random initialization}
    \makebox[1.6in]{Visibility initialization}\\
    \caption{Histograms of the number of iterations on the AIM$@$SHAPE dataset. The horizontal axis is the number of iterations that iPSR takes.
    }
    \label{fig:iterations_histogram}
\end{figure}

\begin{figure}
    \centering
    \includegraphics[width=0.995\linewidth]{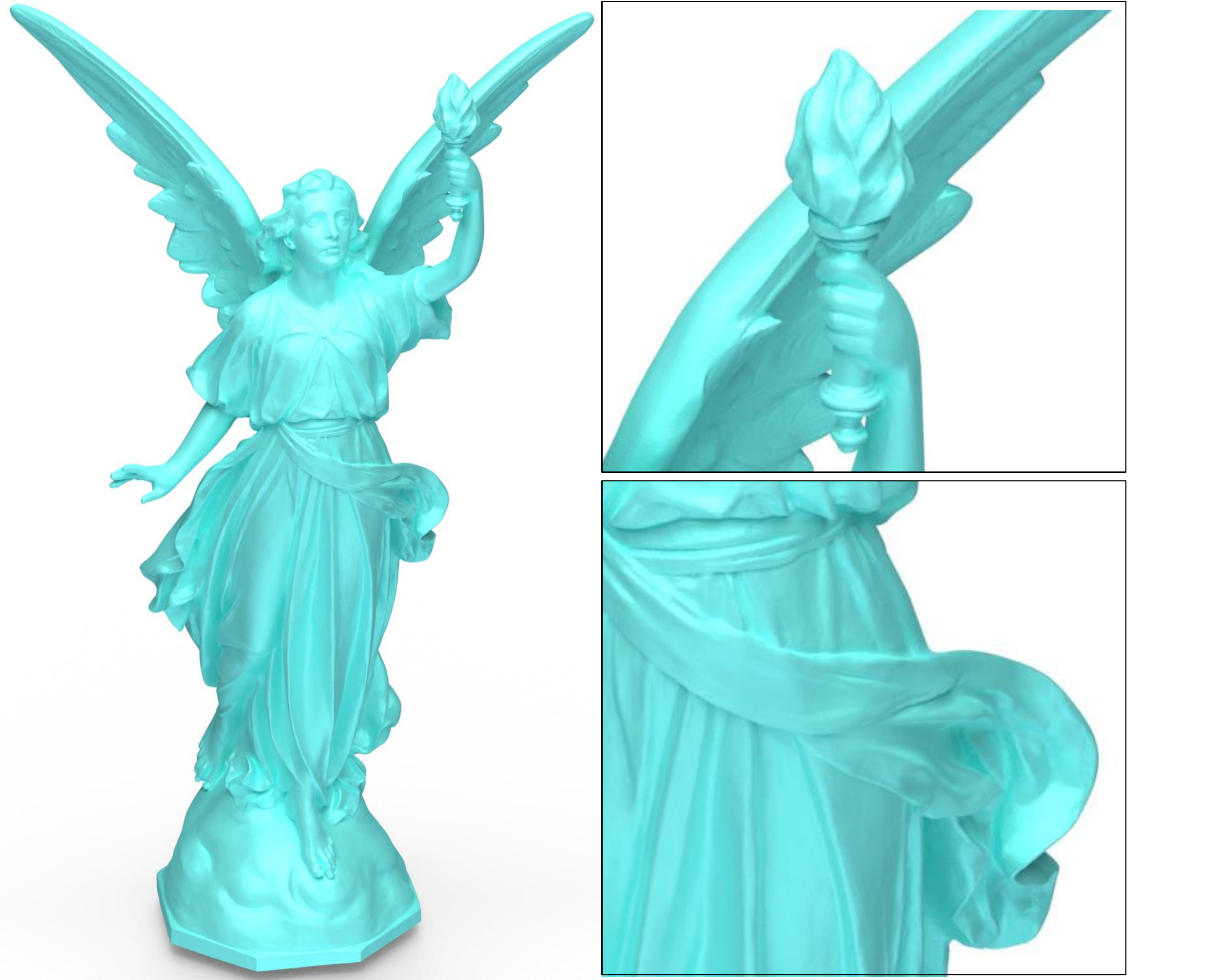}\\
    \caption{Reconstruction of Lucy with 14 million unoriented points.
    Setting the maximum octree depth $D=14$ yields 20 million points in the reconstructed surface with fine detail.
    Using the visibility-based initialization, iPSR converges in only 5 iterations and takes about 1.2 hours.}
    \label{fig:lucy}
\end{figure}

\section{Experimental Results}
\label{sec:results}
\paragraph{Experimental setup.} We implemented iPSR in C++ and tested it on a workstation with Intel Core i9-11900K CPU, 64 GB RAM and Nvidia GeForce RTX 3090 GPU with 24GB memory. The visibility based normal initialization~\cite{Katz2007} was implemented in Matlab and we simply used files to exchange data. The performance of visibility initialization could be further improved if implemented in C++. We carried our experiments on the AIM$@$SHAPE dataset\footnote{\url{http://visionair.ge.imati.cnr.it/ontologies/shapes/link.jsp}}. Since PSR reconstructs watertight surfaces, we removed open meshes and kept the remaining 351 closed models in our test. For each mesh model, we simply used the mesh vertices as the raw points. For models with multiple connected components, we kept only the largest one. After computing the signed distance field $\chi$, we applied the marching cube algorithm ~\cite{Lorensen1987,Wilhelms1992} to extract the iso-surface with iso-value $\frac{1}{n}\sum_{i=1}^n\chi(s_i)$. Then we used the Metro tool~\cite{Cignoni1998} to measure the distance $\varepsilon$ between the reconstructed mesh and the original mesh as the quality measure. To make the measure unitless, the measure is divided by the model scale.

\paragraph{Parameters.}
iPSR has 4 parameters, which are the maximum octree depth $D$, the screened PSR weight $\alpha$, the convergence threshold $\delta$, and the neighborhood parameter $k$. Among them, the octree depth $D$ and weight $\alpha$ are the parameters of screened PSR~\cite{Kazhdan2013}.
The screened PSR weight $\alpha$ trades off the importance of fitting the gradients and fitting the values~\cite{Kazhdan2013}. We set the default weight $\alpha=10$ and keep it a constant in the iterative procedure of iPSR. 

We determine the convergence threshold $\delta$ as follows. We notice that after a few iterations, the normals of most of the samples become stable, but there are still some regions requiring further \textit{local} improvement. Such regions might not receive enough care if using a \textit{global} average of normal variance. So in our implementation, we use the average of the top $0.1\%$ of the $L_2$ normal variance and empirically set the convergence threshold $\delta = 0.175$. For  noisy and/or incomplete models, we also set a maximal iteration number 30 to prevent excessive iterations. 

In iPSR, the neighborhood parameter $k$, which is the number of neighboring samples associated to each triangular face, is a pre-defined, fixed constant. On one hand, the value of $k$ should not be too small, otherwise we may not obtain any candidate points for a sample. On the other hand, $k$ cannot be too large, since a large $k$ means searching is not within a local region. We tested the influence of $k$ on three representative models, Bunny (Figure~\ref{fig:bunny_ear}, row 4), Walrus (Figure~\ref{fig:robustness}, row 6, left) and Metatron (Figure~\ref{fig:robustness}, row 2, left) with $k$ ranging from $5$ and $50$. The running time of iPSR is linearly proportional to the number of iterations. Walnus is sparse and structured, its reconstruction time decreases as $k$ increasing, since the total number of iterations decreases. 
For uniformly sampled models (Bunny and Metatron), we did not observe a strong relation between $k$ and the number of iterations. For example, for all values of $k$, it takes iPSR 4 iterations to converge on Bunny, and 9-11 iterations on Metatron.
Also, for Bunny and Metatron, the value of $k$ has little effect on the reconstruction quality. However, for Walnus, we observed that the optimal range is $k\in[10,20]$.  See Figure~\ref{fig:various_neighbor} (row 1). Thus, we set $k=10$ in our experiments.

The user-specified parameter $D$ is the maximum octree depth for domain discretization. The actual depth, which depends on geometric complexity of the input model, may be smaller than $D$. The parameter $D$ plays a critical role in reconstruction accuracy. In general, the deeper the depth of the octree, the more accurate results iPSR yields, and of course the longer computation time that it takes. The typical range of $D$ is between 8 and 12. We tested the effect of $D$ on three representative models, Bunny ($m=36K$), Raptor ($m=1.5M$) and Walnus ($m=1K$) in Figure~\ref{fig:various_neighbor} (row 2). Notice that Walnus is highly sparse and its actual octree depth never exceeds 8, since octree decomposition stops when each cell contains only a single input point. As a result, increasing the depth parameter $D$ does not increase the actual octree depth, thereby having no effect on the reconstruction quality. Since Bunny is also small and smooth, its reconstruction quality remains almost unchanged for $D>8$. In contrast to Bunny and Walnus, the Raptor model (Figure~\ref{fig:compare_psr}, row 2) has very rich geometric detail, therefore a shallow depth $D=8$ is not sufficient to produce enough octree nodes to represent the shape. We must increase $D$ to reduce the reconstruction error, but at the cost of longer computation time and higher memory consumption. In our implementation, the default octree depth is 10, but the user can increase or decrease the value $D$ by judging the geometric complexity of the input model.

In our implementation, we only use the Dirichlet energy rather than Neumann boundary condition in screened PSR so that we can ensure the reconstructed surface does not exceed the domain boundary~\cite{Kazhdan2013}.

\begin{figure}[!htbp]
    \centering
    \includegraphics[width=0.24\linewidth]{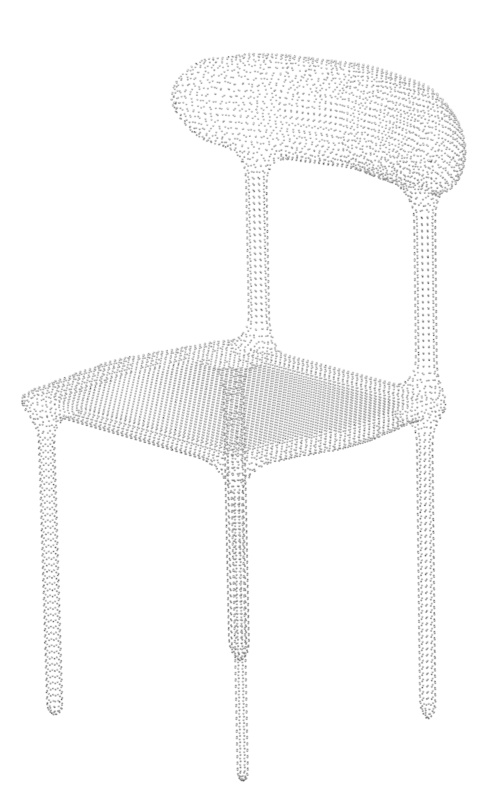}
    \includegraphics[width=0.24\linewidth]{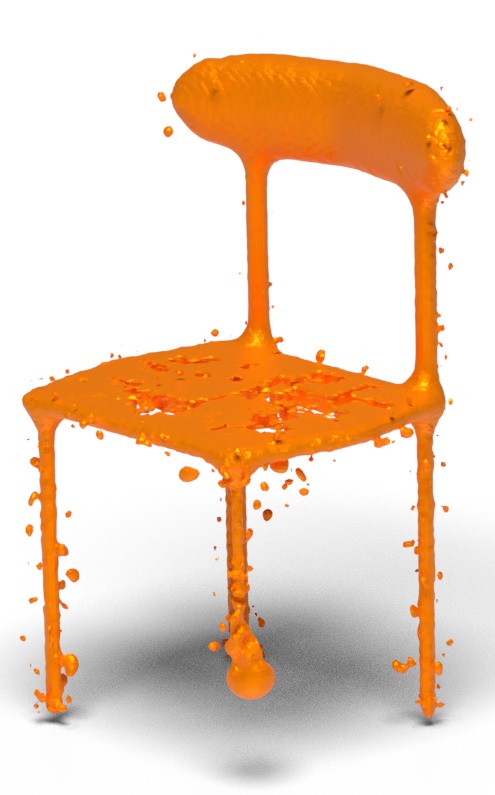}
    \includegraphics[width=0.24\linewidth]{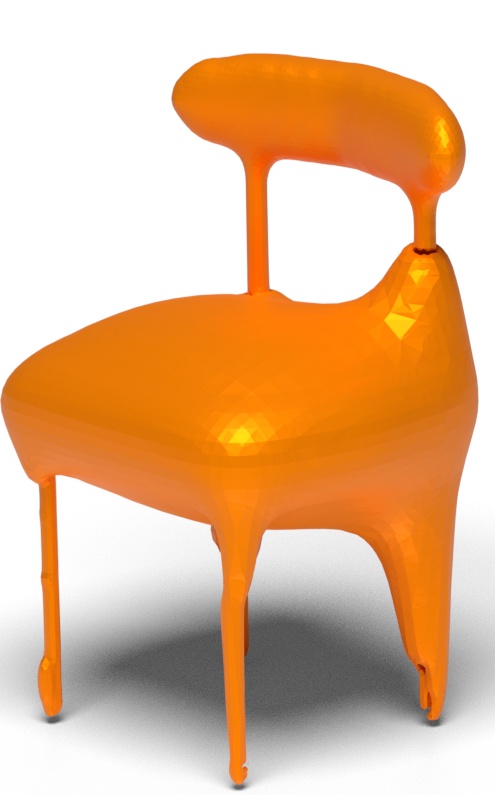}
    \includegraphics[width=0.24\linewidth]{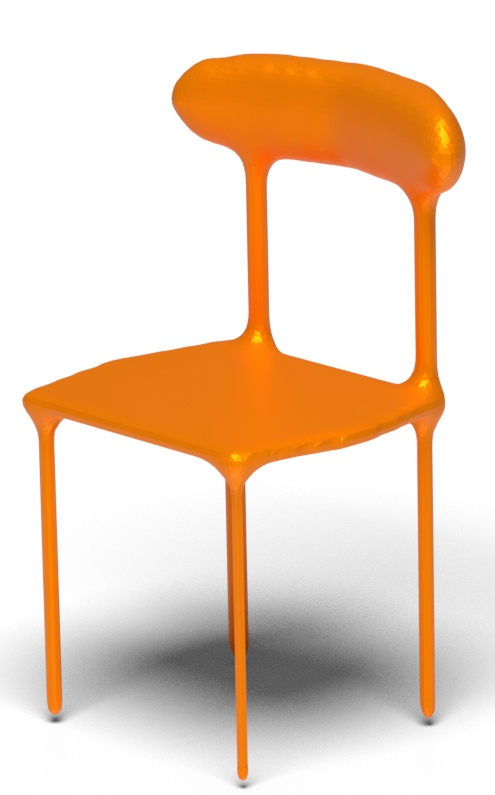}\\
    \includegraphics[width=0.24\linewidth]{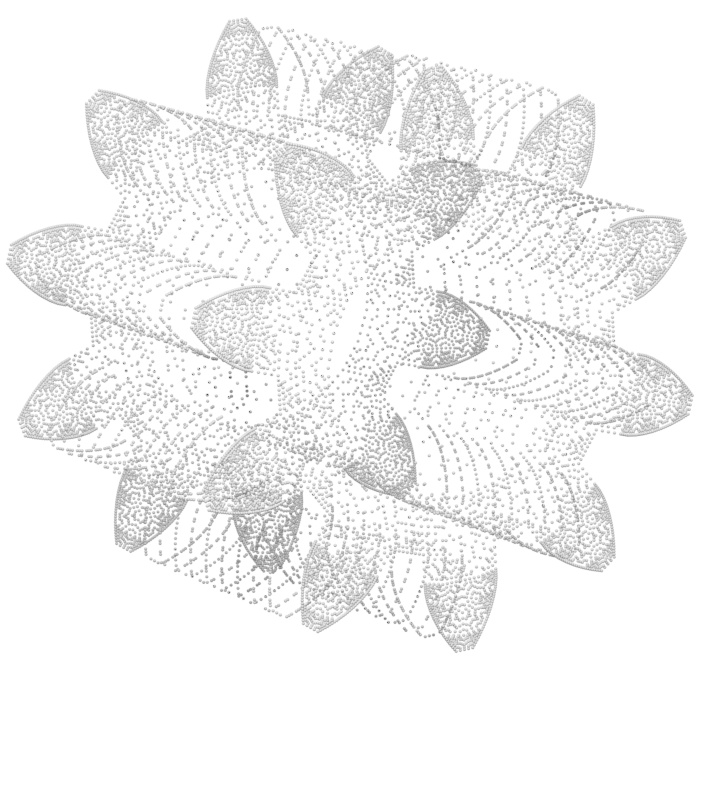}
    \includegraphics[width=0.24\linewidth]{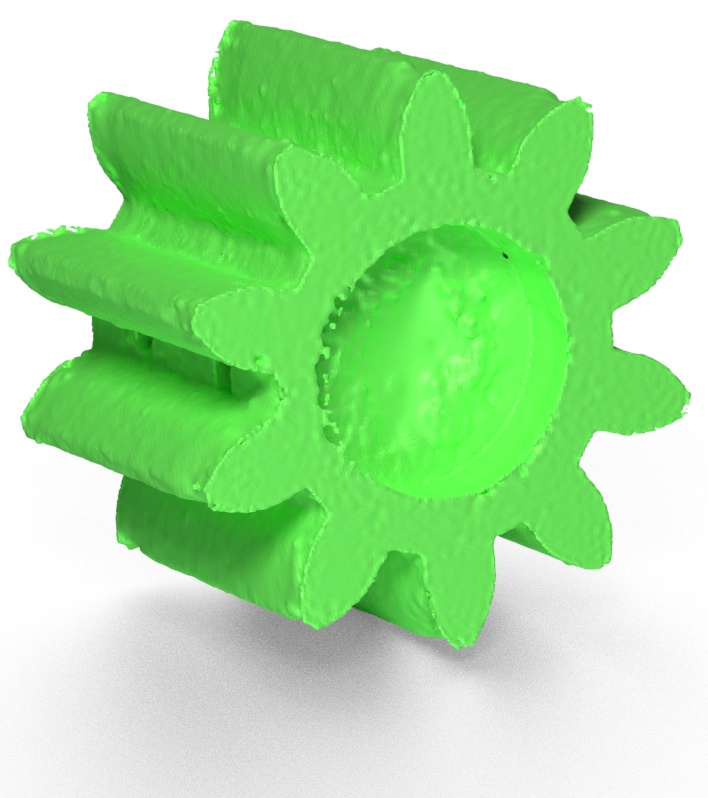}
    \includegraphics[width=0.24\linewidth]{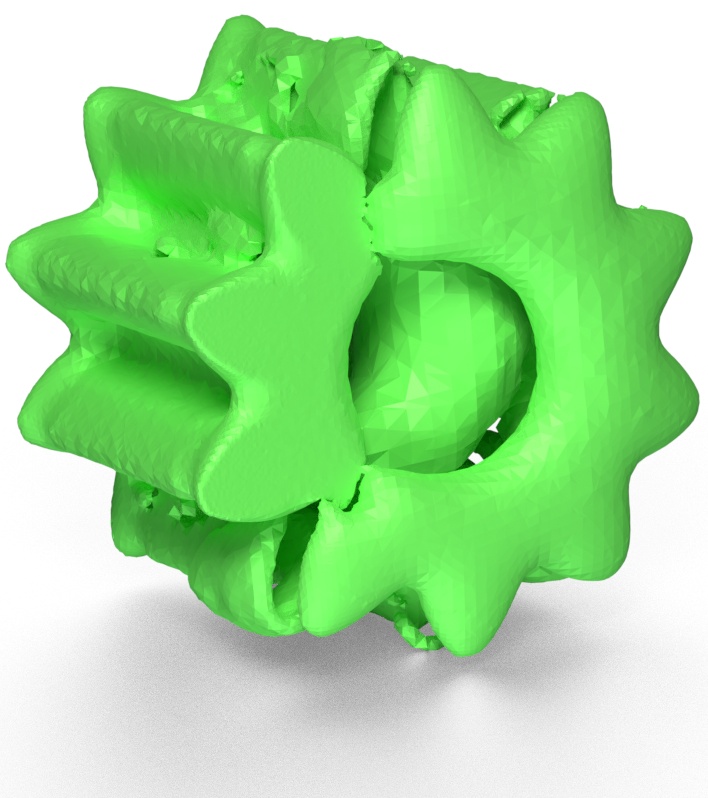}
    \includegraphics[width=0.24\linewidth]{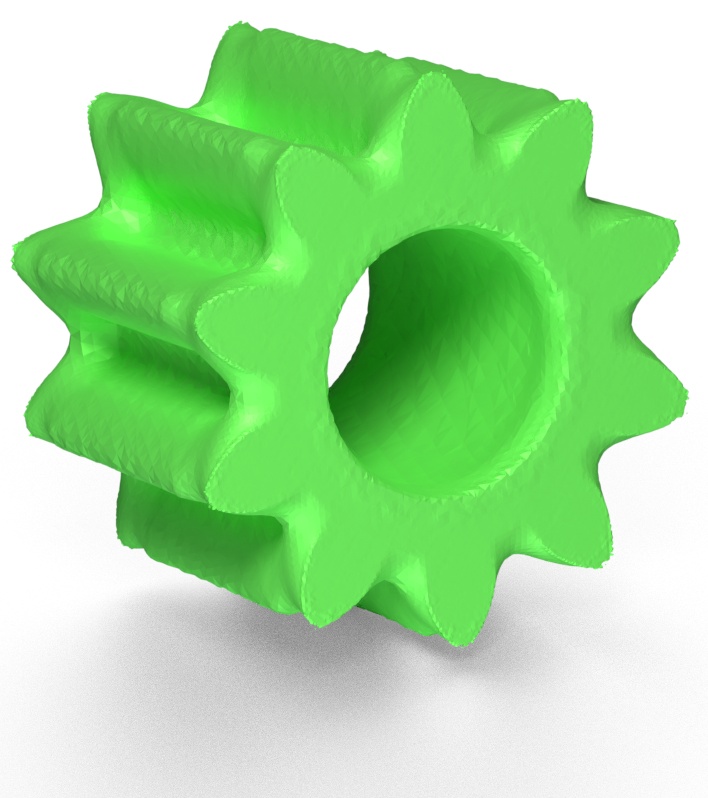}\\
    \includegraphics[width=0.24\linewidth]{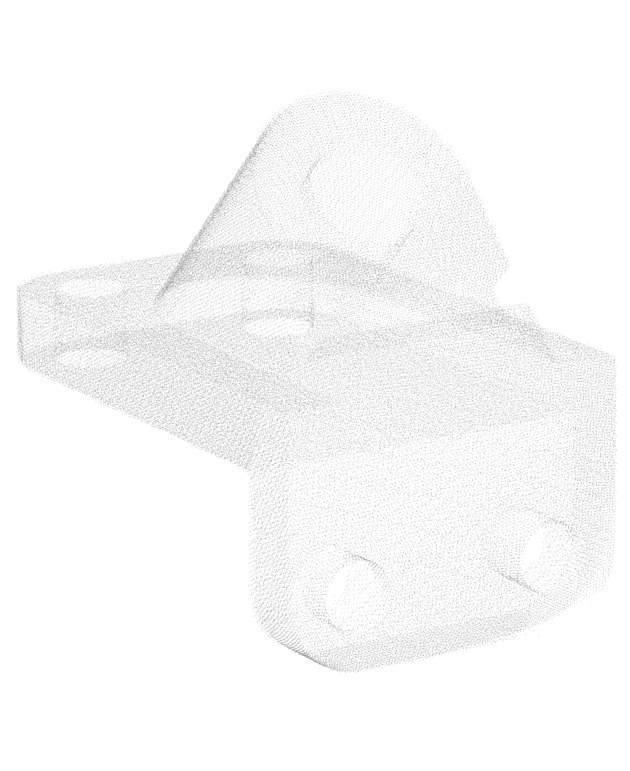}
    \includegraphics[width=0.24\linewidth]{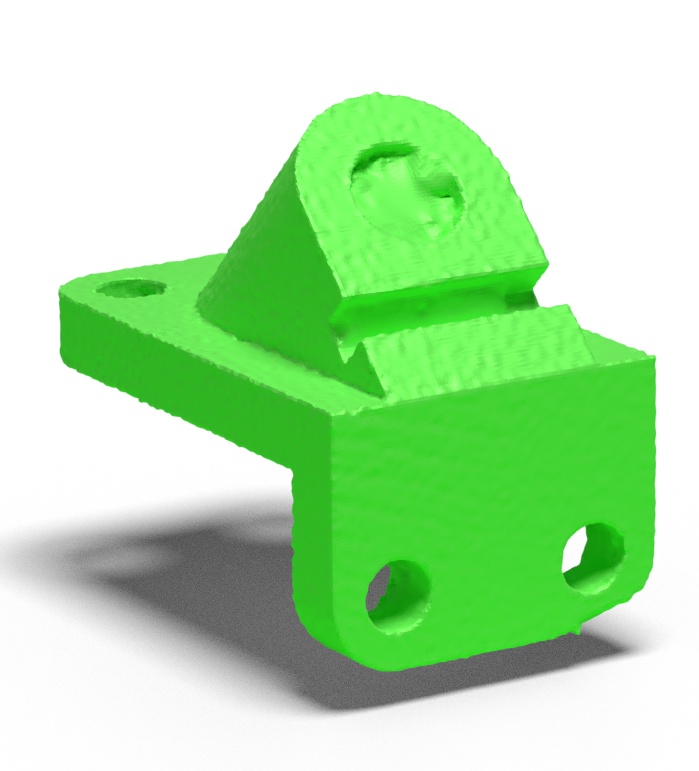}
    \includegraphics[width=0.24\linewidth]{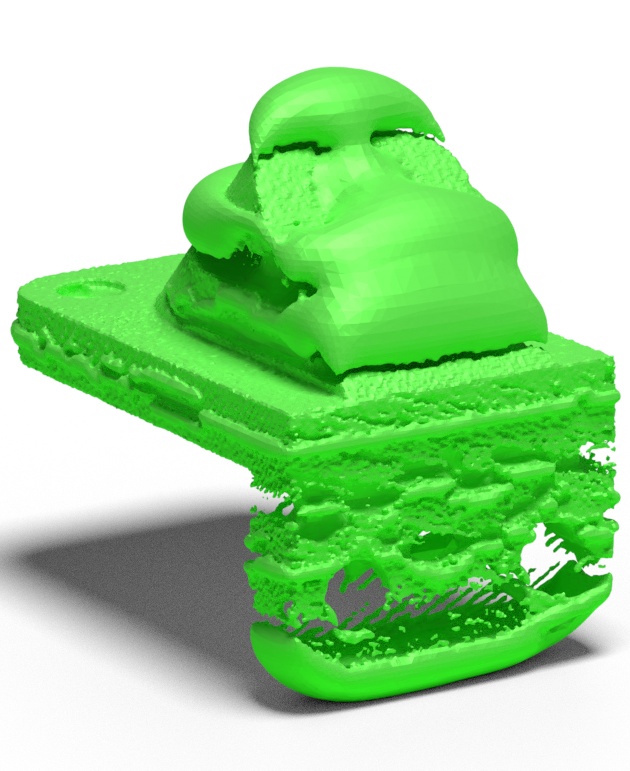}
    \includegraphics[width=0.24\linewidth]{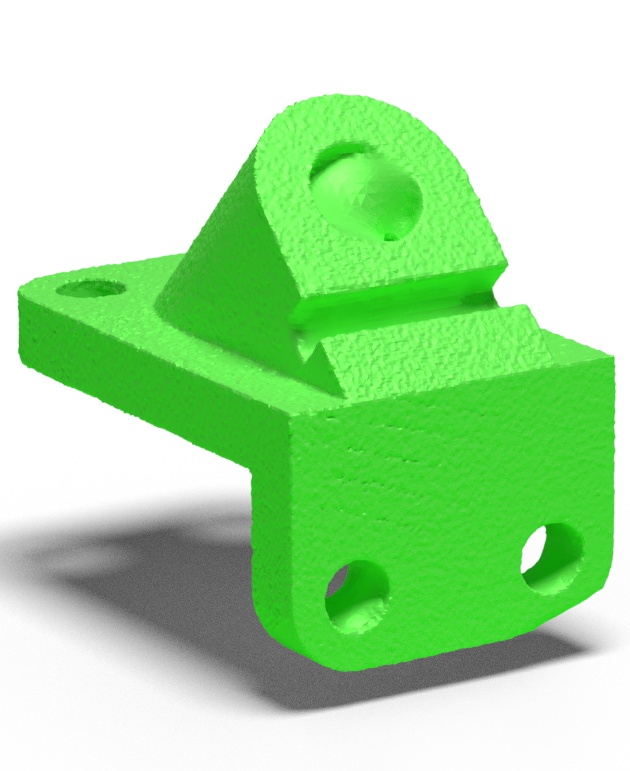}\\
    \makebox[0.24\linewidth]{Input}
    \makebox[0.24\linewidth]{DPSR}
    \makebox[0.24\linewidth]{Dipole}
    \makebox[0.24\linewidth]{iPSR}
    \caption{Man-made models.}
    \label{fig:manmade}
\end{figure}

\begin{figure}[!htbp]
    \centering
    \includegraphics[width=1.0in]{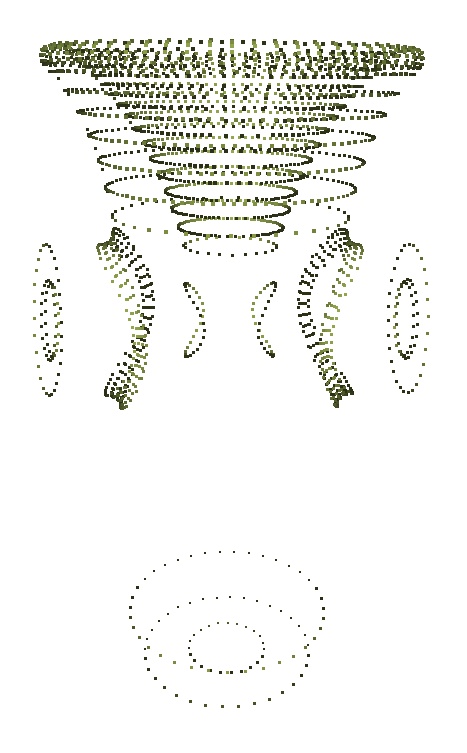}
    \includegraphics[width=1.0in]{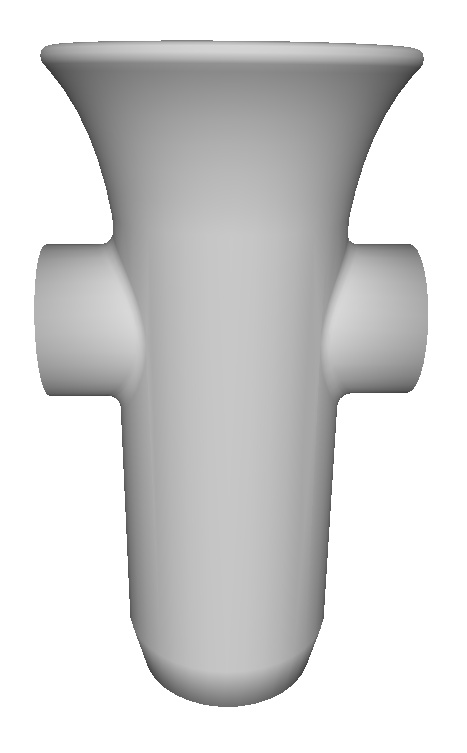}
    \includegraphics[width=1.0in]{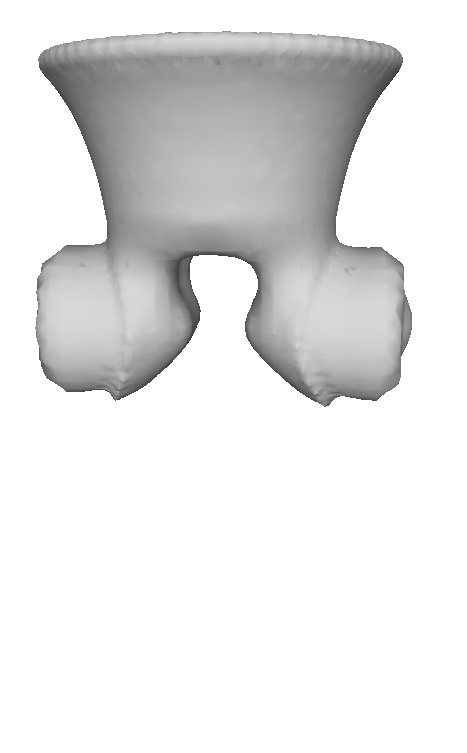}\\
    \includegraphics[width=1.0in]{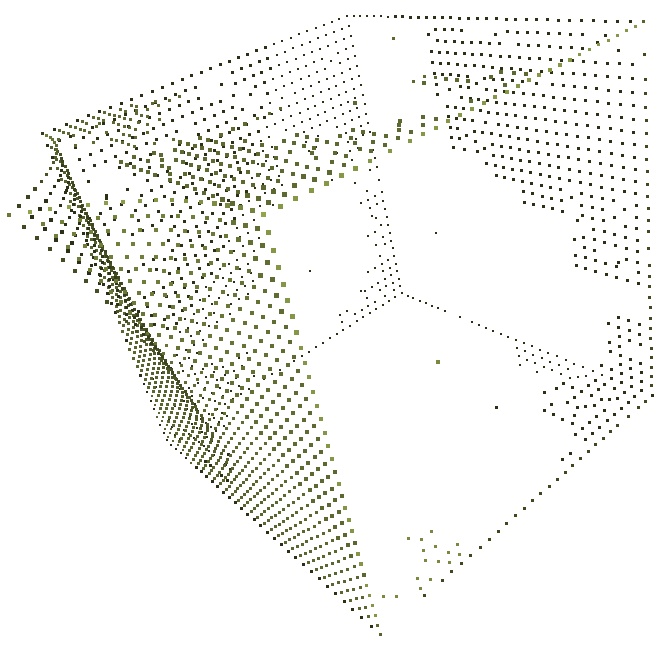}
    \includegraphics[width=1.0in]{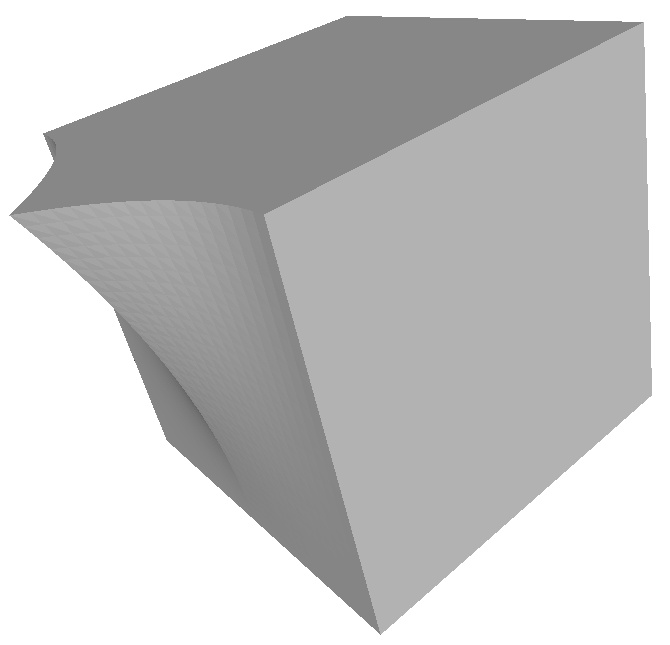}
    \includegraphics[width=1.0in]{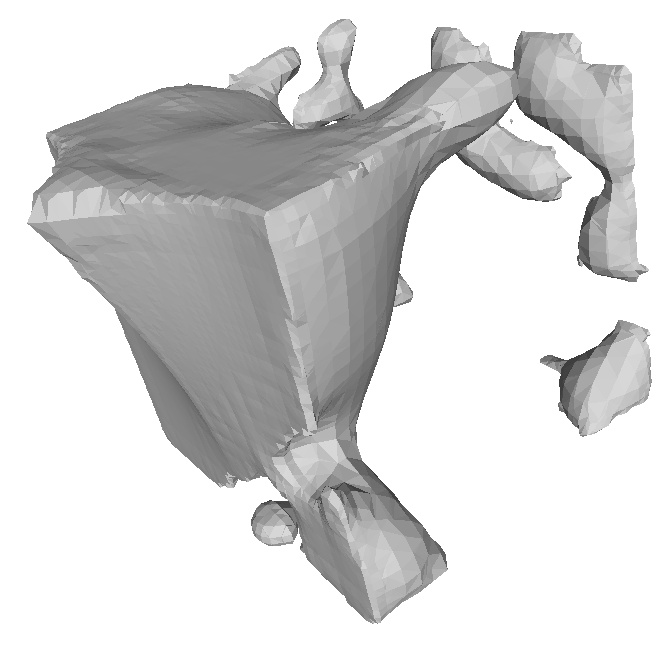}\\
    \makebox[1.0in]{Input}
    \makebox[1.0in]{GT mesh}
    \makebox[1.0in]{iPSR+random}
    \caption{Failed cases. In the first example, the lower part is treated wrongly as outliers due to lack of samples connecting the bottom and the top parts. In the second example, 5 out of the 6 sides of the cube-like shape are sampled in an very uneven manner.  }
    \label{fig:error_models}
\end{figure}

\paragraph{Results.} Figure~\ref{fig:iterations_histogram} shows the histogram of the number of iterations on the AIM$@$SHAPE dataset. We observe that iPSR converges quickly with random initialization. It takes no more than 6 iterations for 58\% of the testing models and the median is 6. There are also 39 models that requires iPSR more than 30 iterations to converge with random initialization. Since the improvements are not visually significant in the final iterations, we simply force iPSR to stop at the 30th iteration.  Visibility initialization is effective for large models and can further reduce 5 iterations on average for models with more than $100$K points. We notice that there are a few test models with very poor and/or uneven sampling rates. iPSR cannot recover geometry if the sample rates are too low. Figure~\ref{fig:error_models} shows two such failed examples. 

\begin{figure}
    \centering
    \includegraphics[width=0.45\linewidth]{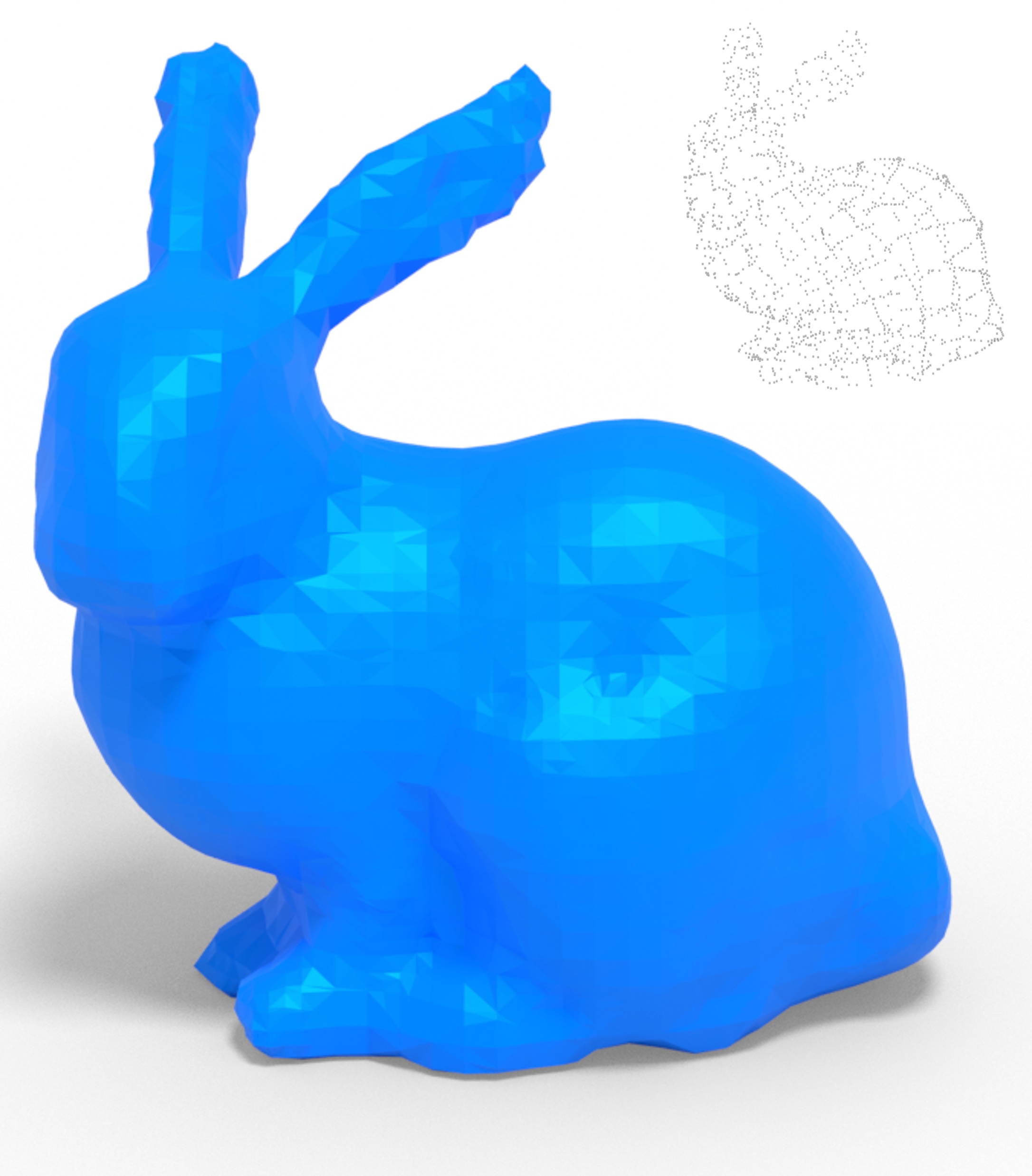}
    \includegraphics[width=0.45\linewidth]{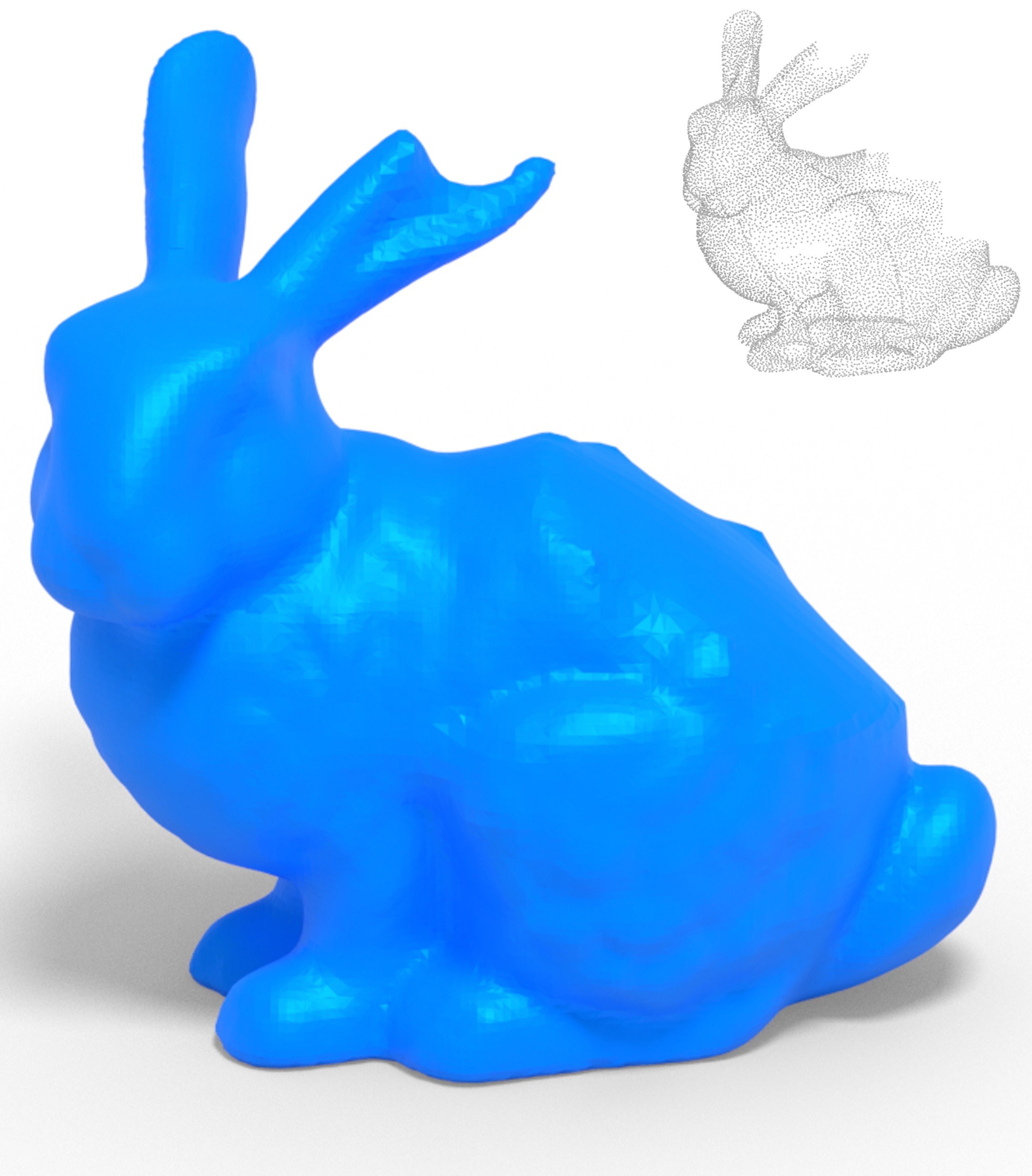}\\
    \caption{Left: Given a structured, sparse input with 1.5K points, iPSR can fill in the gaps between wireframes and produce a smooth shape. Right: The input is un-structured and incomplete, iPSR cannot recover the missing geometry.}
    \label{fig:bunny_frame}
\end{figure}

\paragraph{Large-scale data.} We applied iPSR to the Lucy model, which contains 14 million samples. As shown in Figure~\ref{fig:lucy}, setting the maximum octree depth $D=14$, iPSR can preserve fine detail well in the reconstructed surface. We also tested our method on indoor and outdoor scenes~\cite{Park2017,Chang2015}. Figure~\ref{fig:teaser} illustrates two examples of large-scale 3D scene reconstruction using iPSR. The indoor scene is scanned by a Lidar scanner~\cite{Park2017}, and we used the merged and resampled points as our input. The outdoor scene is scanned by a RGBD camera~\cite{Choi2016}. As only raw depth images and the reconstructed surface mesh are available, we took the mesh vertices as the input to iPSR. Note that the scene models have various types of defects, such as noise, outliers, missing parts, and non-uniform sampling. Nevertheless, iPSR can produce visually pleasing results. 

\paragraph{Structured and sparse data.} iPSR inherits the smooth approximation property of screened PSR and can generate a smooth water-tight surface well approximating given points. As a result, iPSR works for sparse but structured inputs. Figure~\ref{fig:bunny_frame} (left) shows a sparse Bunny model, whose points are sampled from the edges of a quadrilateral tessellation. Since the points are structured, iPSR can generate smooth patches to fill in the gaps between wires. However, if the input model is incomplete and unstructured, iPSR is not intelligent enough to figure out the missing shape. See Figure~\ref{fig:bunny_frame} (right).

\section{Comparison}
\label{sec:comparison}

We compared our method with a few recent works for reconstructing from \textit{unoriented} points, including Dipole~\cite{Metzer2021}, differentiable Poisson surface reconstruction (DPSR)~\cite{Peng2021} and variational implicit point set surfaces (VIPSS)~\cite{Huang2019}. DPSR has two versions, which are based on optimization and deep learning, respectively. To make fair comparison to our work, we used the optimization-based DPSR in the paper.

\begin{figure}[!htbp]
    \centering
    \includegraphics[width=1.6in]{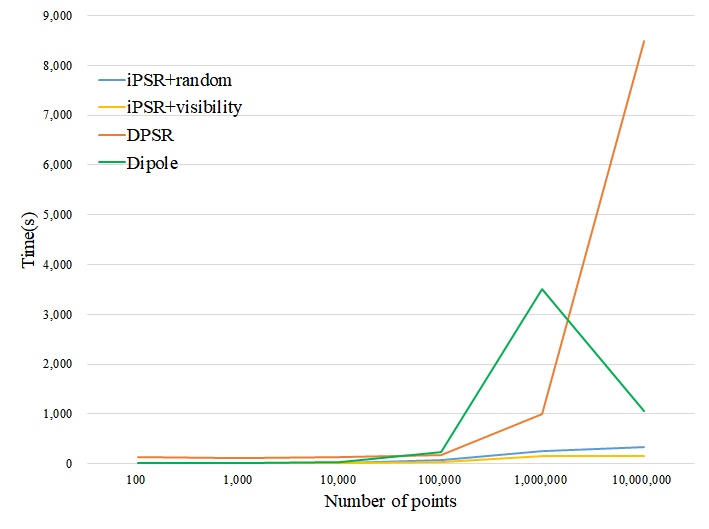}
    \includegraphics[width=1.6in]{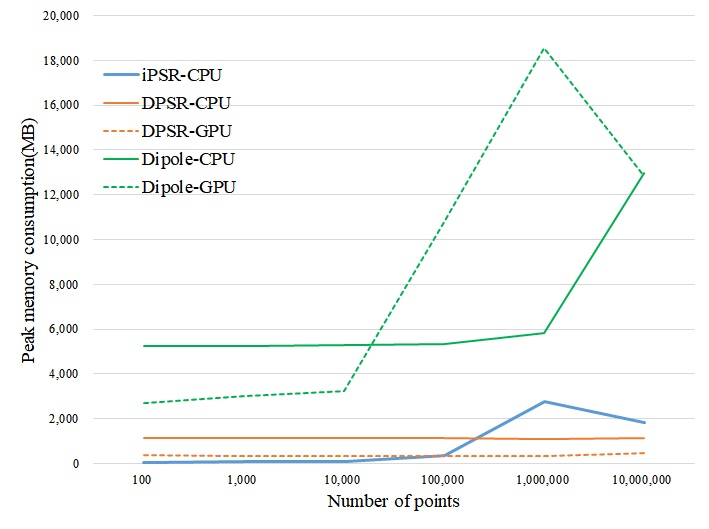}\\
    \makebox[1.6in]{Running time}
    \makebox[1.6in]{Memory}\\
    \caption{Scalability. We measure the running time and memory consumption on models with 100, 1K, 10K, 100K, 1M and 10M points, respectively. We set the maximum octree depth $D=10$ for iPSR. For the 10M-point-model, Dipole orients normals for representative points only, so there is a sudden drop in GPU memory consumption the plot. DPSR uses fixed resolution to voxelize the indicator function, so its memory consumption is almost a constant. }
    \label{fig:scalability}
\end{figure}

\begin{table*}[!htbp]
    \centering
     \caption{\label{tab:performance}Statistics. $m$: the number of input points; $T/T_r/T_v$: the running time in seconds; $I_r/I_v$: the number of iterations of iPSR, where the subscripts r and v indicate random and visibility initialization, respectively; $M_\textrm{CPU}$ and $M_\textrm{GPU}$: the peak memory consumption in MB for CPU and GPU respectively; $\varepsilon$ (\%): the normalized reconstruction error in percentage measured by Metro ~\cite{Cignoni1998}. Asterisk $*$ indicates models whose errors are measured by the Hausdorff distance between points due to lack of ground-truth meshes. We set the maximum octree depth $D=10$ for all models. Random initialization was used when measuring the peak memory $M_\textrm{CPU}$ and the reconstruction error $\varepsilon$ of iPSR.
    }
    \begin{scriptsize}
    \begin{tabular}{c||cccc|cccc|ccc|cccccc}
        \hline
        \multirow{2}{*}{Model ($m$)} &\multicolumn{4}{c|}{DPSR}&\multicolumn{4}{c|}{Dipole}&\multicolumn{3}{c|}{VIPSS}&\multicolumn{6}{c}{iPSR}\\
        \cline{2-18}
         & $T$ & $M_\textrm{CPU}$ & $M_\textrm{GPU}$ & $\varepsilon$ (\%) & $T$ & $M_\textrm{CPU}$ & $M_\textrm{GPU}$ & $\varepsilon$ (\%)& $T$ & $M_\textrm{CPU}$ & $\varepsilon$ (\%) & $T_r$ & $I_r$ & $T_v$ & $I_v$ & $M_\textrm{CPU}$  & $\varepsilon$ (\%) \\
         \hline
         Fertility~(4.4K, Fig.~\ref{fig:robustness}) & 140.2&1146.1& 327.6& 9.62& 14.8& 5269.1& 3031.8& 8.98& 1161.6& 13184.3& 3.08& 16.9& 14&15.6&11&93.1&6.70\\
         Noisy Hand 1~(8.7K, Fig.~\ref{fig:robustness}) & 148.2 & 1146.8& 288.7& 0.72& 26.5& 5271.4& 3232.8&  6.77& 8650.7& 51855.4& 86.89& 9.4& 6 & 11.6& 3& 119.7 & 0.91\\
         Noisy Hand 2~(8.7K, Fig.~\ref{fig:robustness})& 153.3& 1146.1& 245.8& 2.17& 27.6& 5267.6& 3356.7& 5.84& 8920.3& 51867.6& 67.54& 14.4& 11& 16.0& 8& 119.4 &1.76\\
         High Genus 2~(9.5K, Fig.~\ref{fig:robustness}) & 186.3& 1145.9& 346.8& 8.61& 28.0& 5270.1& 3212.7& 13.15& 10782.3&61942.9& 1.45& 8.0& 5&8.8&3&105.8 &0.42 \\
         Chair~(10K, Fig.~\ref{fig:manmade}) & 128.6&1145.3& 245.8& 22.30& 28.5& 5270.2& 3386.3& 13.87& - &- &- &35.5& 30&38.6& 30&109.6& 0.18\\
         Woodfish~(12.9K, Fig.~\ref{fig:robustness}) & 131.7& 1145.2& 245.8& 2.85& 36.2& 5264.6&3276.4&8.67&- &- &- &18.8& 12& 15.3&7&128.8 &1.56\\
         Pinion~(15.6K, Fig.~\ref{fig:manmade}) & 176.7& 1145.3&290.7&11.82& 34.9&5274.1&3138.8&11.40&- &- &- &33.2&19&34.5&18&133.0&2.50\\
         320-wrl~(27.4K, 1\% outliers, Fig.~\ref{fig:robustness}) & 146.1& 1145.8& 245.8& 1.38& 65.9& 5280.2& 3575.8& 1.10& - &- &- &11.9& 6&11.1&3& 141.8 &0.48\\
          Bimba~(50.5K, non-uniform, Fig.~\ref{fig:robustness}) & 154.0& 1145.9& 288.7& 1.52& 101.3& 5291.7& 4587.5& 3.92& - &- &- &33.2& 10& 19.8& 4& 220.4& 1.72\\
         High Genus 1~(69.2K, Fig.~\ref{fig:robustness}) & 273.0& 1145.7&245.8& 4.88& 166.3& 5303.9& 3784.5& 11.38& -&- & -&42.8& 10& 21.5& 3&312.6 &0.31  \\
         Bimba~(80.8K, 1\% outliers, Fig.~\ref{fig:robustness}) & 175.6& 1146.1& 245.8& 4.54& 163.3& 5309.0& 5241.1& 0.70&- &- &- &32.2& 7& 23.1& 3&320.4 &0.36\\
         Anchor~(85.1K, Fig.~\ref{fig:manmade}) &182.6& 1146.5&245.8&13.27*&170.9&5312.8&4205.3&7.83*& - &- &- &43.1&10&36.1&7&423.8&9.47*\\
         Four Children~(660K, Fig.~\ref{fig:robustness})& 597.7& 1143.0&545.5  & 2.74& 2578.0& 5633.6&12719.7 & 8.82&- &- &-& 259.1& 10& 160.8& 4& 2125.9 & 1.45\\ 
         \hline
    \end{tabular}
    \end{scriptsize}
\end{table*}

\subsection{Efficiency and Scalability}
VIPSS is an elegant method for reconstructing smooth surfaces from sparse input. However, it works only for input with up to a few thousand points due to high memory consumption. 
iPSR and VIPSS run on CPUs, whereas Dipole and DPSR perform most of the computational tasks on GPUs. 
Therefore, the RAM consumption of Dipole and DPSR is loosely related to the input size, but their CUDA memory consumption increases rapidly when the number of input points goes up. Moreover, DPSR fixes the resolution of indicator functions $256\times 256\times 256$, since the current GPUs cannot afford higher resolutions.
To deal with large models (e.g., with more than 1 million points), Diople adopts a strategy that processes representative points instead of all input points due to the CUDA memory constraint.
iPSR can work for large-scale models and recover fine geometry detail thanks to the efficiency and scalability of screened PSR. See  Figure~\ref{fig:scalability}.

\begin{figure}[!htbp]
    \centering
    \includegraphics[width=3.0in]{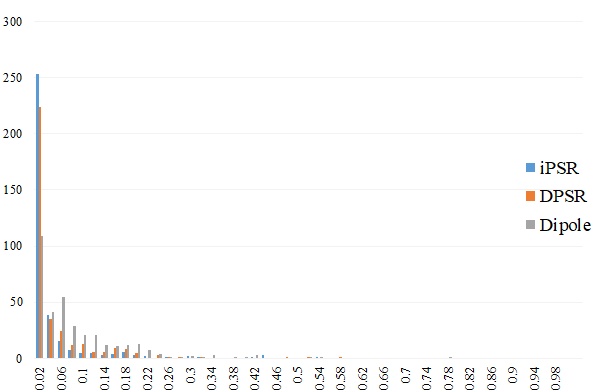}
    \vspace{1pt}
\begin{tabular}{c|c|c|c}
    \hline
        Method & \% of models with $\varepsilon$ < 2\% & Mean & Median\\  \hline
        iPSR & 72.1 & 3.18\% & 0.71\%\\
        DPSR & 63.8 & 3.83\% & 0.60\%\\
        Dipole & 31.1 & 7.75\% & 4.76\% \\
        \hline
    \end{tabular}
    \caption{Histogram of reconstruction error $\varepsilon$ on the AIM$@$SHAPE dataset. The horizontal axis shows the normalized error and the vertical axis is the frequency. Our method with random initialization is more accurate than DPSR and Dipole.}
    \label{fig:histogram}
\end{figure}

\begin{figure*}[!htbp]
    \centering
    \begin{scriptsize}
    \setlength\tabcolsep{0pt}
    \begin{tabular}{cccccccccc}
    \includegraphics[width=0.64in]{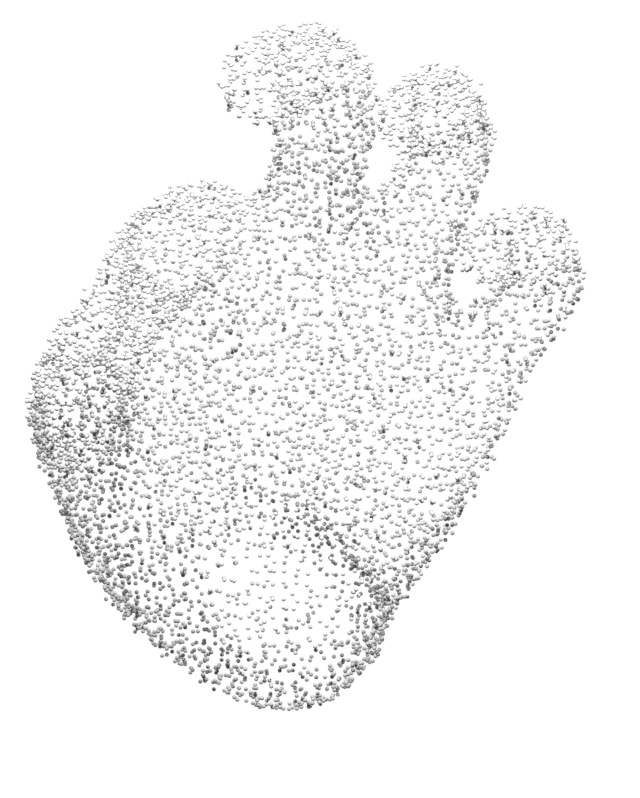} &
    \includegraphics[width=0.64in]{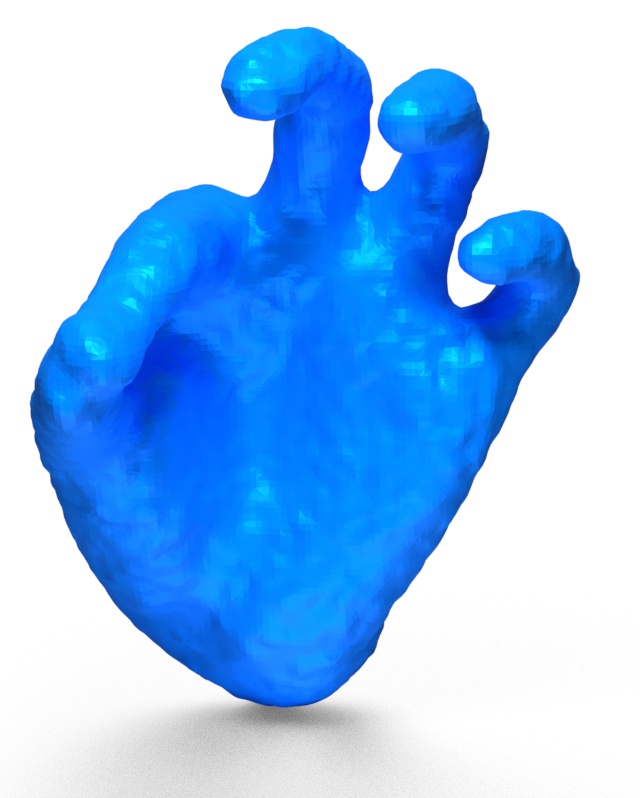}&
    \includegraphics[width=0.64in]{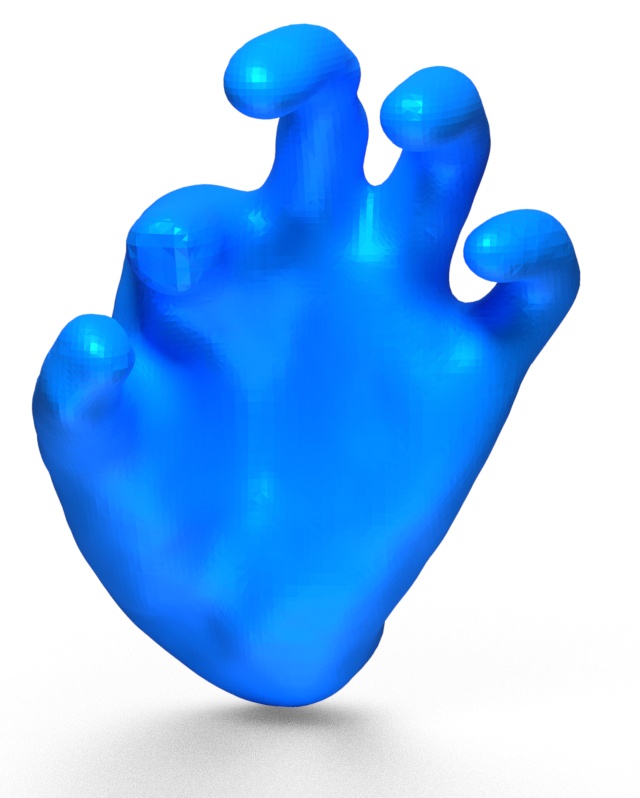}&
    \includegraphics[width=0.64in]{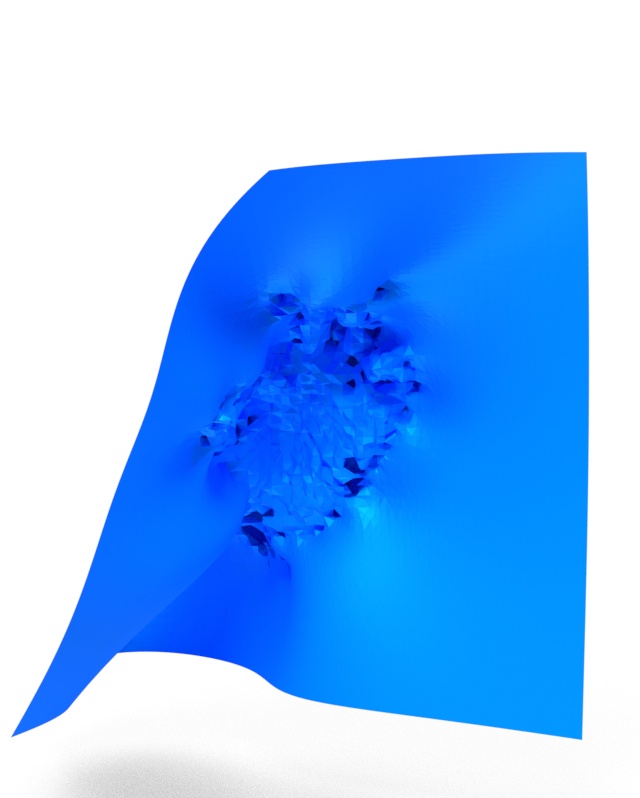}&
    \includegraphics[width=0.64in]{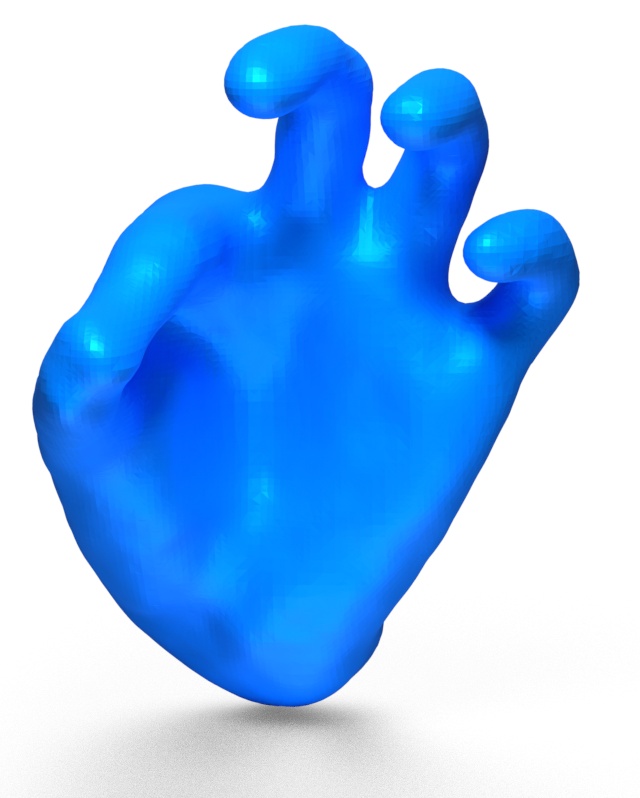}&
    \includegraphics[width=0.64in]{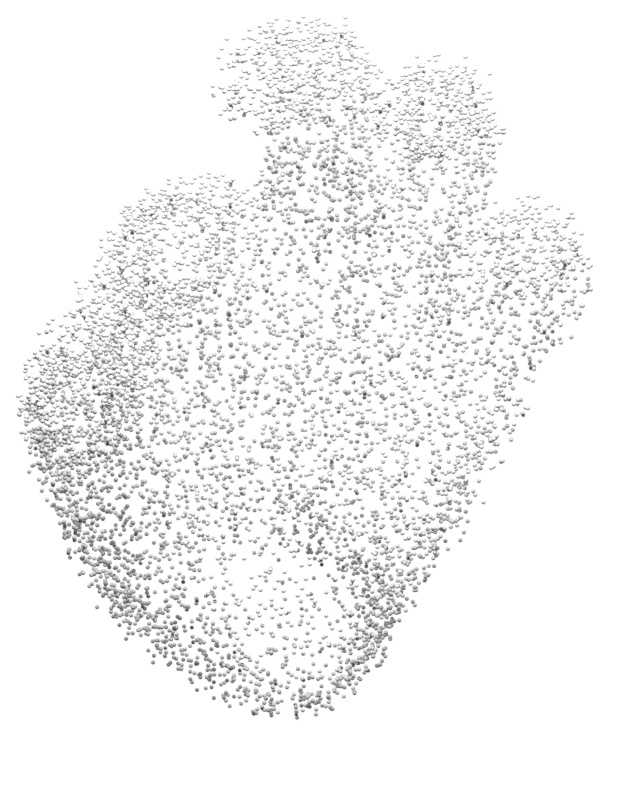}&
    \includegraphics[width=0.64in]{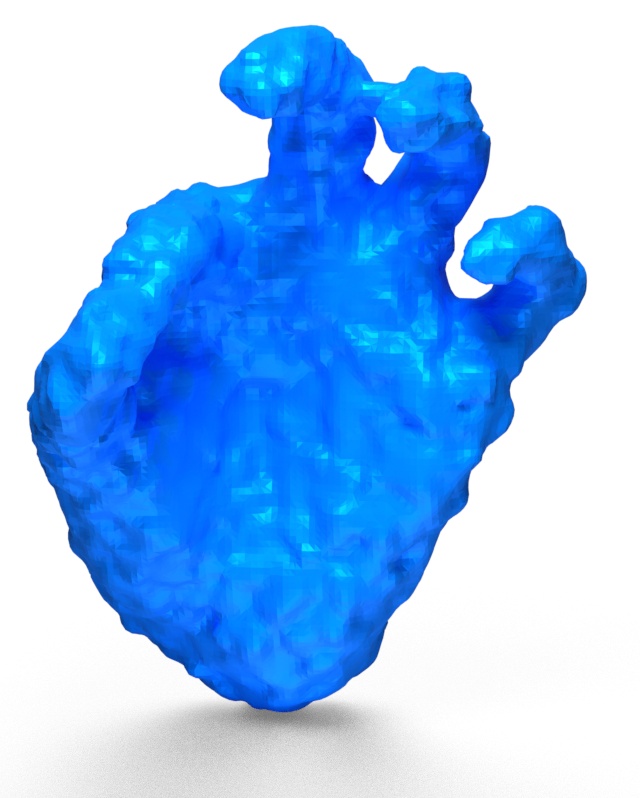}&
    \includegraphics[width=0.64in]{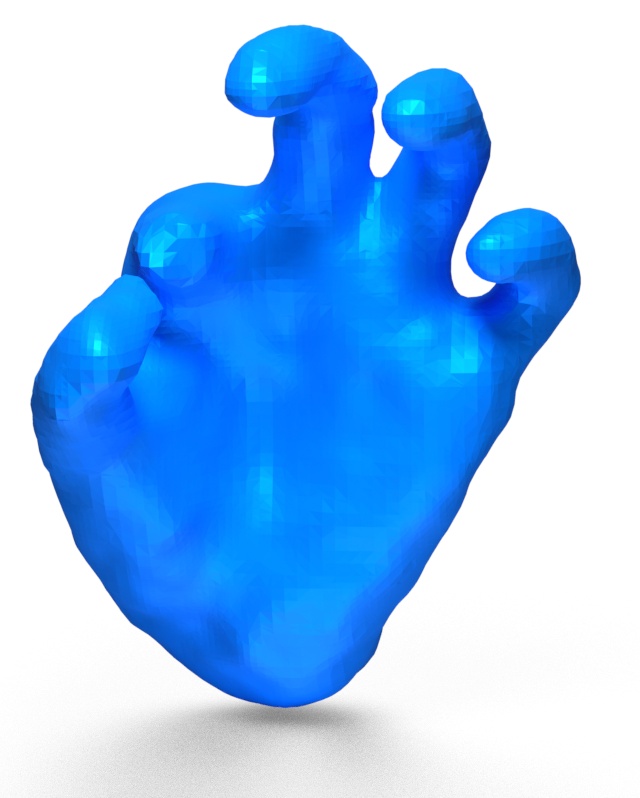}&
    \includegraphics[width=0.64in]{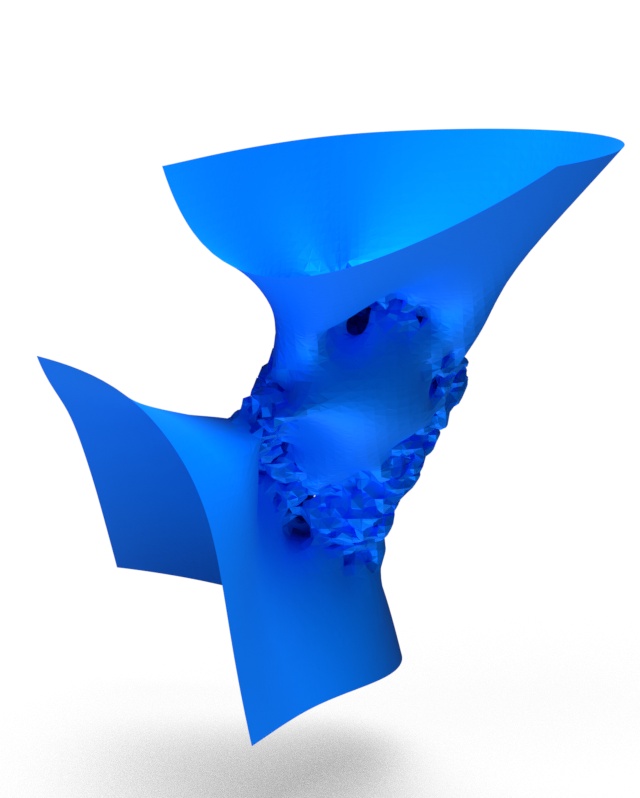}&
    \includegraphics[width=0.64in]{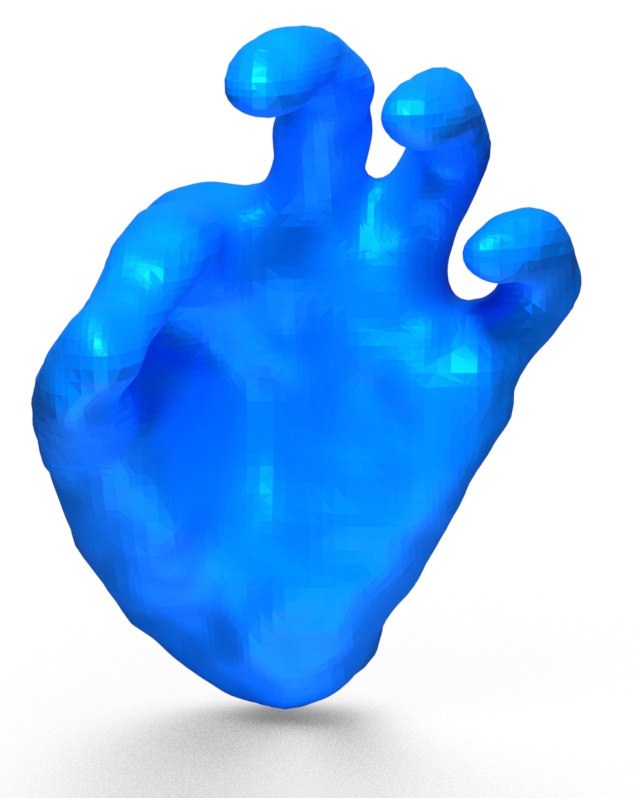}\\
    \includegraphics[width=0.64in]{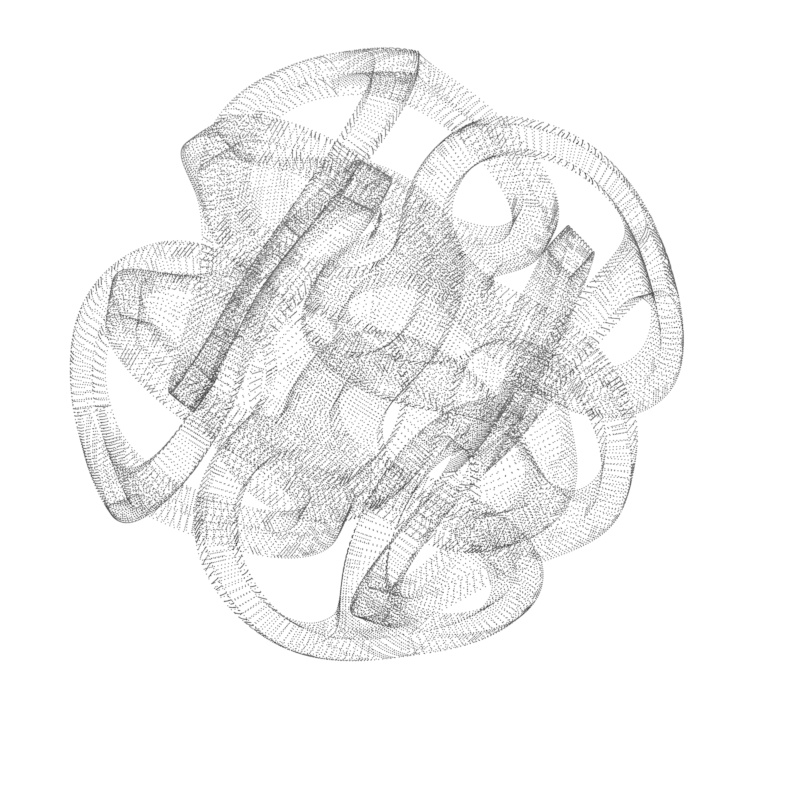}&
    \includegraphics[width=0.64in]{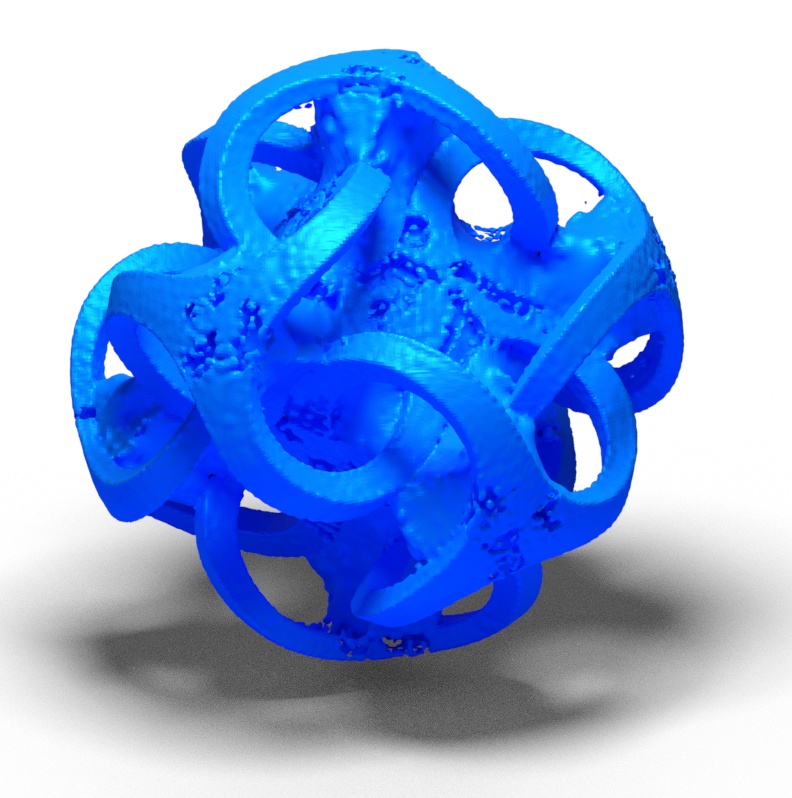}&
    \includegraphics[width=0.64in]{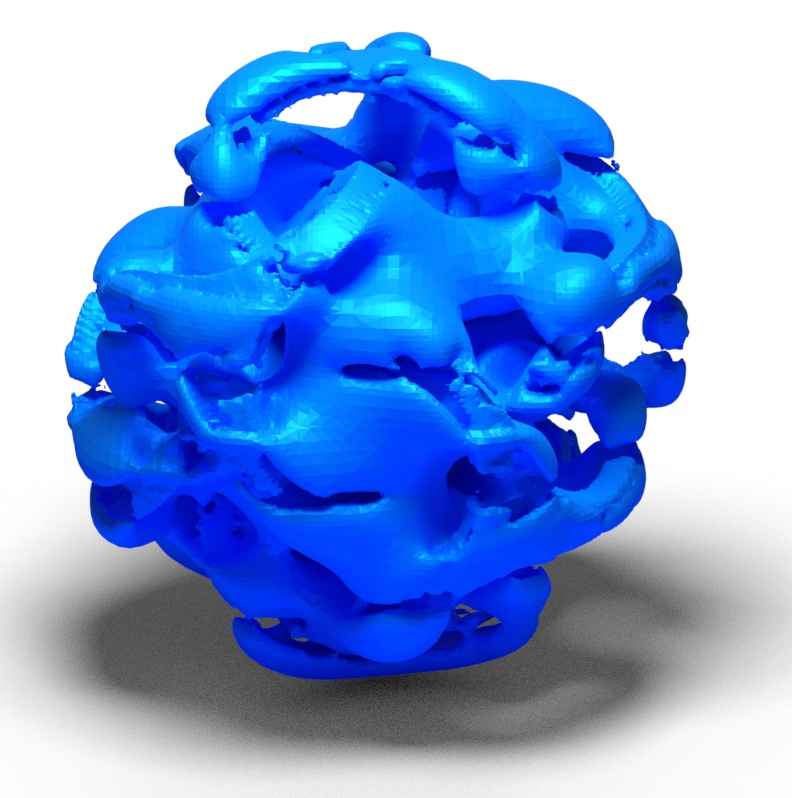}&
    \includegraphics[width=0.64in]{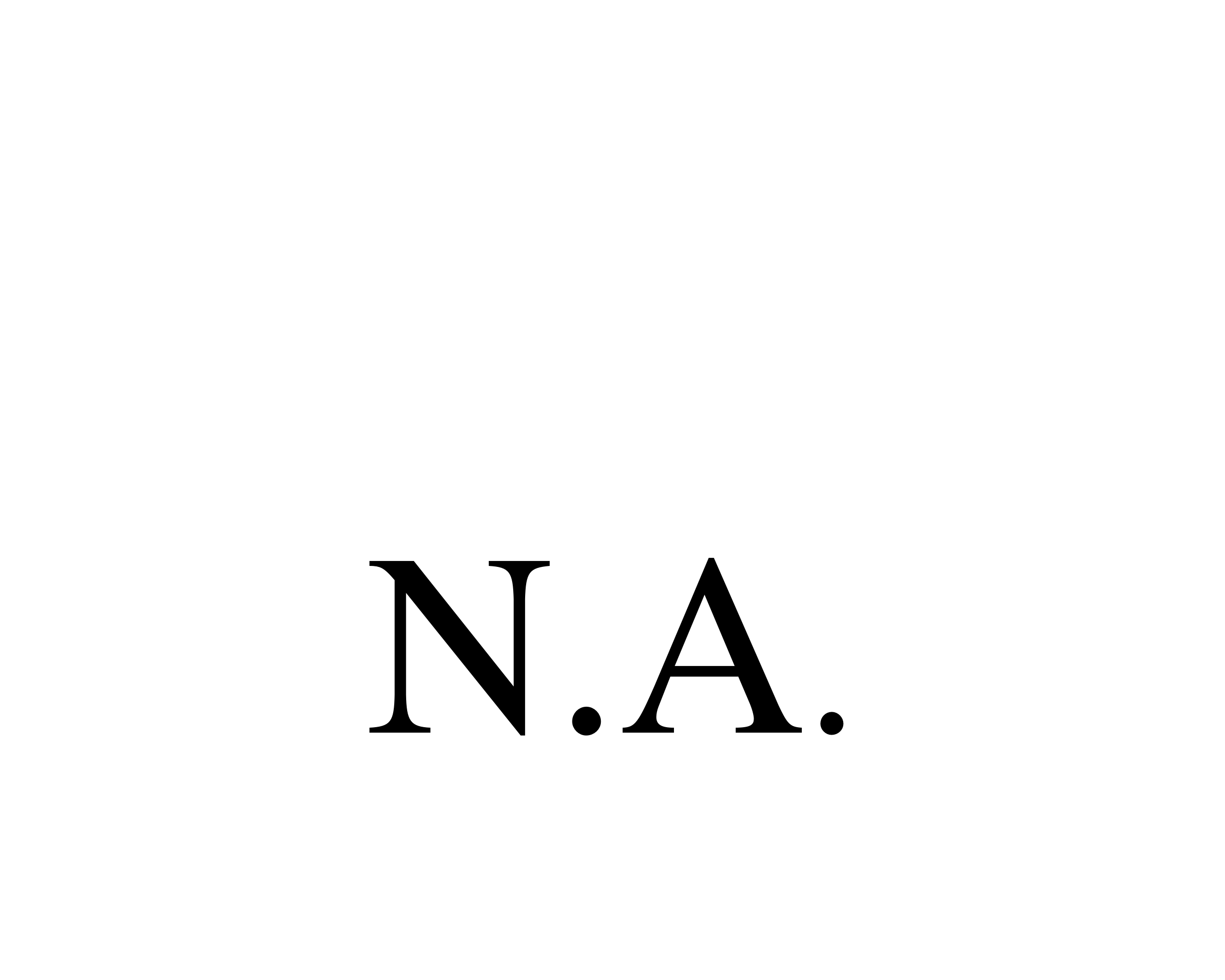}&
    \includegraphics[width=0.64in]{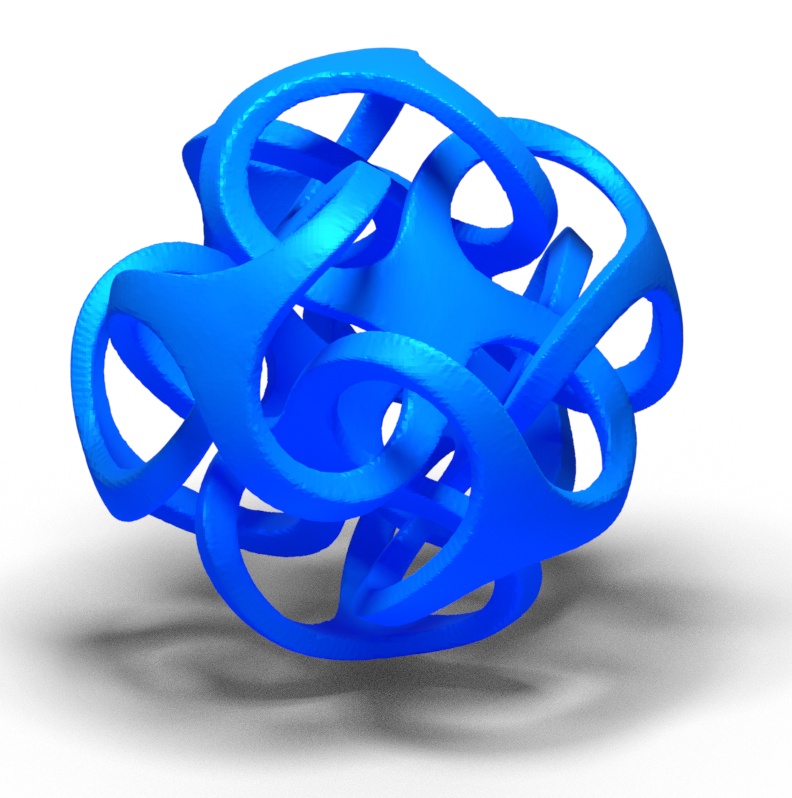}&
    \includegraphics[width=0.64in]{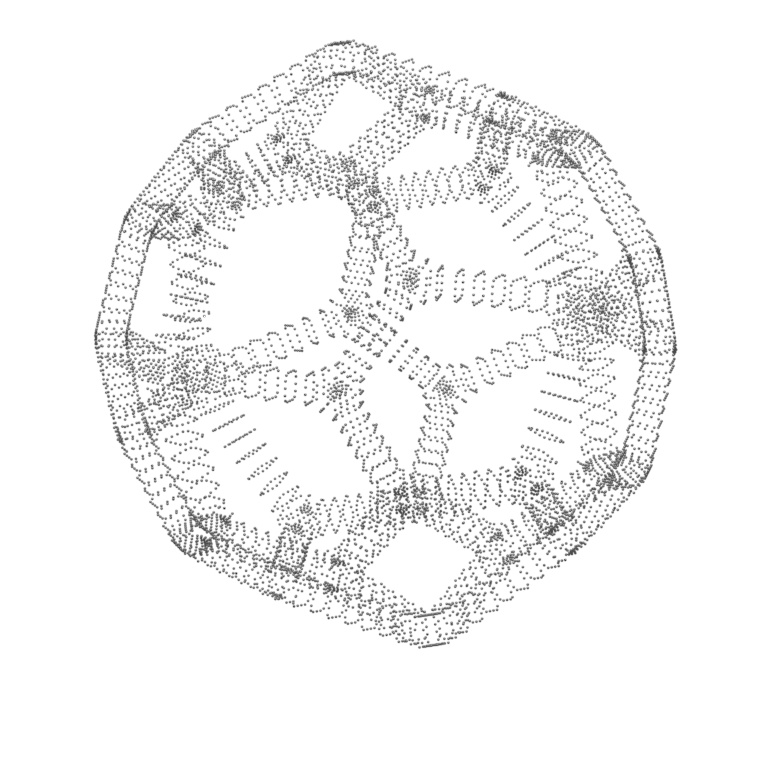}&
    \includegraphics[width=0.64in]{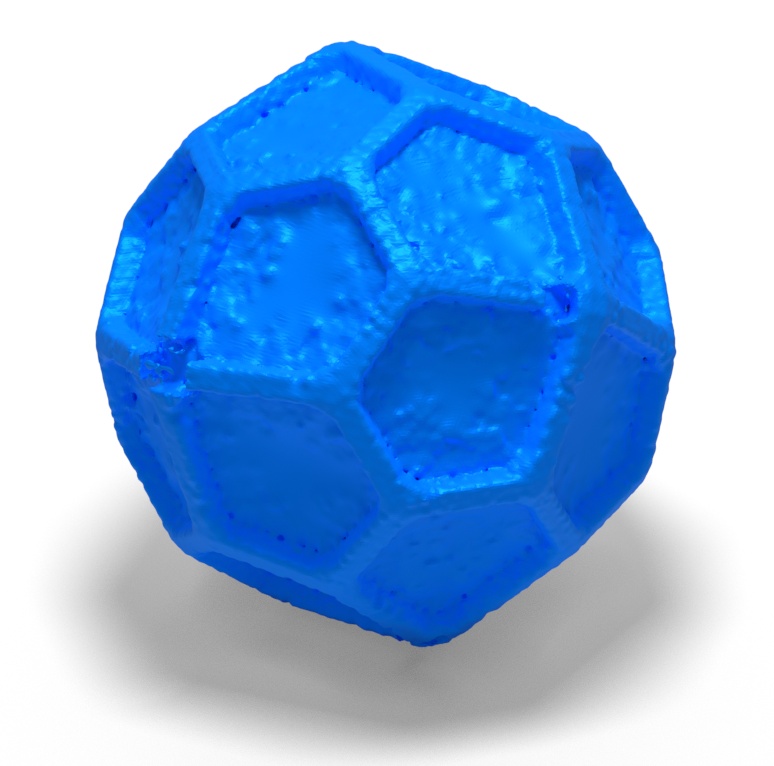}&
    \includegraphics[width=0.64in]{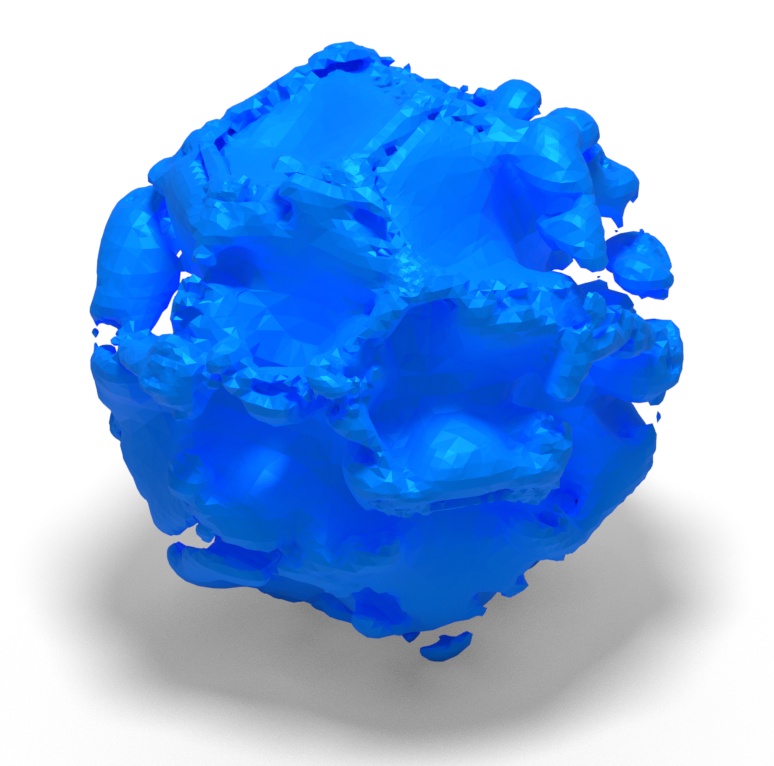}&
    \includegraphics[width=0.64in]{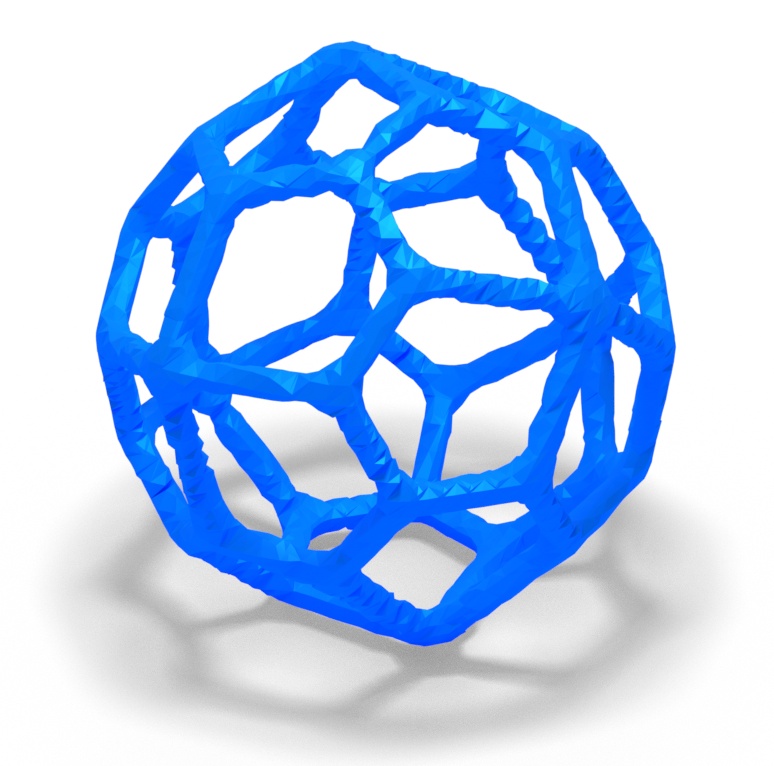}&
    \includegraphics[width=0.64in]{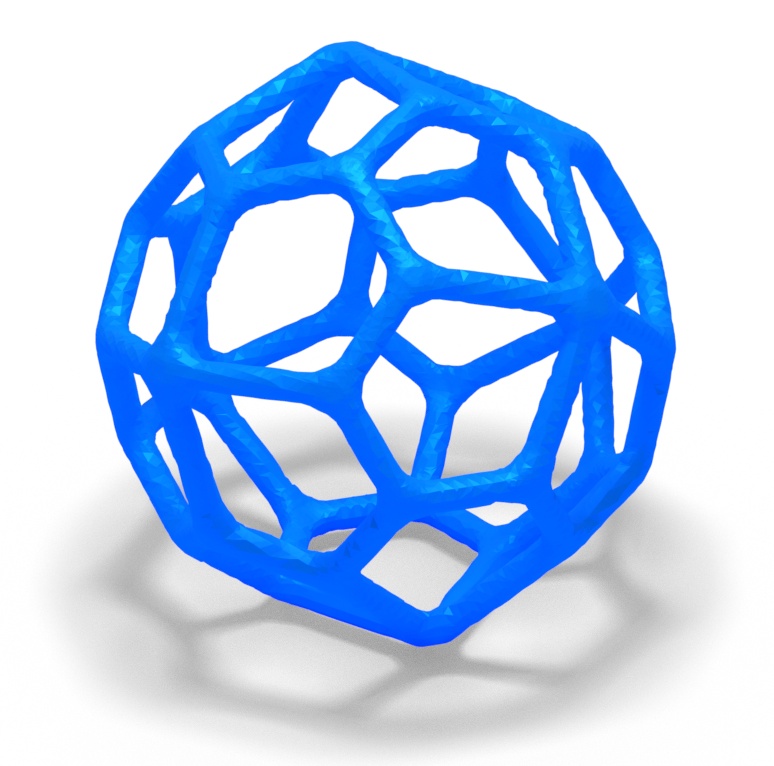}\\
    \includegraphics[width=0.64in]{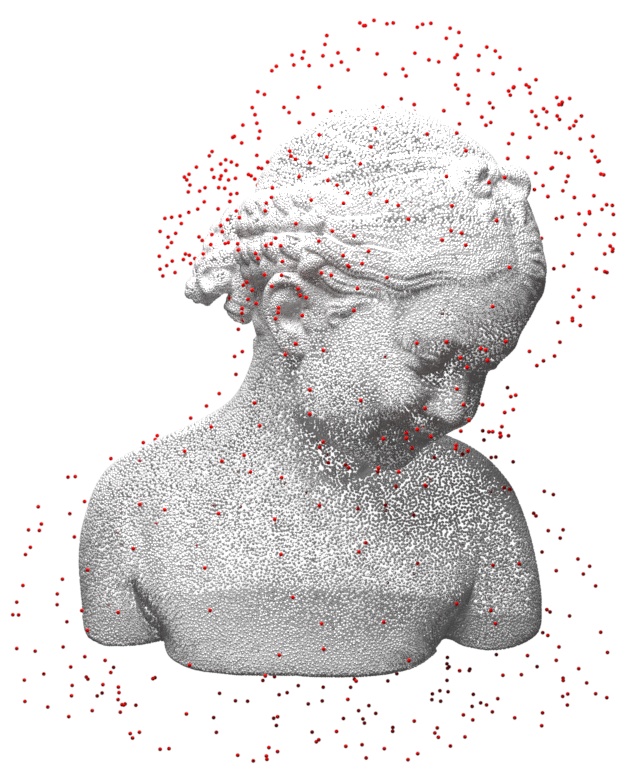} &
    \includegraphics[width=0.64in]{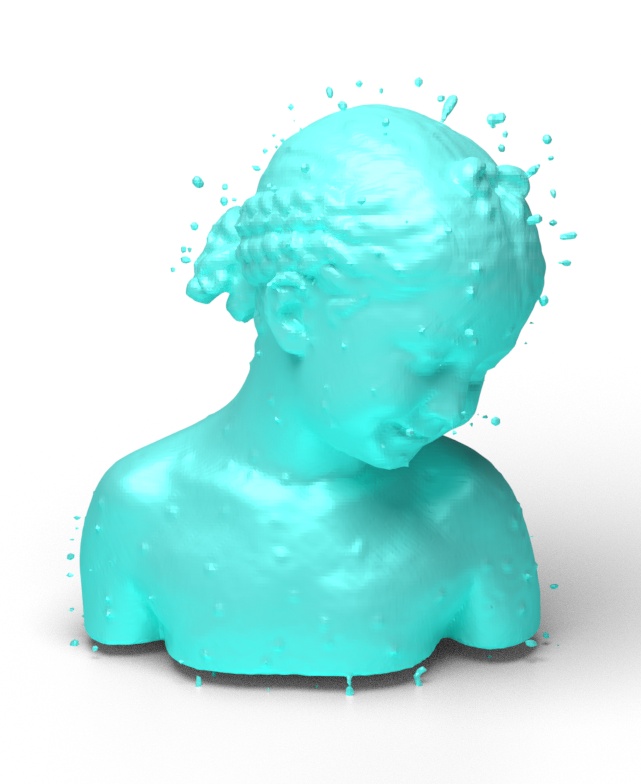} &
    \includegraphics[width=0.64in]{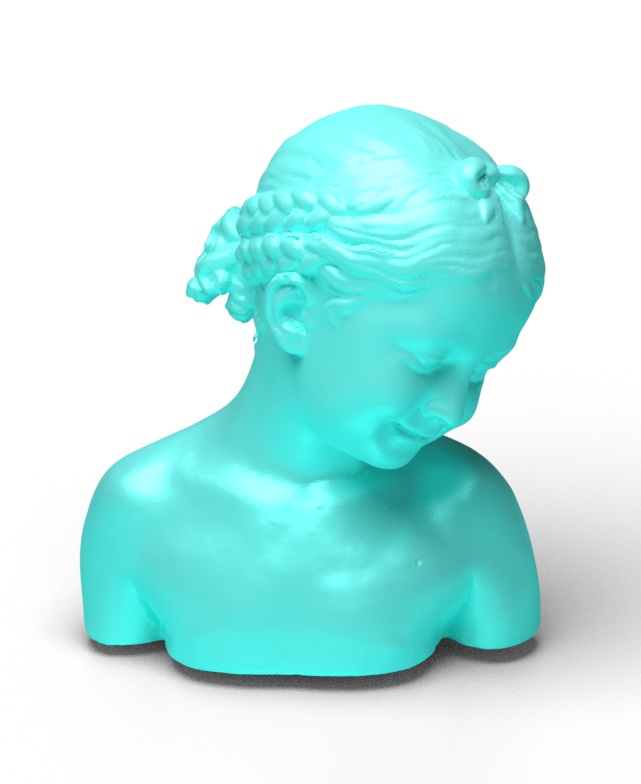} &
    \includegraphics[width=0.64in]{./figures/NA.pdf} &
    \includegraphics[width=0.64in]{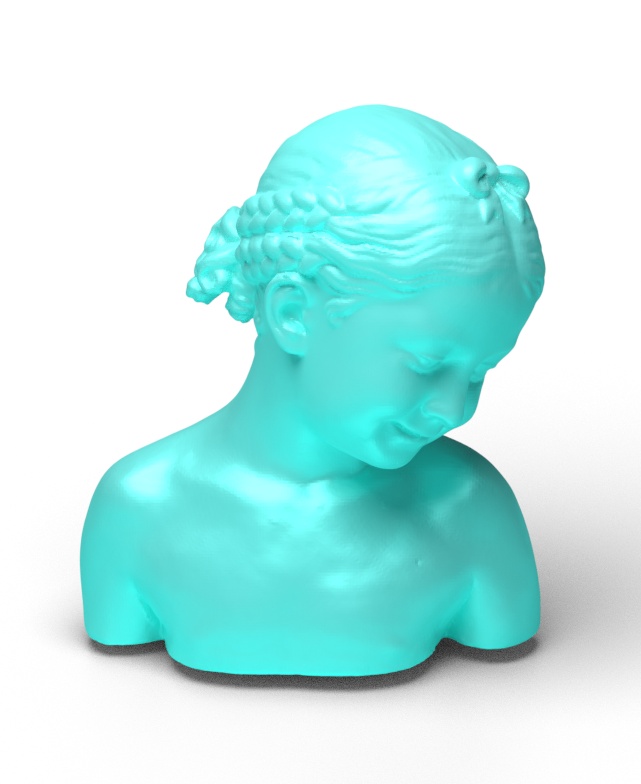} &
    \includegraphics[width=0.78in]{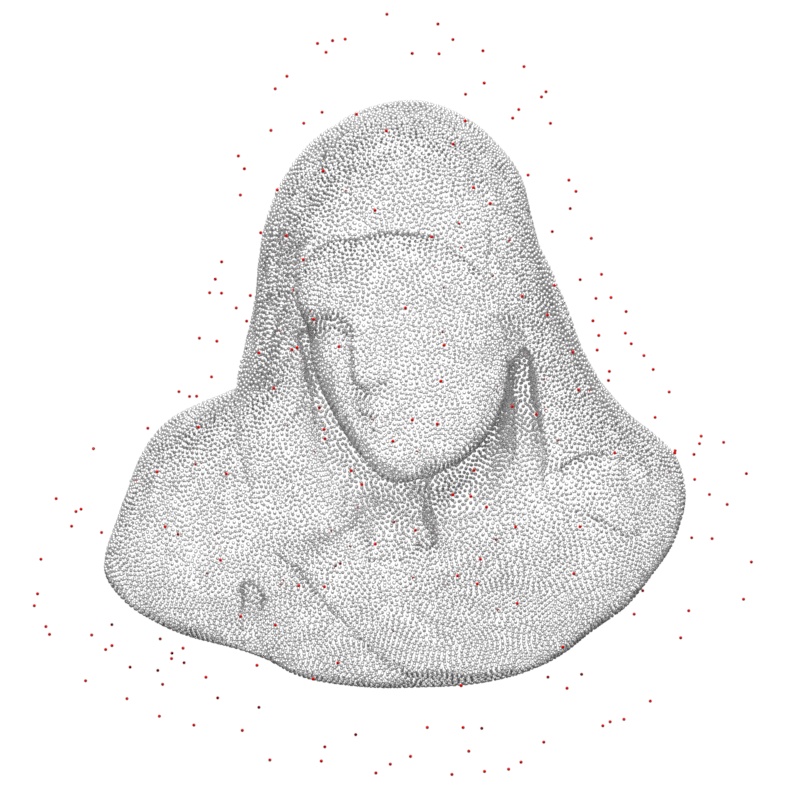} &
    \includegraphics[width=0.78in]{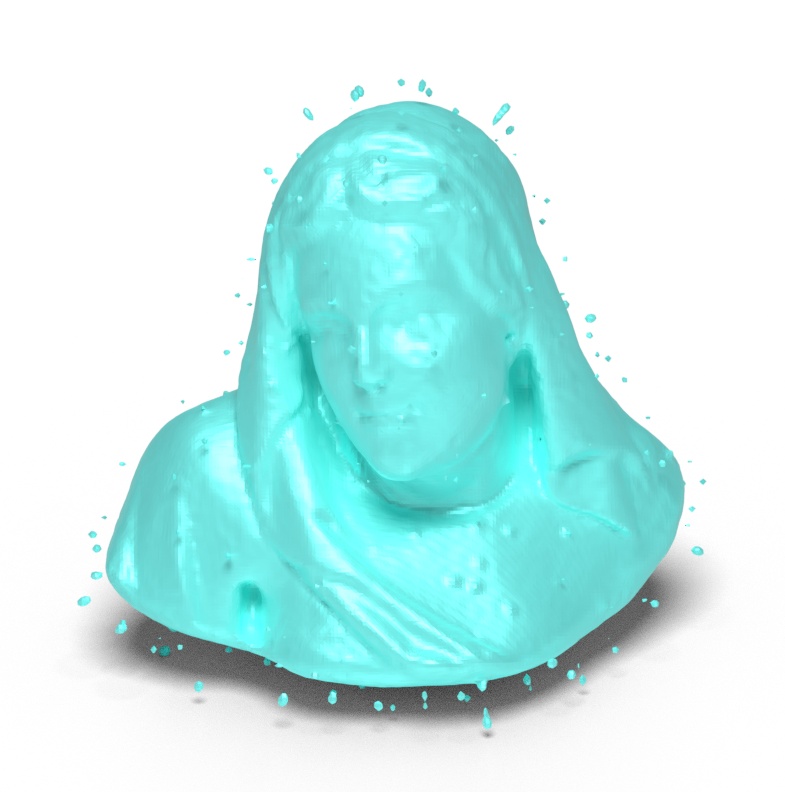} &
    \includegraphics[width=0.78in]{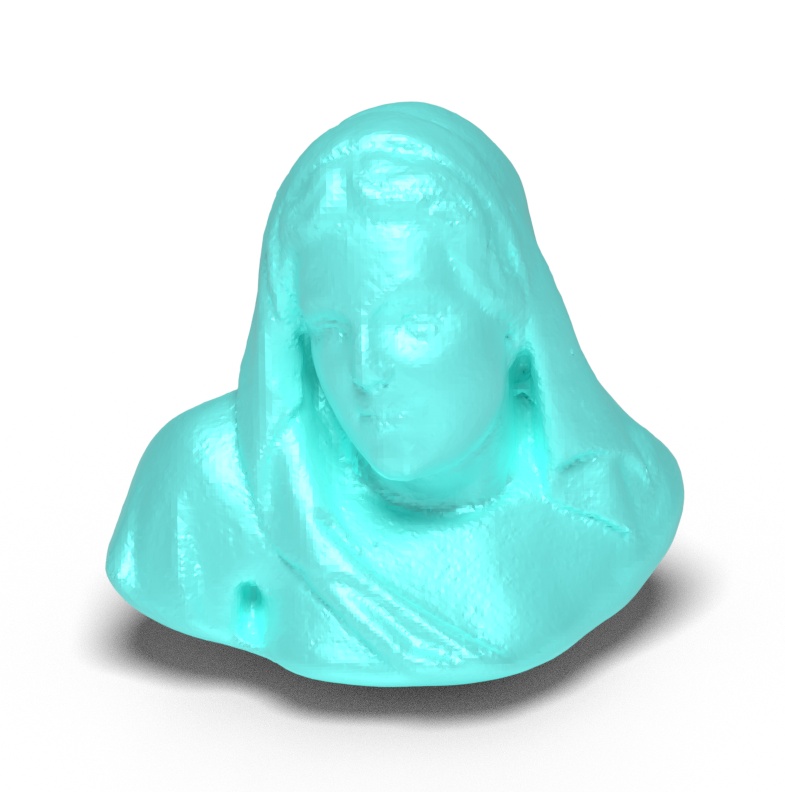} &
    \includegraphics[width=0.64in]{./figures/NA.pdf} &
    \includegraphics[width=0.78in]{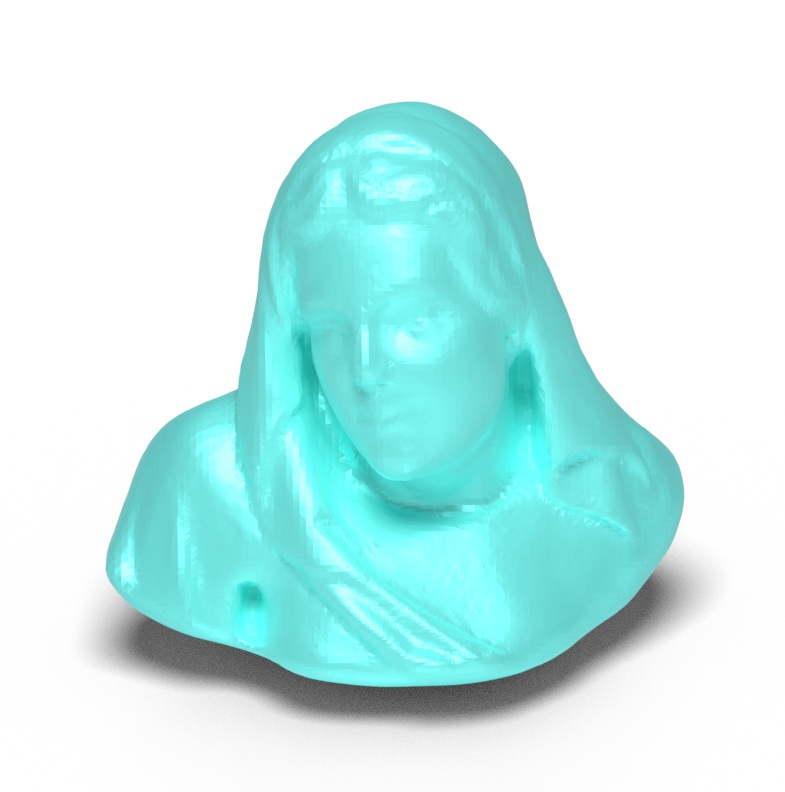}\\
    \includegraphics[width=0.64in]{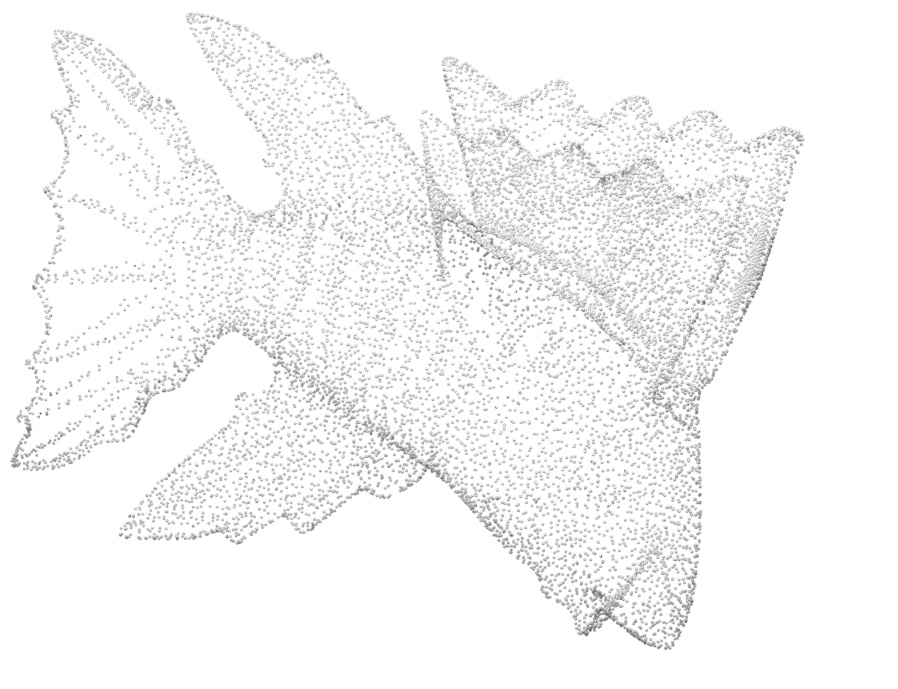}&
    \includegraphics[width=0.7in]{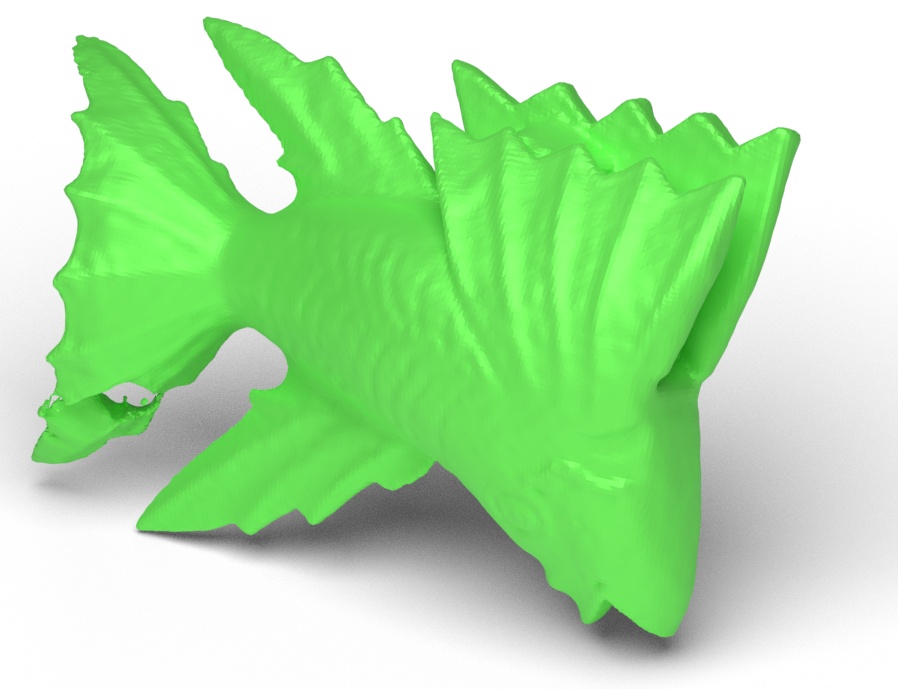}&
    \includegraphics[width=0.7in]{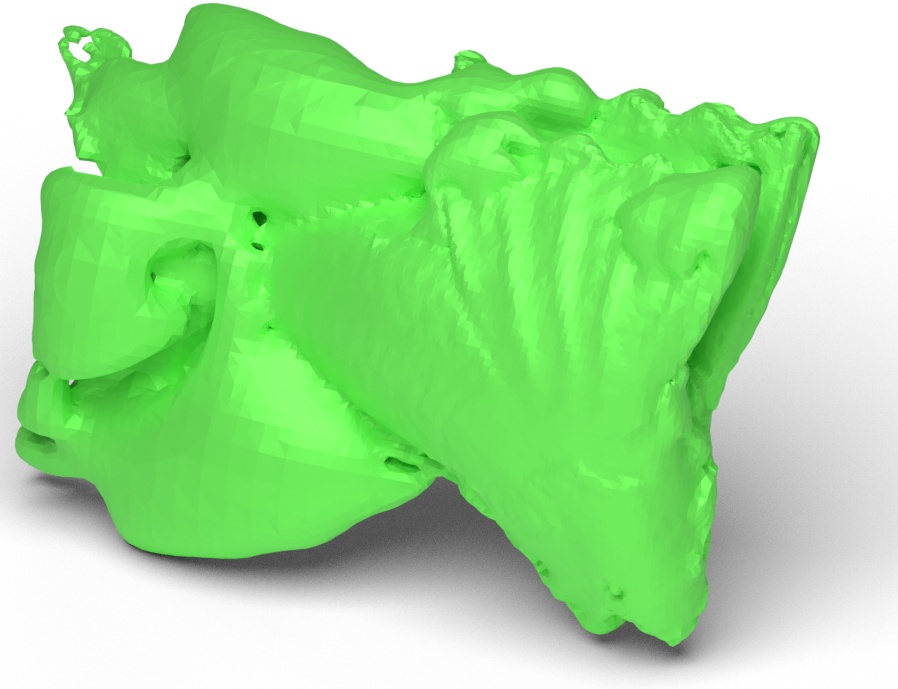}&
    \includegraphics[width=0.64in]{./figures/NA.pdf}&
    \includegraphics[width=0.7in]{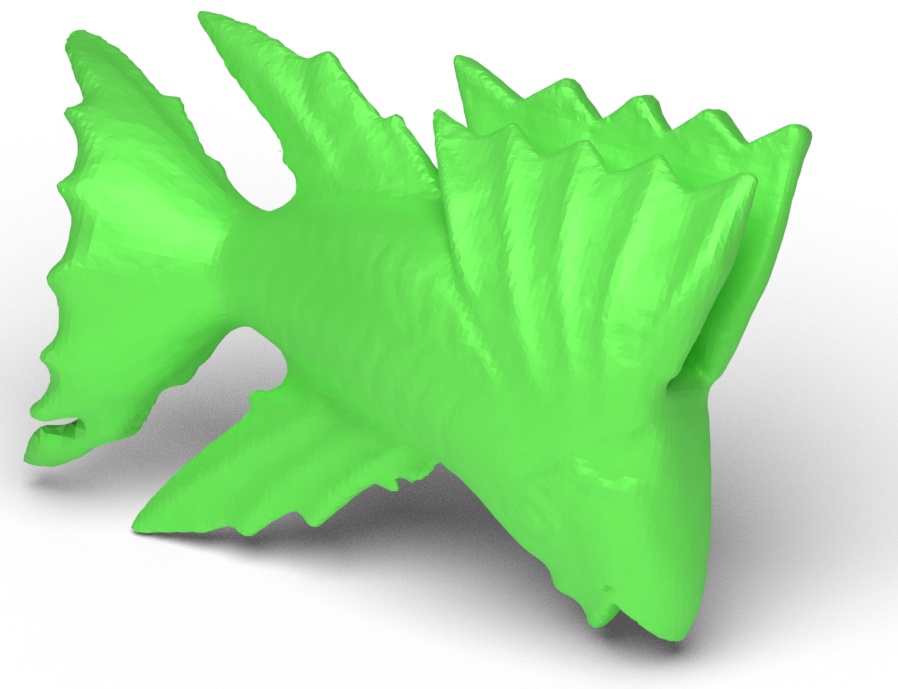}&
    \includegraphics[width=0.6in]{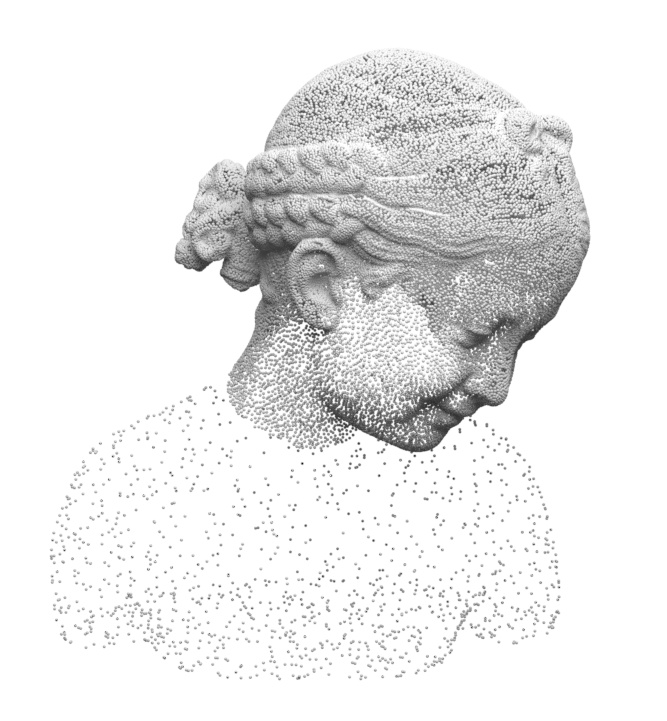}&
    \includegraphics[width=0.6in]{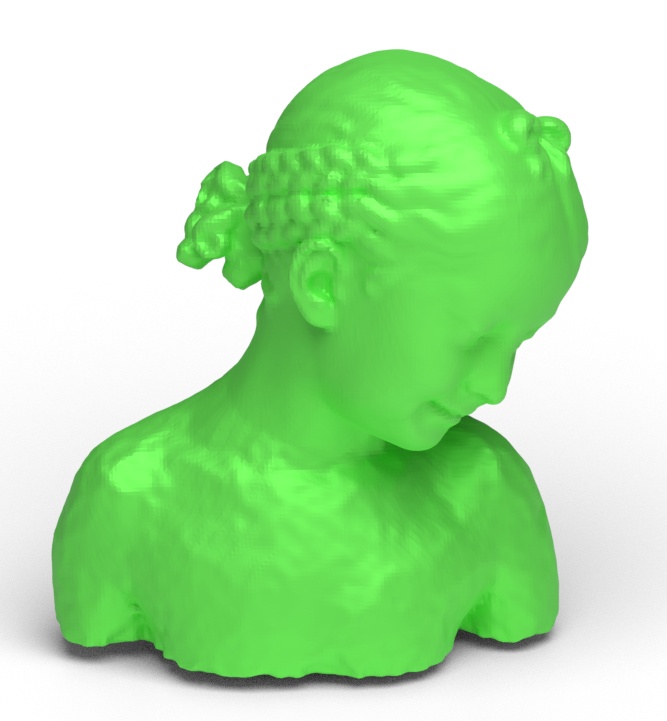}&
    \includegraphics[width=0.6in]{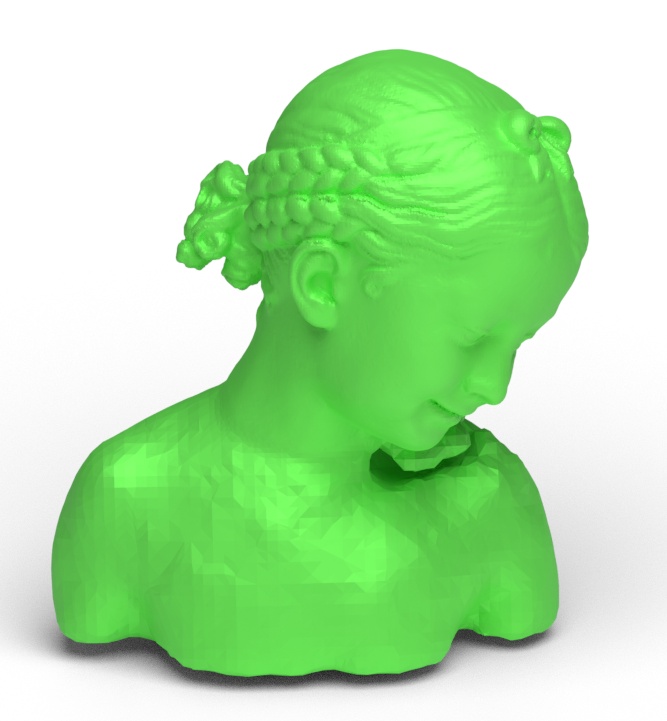}&
    \includegraphics[width=0.64in]{./figures/NA.pdf}&
    \includegraphics[width=0.6in]{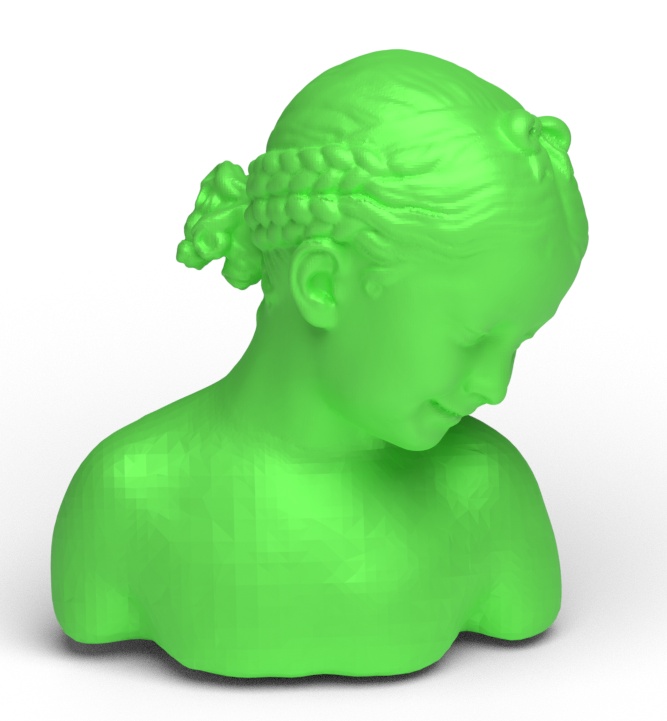}\\
    \includegraphics[width=0.58in]{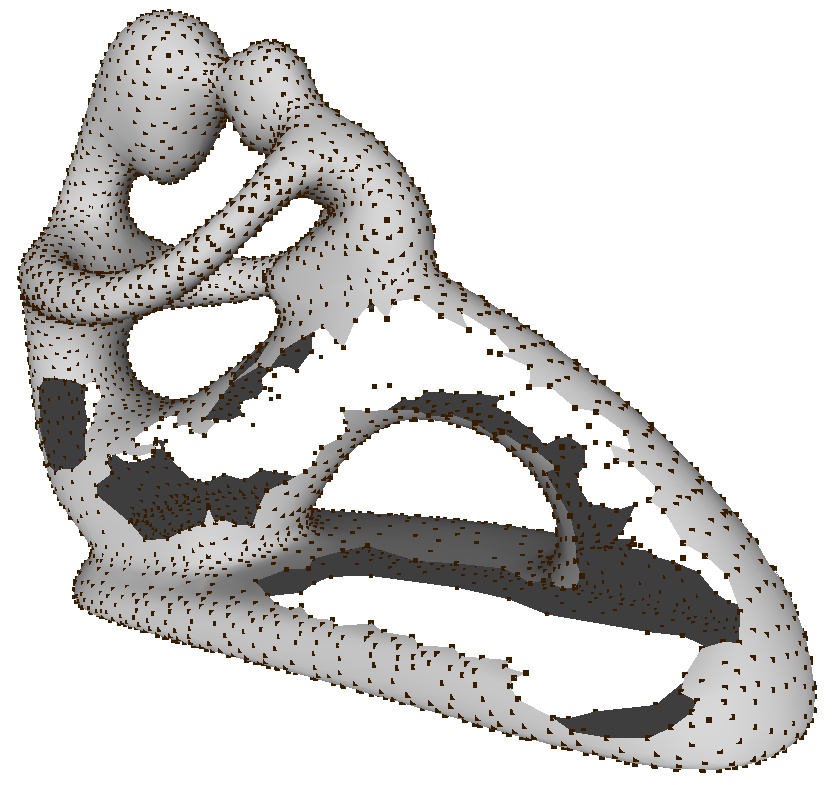}&
    \includegraphics[width=0.58in]{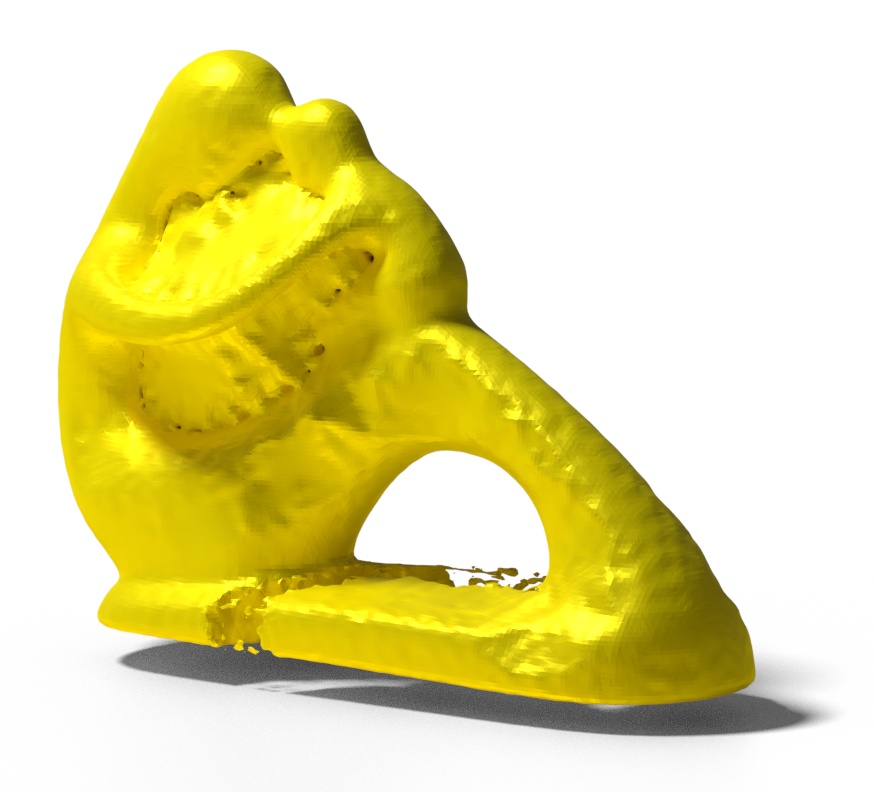}&
    \includegraphics[width=0.58in]{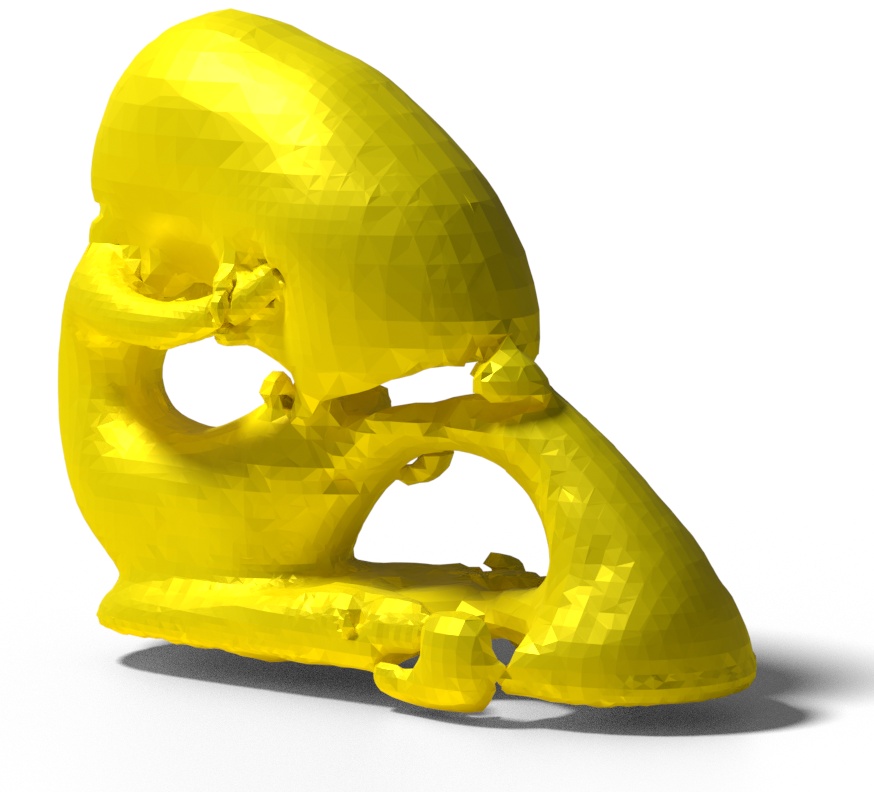}&
    \includegraphics[width=0.58in]{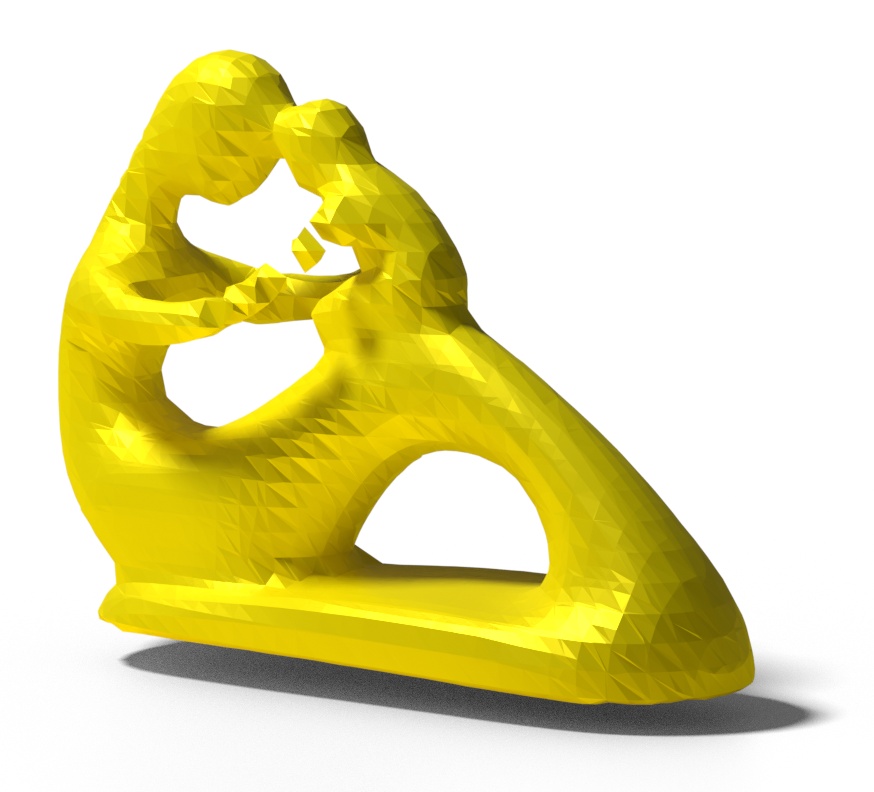}&
    \includegraphics[width=0.58in]{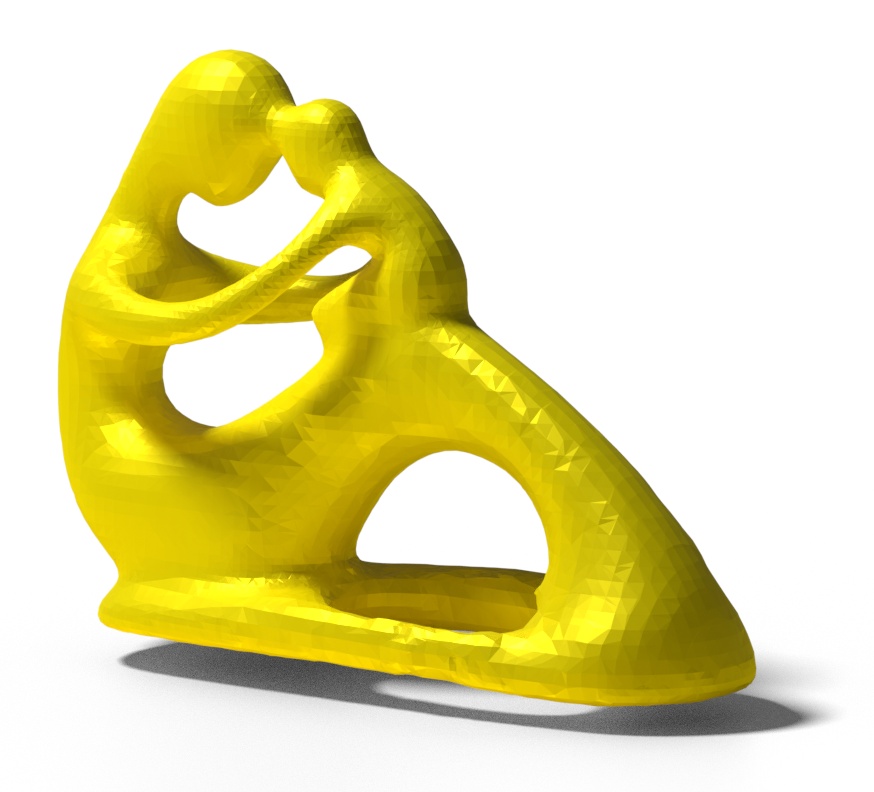}&
    \includegraphics[width=0.58in]{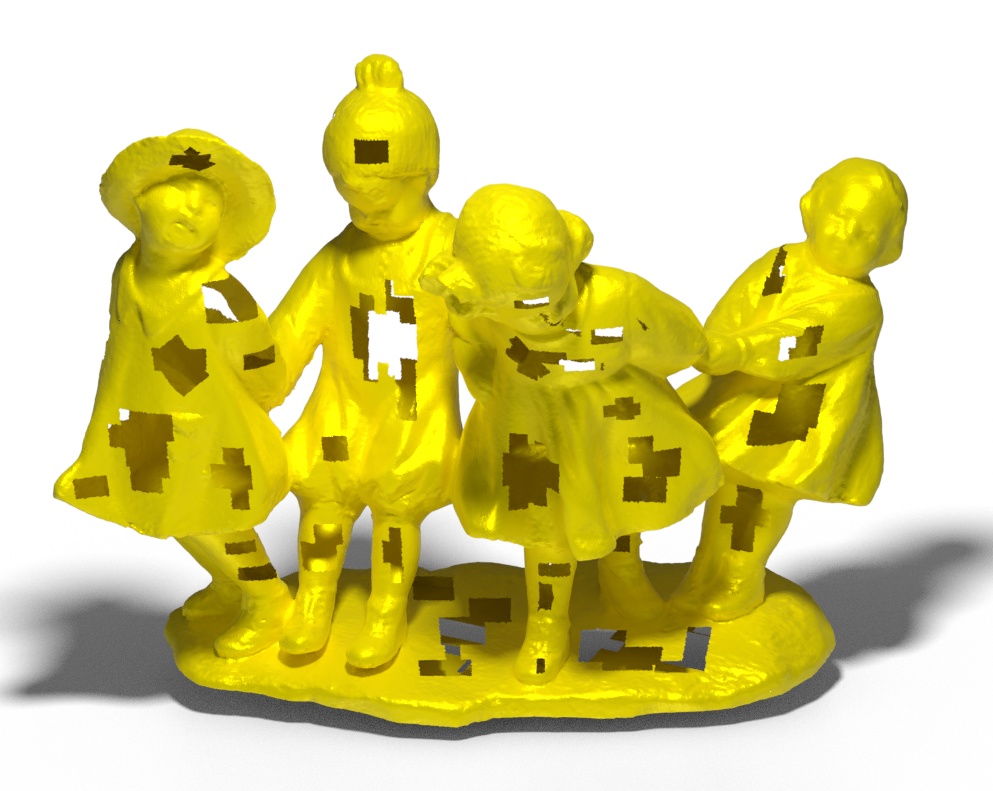}&
    \includegraphics[width=0.64in]{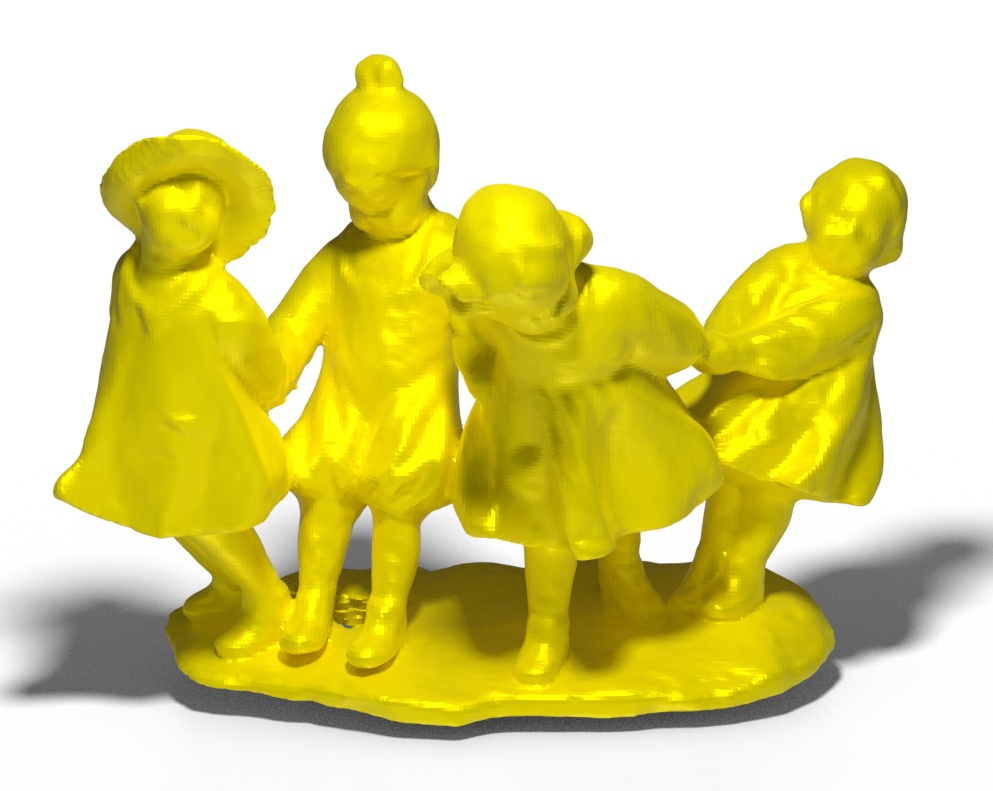}&
    \includegraphics[width=0.64in]{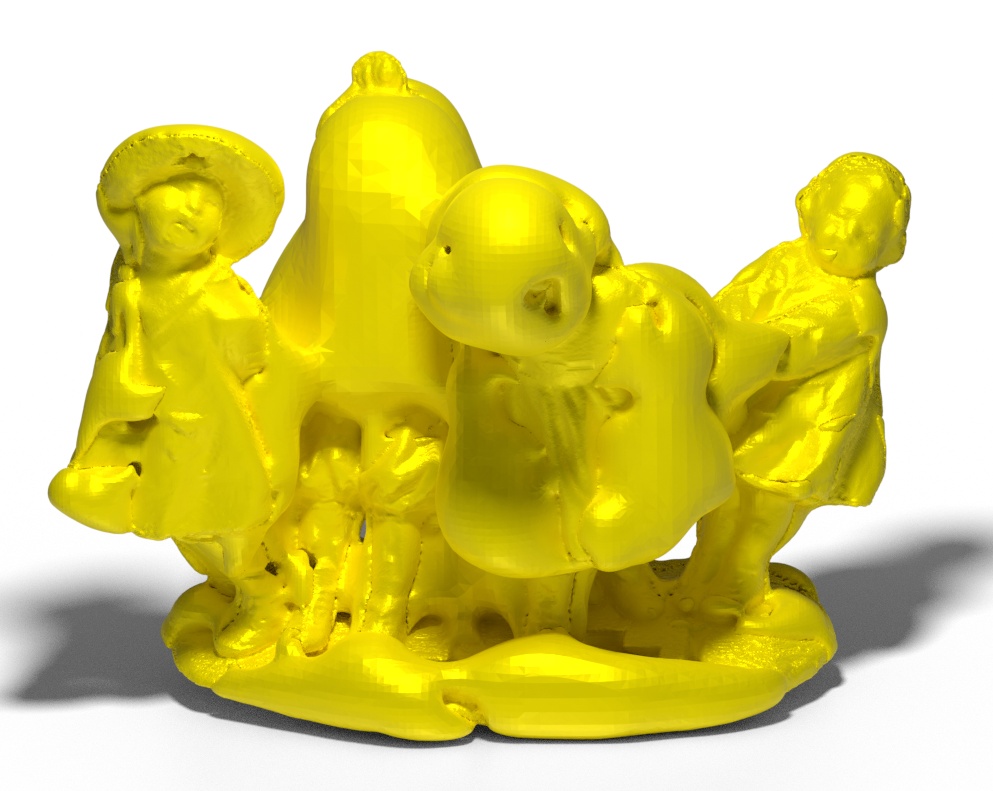}&
    \includegraphics[width=0.64in]{./figures/NA.pdf}&
    \includegraphics[width=0.64in]{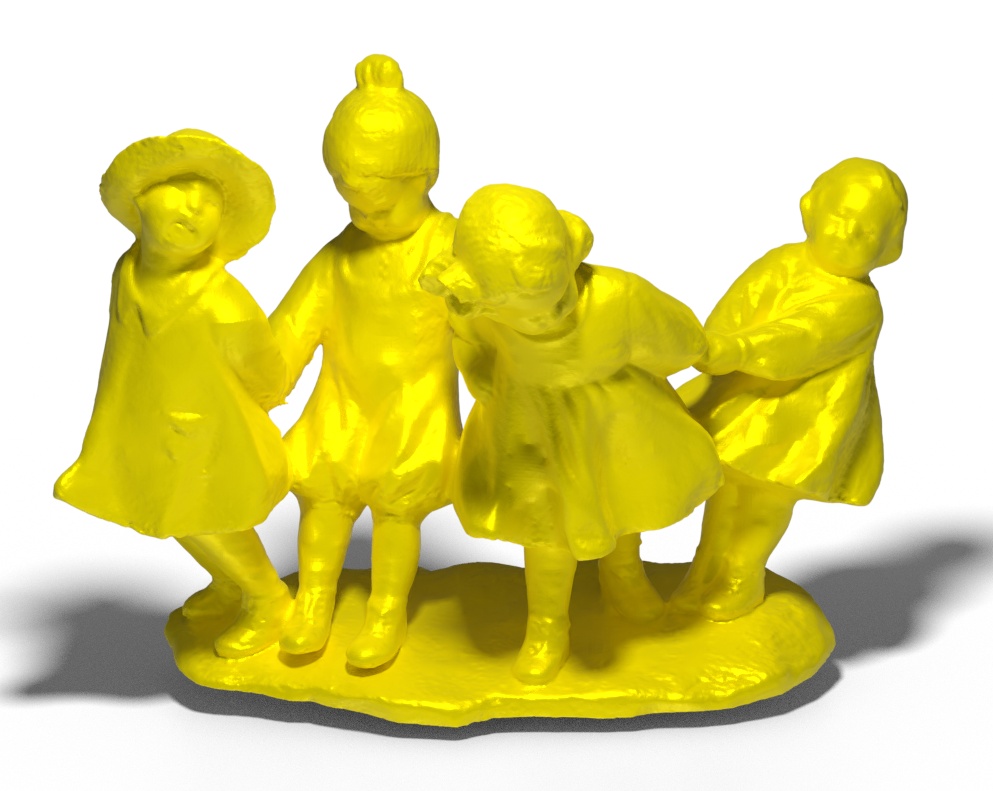}\\
    \includegraphics[width=0.64in]{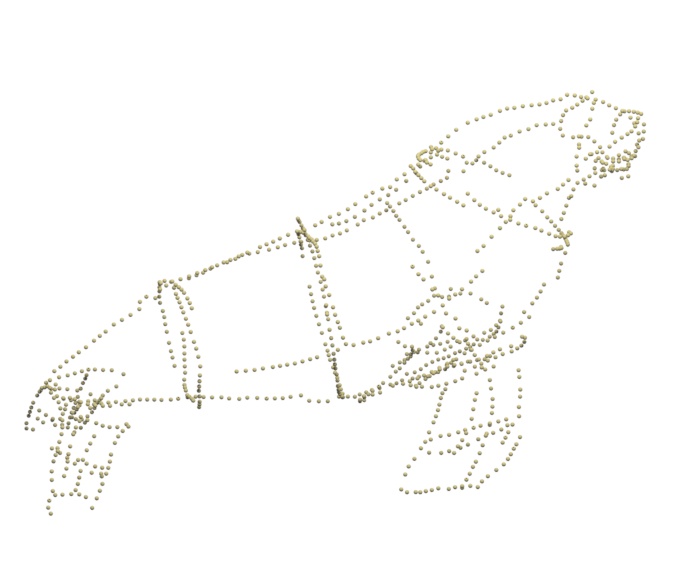}&
    \includegraphics[width=0.64in]{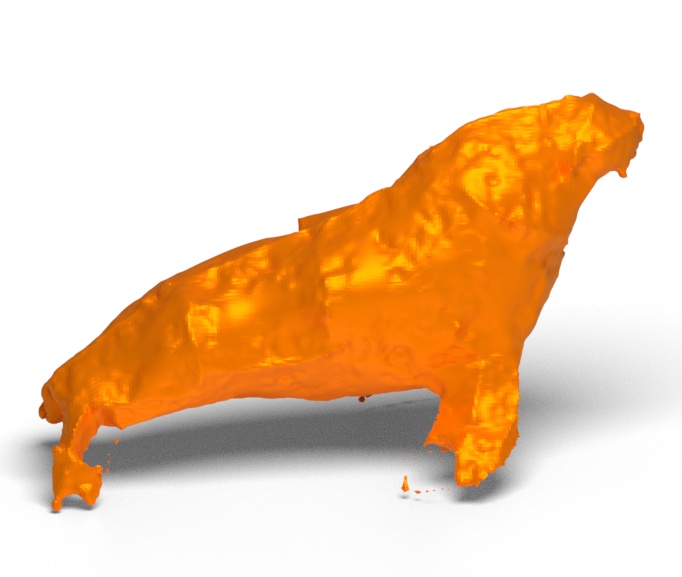}&
    \includegraphics[width=0.64in]{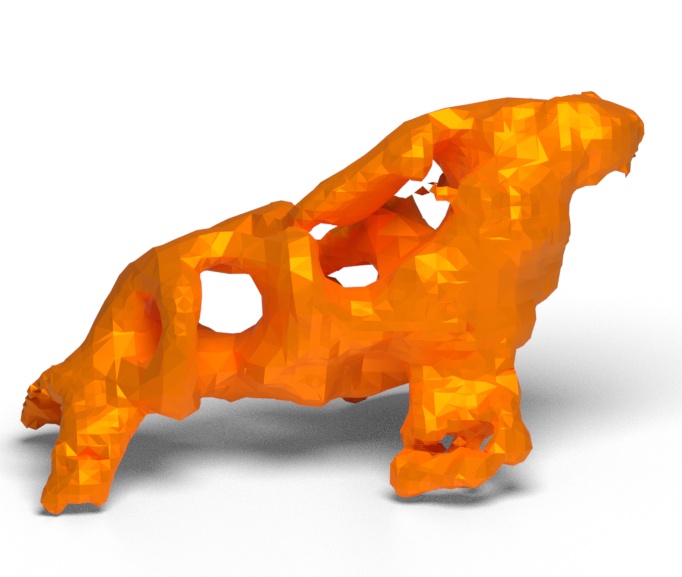}&
    \includegraphics[width=0.64in]{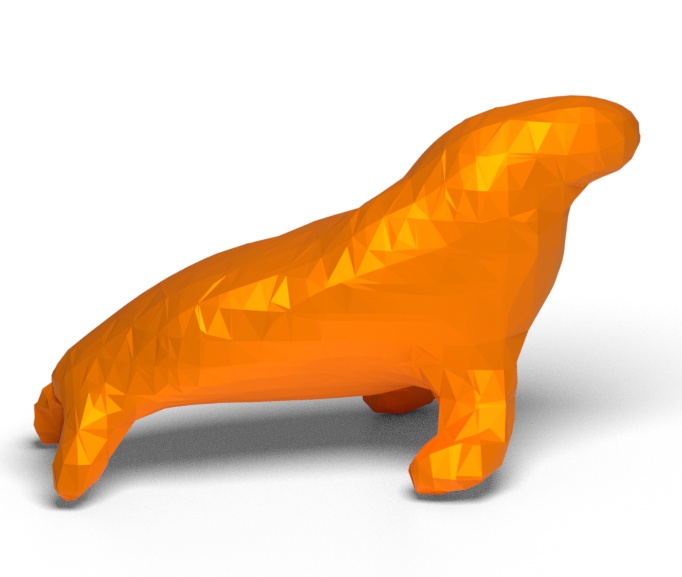}&
    \includegraphics[width=0.64in]{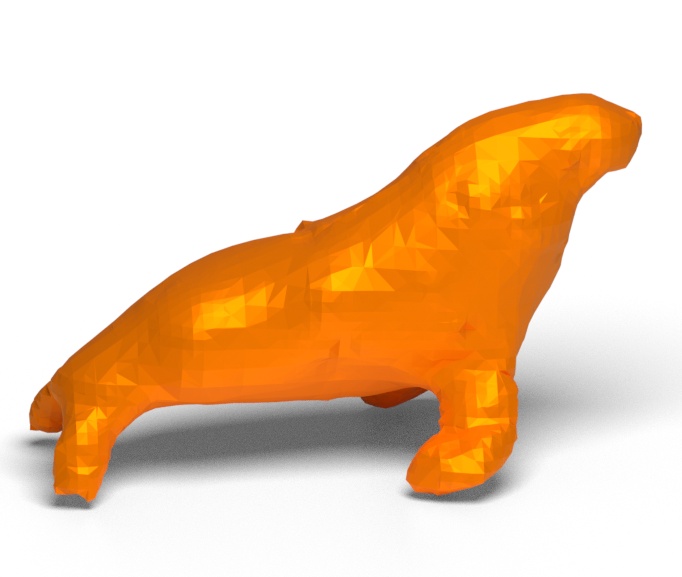}&
    \includegraphics[width=0.64in]{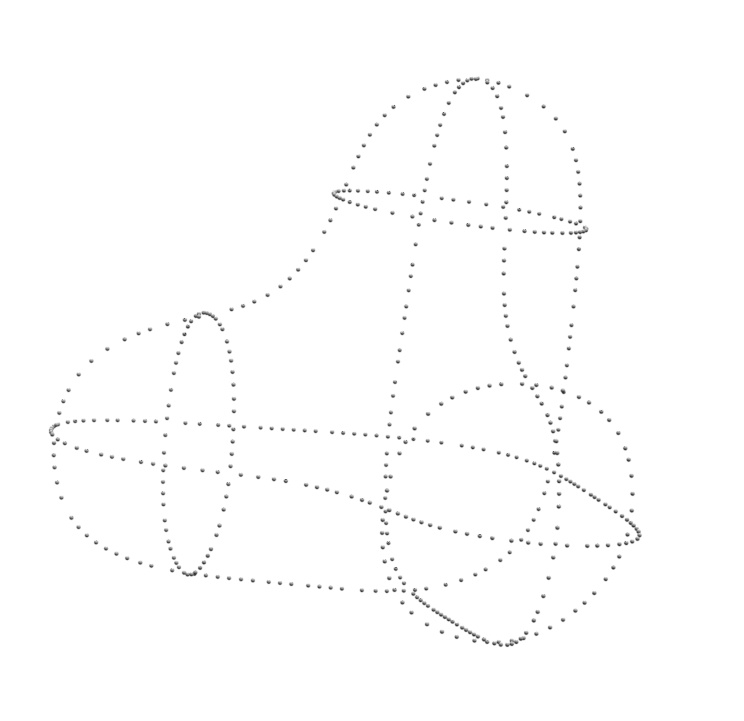}&
    \includegraphics[width=0.64in]{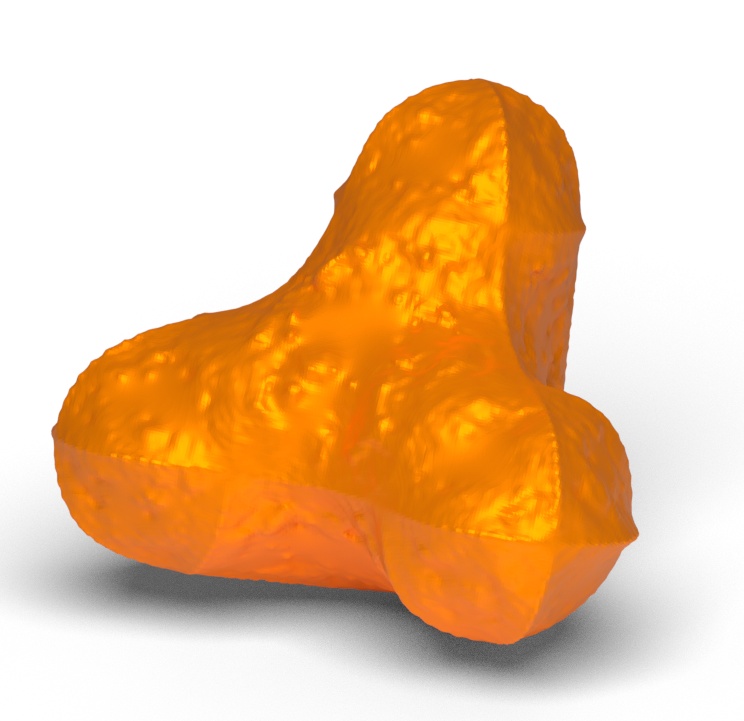}&
    \includegraphics[width=0.64in]{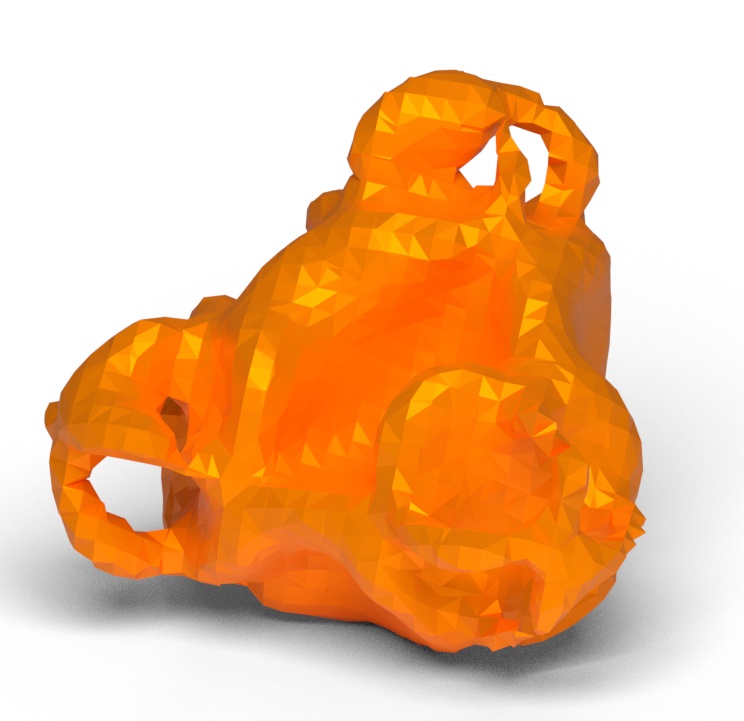}&
    \includegraphics[width=0.64in]{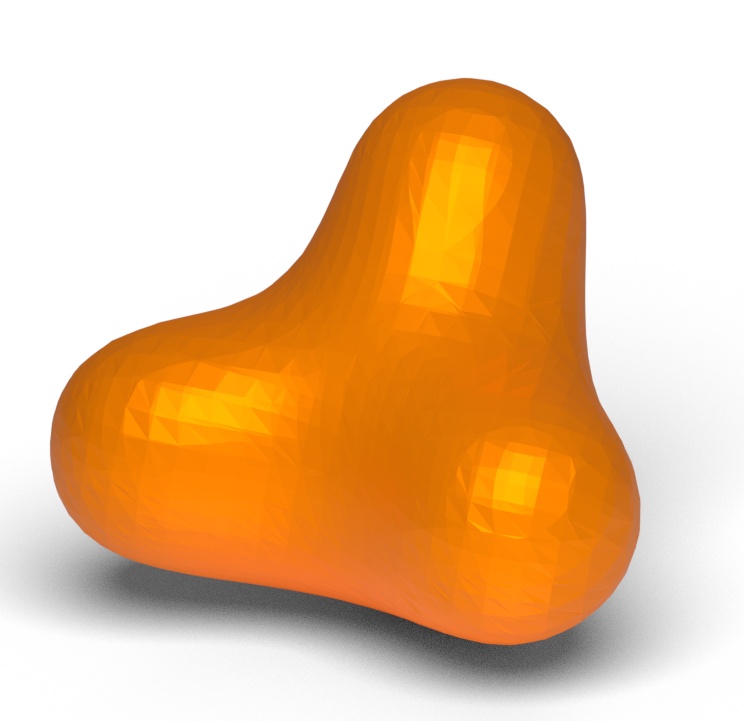}&
    \includegraphics[width=0.64in]{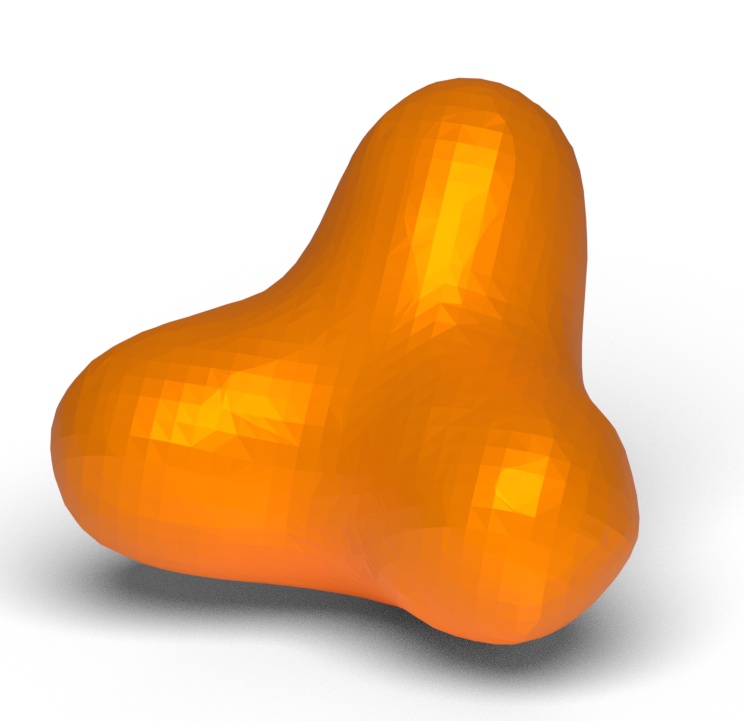}\\
  Input & DPSR & Dipole &  VIPSS & iPSR & Input & DPSR & Dipole & VIPSS & iPSR\\
\end{tabular}
\end{scriptsize}
\caption{Robustness. From top to bottom: noisy data, high-genus models, models with 1\% outliers, non-uniform points, incomplete models and sparse points. VIPSS cannot work for high resolution models due to high memory consumption. Images are rendered in high resolutions, allowing zoom-in examination.}
\label{fig:robustness}
\end{figure*}

\subsection{Reconstruction Quality and Robustness}
We measured the quality of the reconstructed surfaces for all methods using reconstruction error $\varepsilon$. We noticed that most of DPSR results contain multiple connected components even though the input points are samples from a single component. Therefore, we used only the largest component to measure the accuracy of all the methods for fair evaluation.
Table~\ref{tab:performance} reports the statistics for a few representative models.
Figure~\ref{fig:histogram} shows the histogram of reconstruction error of the AIM$@$SHAPE dataset. For 72.1\% of the models, our results have reconstruction error less than 2\%. The mean error 3.18\% and median 0.71\% are also quite low, demonstrating the high quality of the reconstructed surfaces.

\begin{figure}[!htbp]
    \centering
    \begin{scriptsize}
    \includegraphics[width=0.525in]{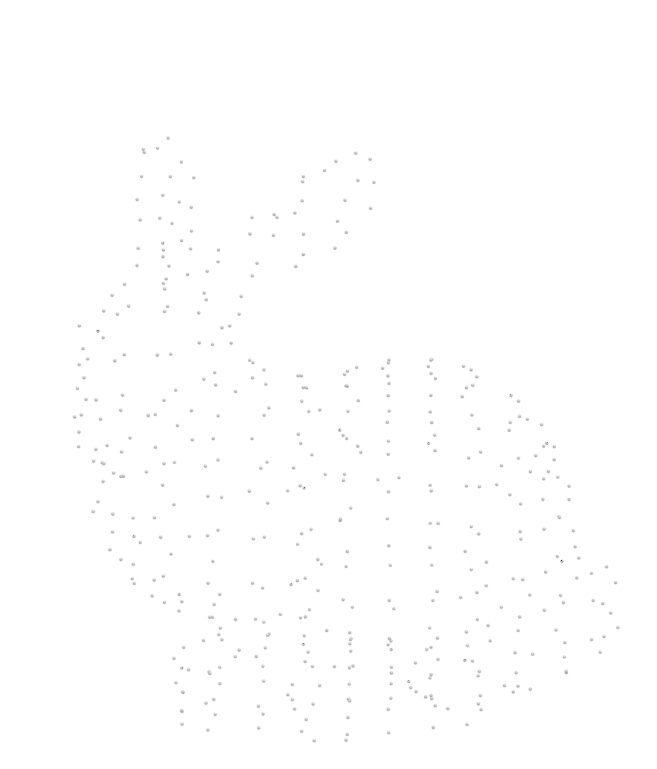}
    \includegraphics[width=0.525in]{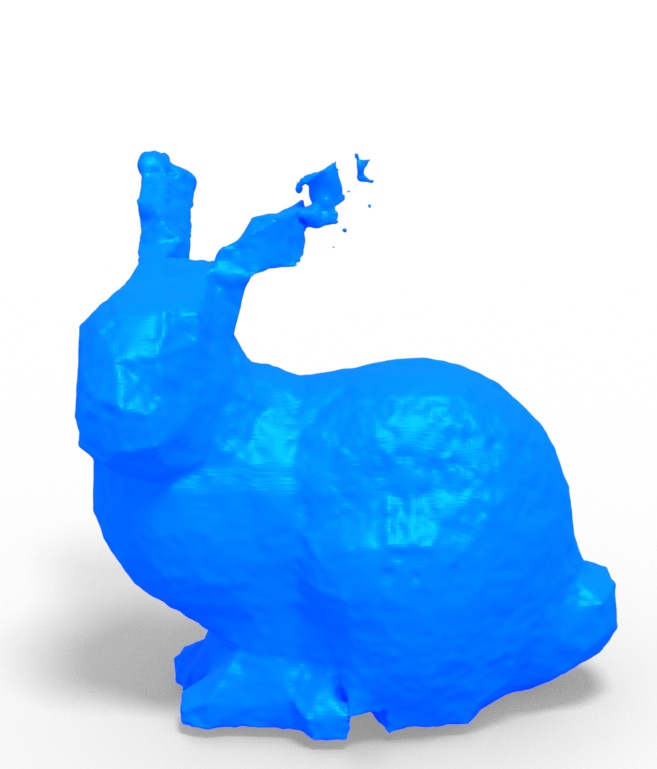}
    \includegraphics[width=0.525in]{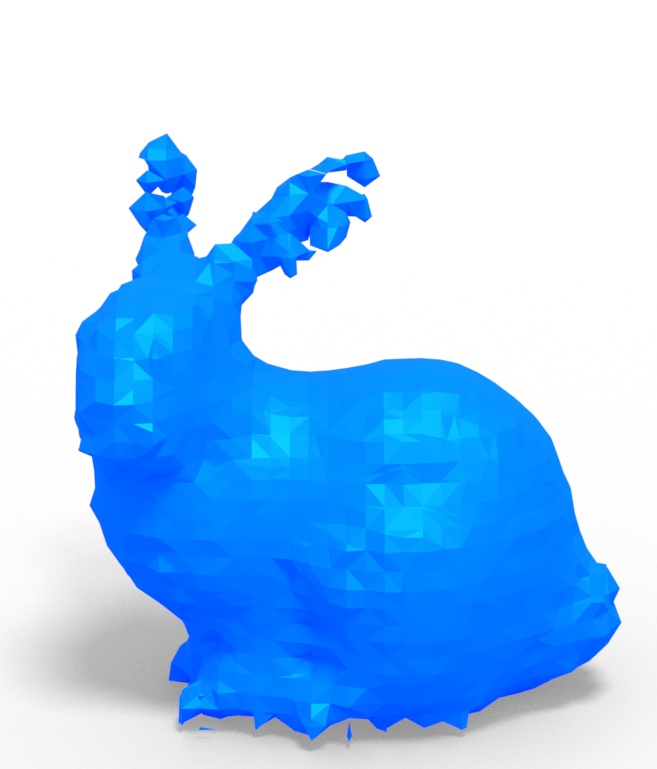}
    \includegraphics[width=0.525in]{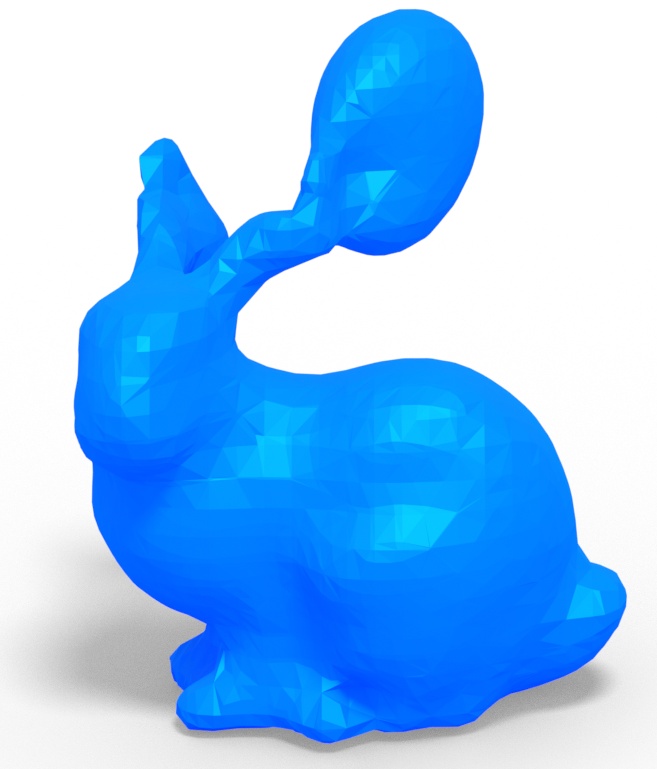}
    \includegraphics[width=0.525in]{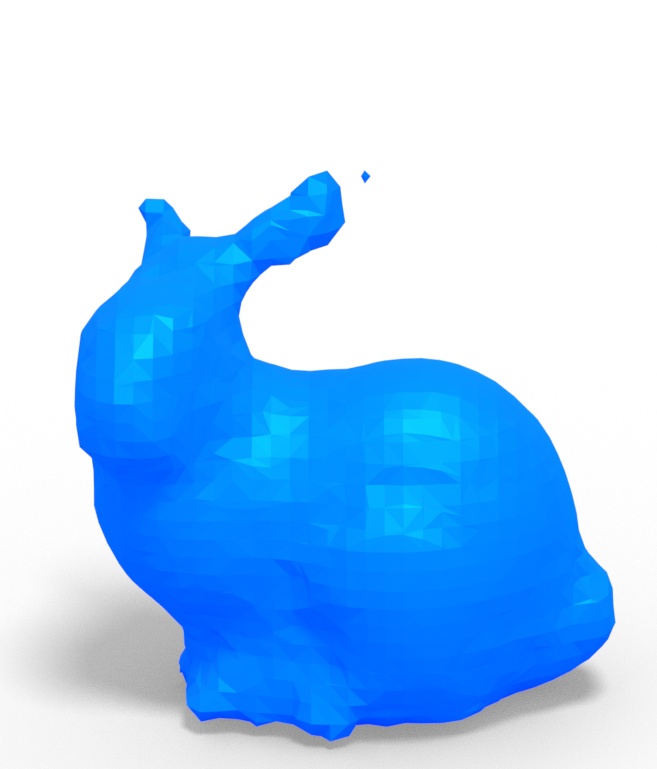}
    \includegraphics[width=0.525in]{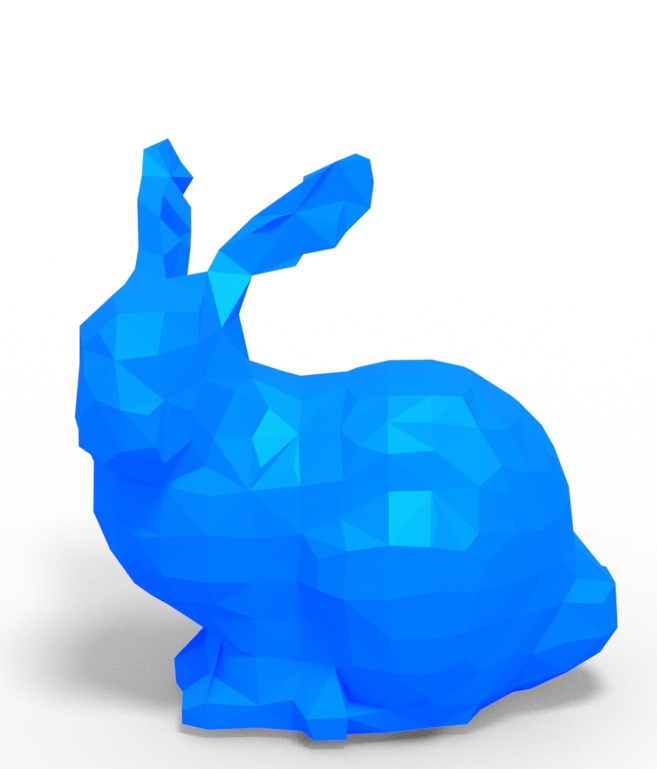}\\
    \makebox[0.525in]{453 pts}
    \makebox[0.525in]{(5.31\%,136.2s)}
    \makebox[0.525in]{(11.96\%,7.9s)}
    \makebox[0.525in]{(23.91\%,6.3s)}
    \makebox[0.525in]{(9.96\%,7.6s)}
    \makebox[0.525in]{ - } \\
        \includegraphics[width=0.525in]{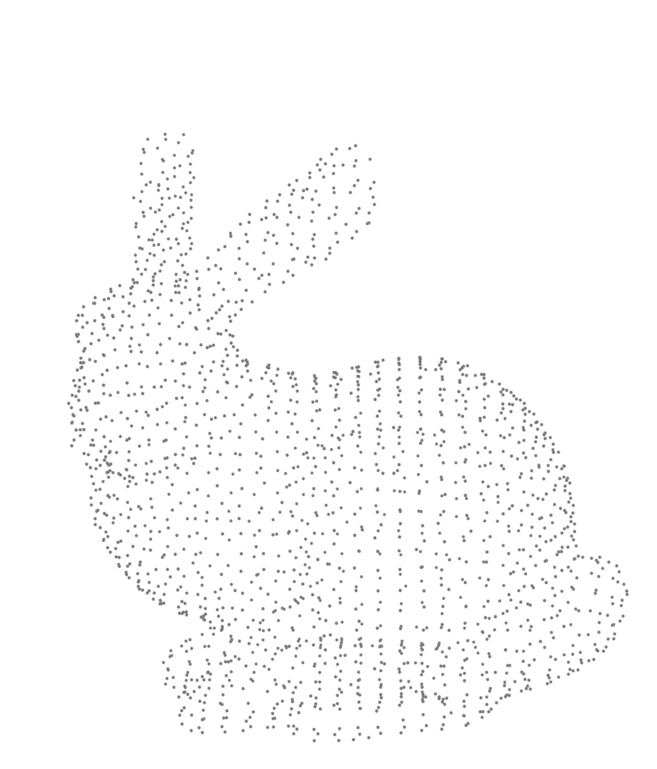}
    \includegraphics[width=0.525in]{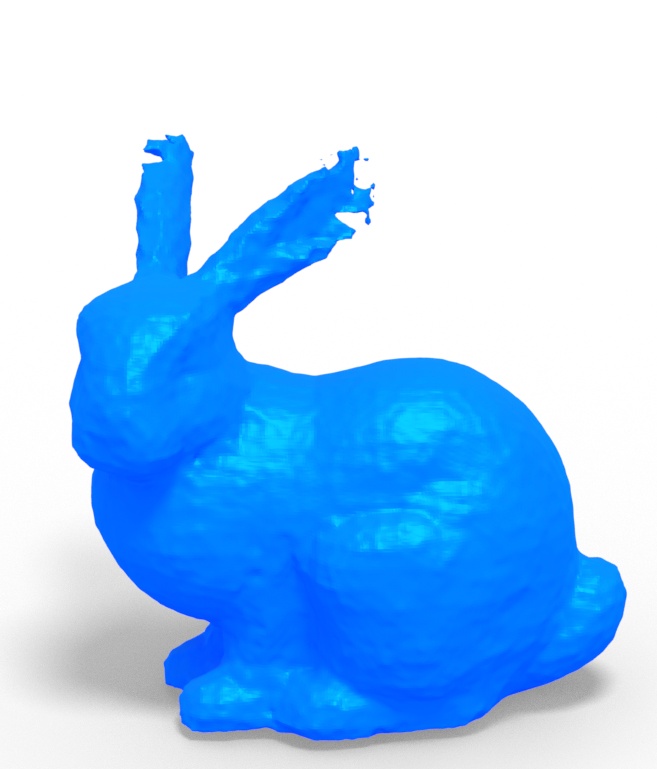}
    \includegraphics[width=0.525in]{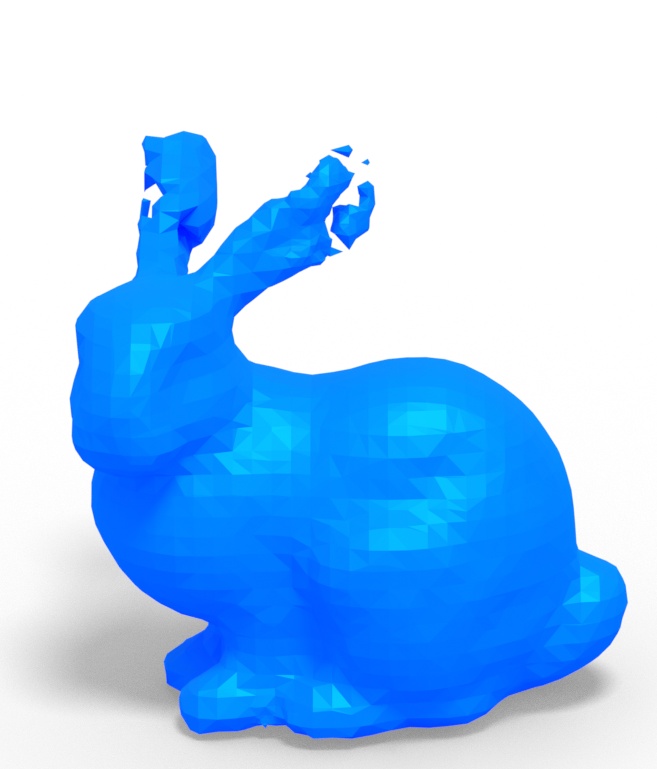}
    \includegraphics[width=0.525in]{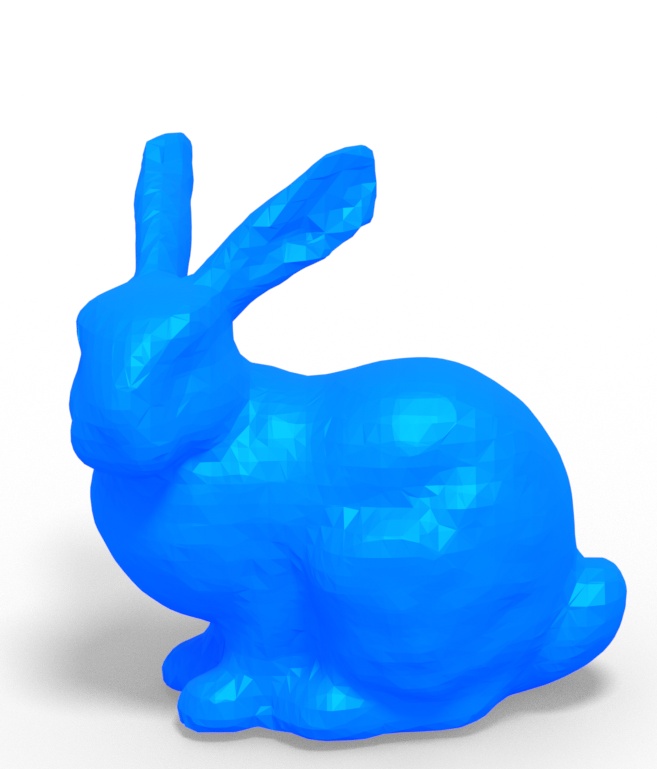}
    \includegraphics[width=0.525in]{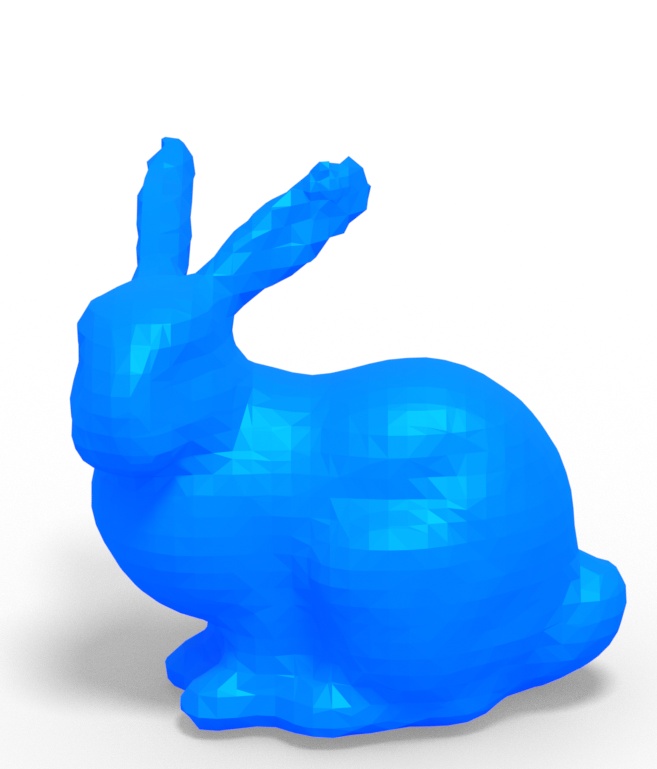}
    \includegraphics[width=0.525in]{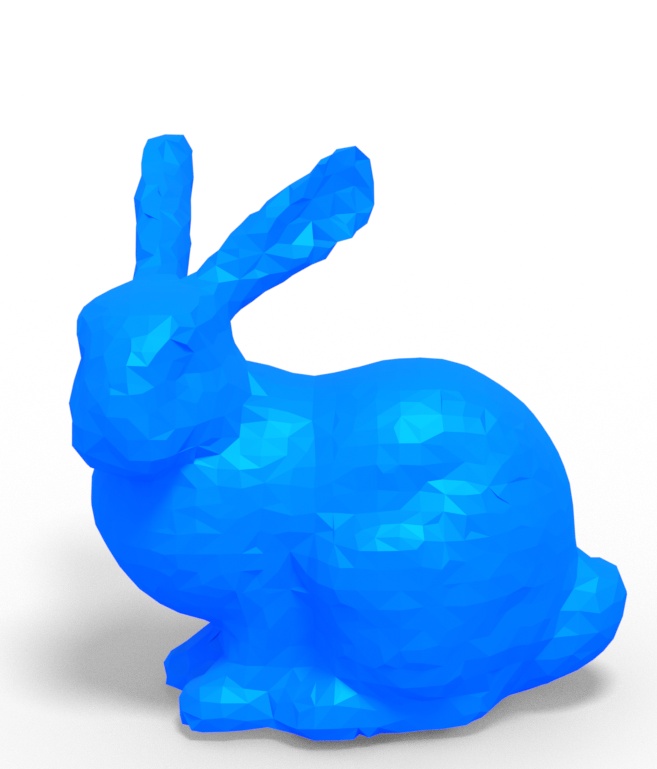}\\
    \makebox[0.525in]{1,889 pts}
    \makebox[0.525in]{(3.77\%,139.2s)}
    \makebox[0.525in]{(4.76\%,10.3s)}
    \makebox[0.525in]{(3.81\%,132.0s)}
    \makebox[0.525in]{(3.00\%,26.8s)}
    \makebox[0.525in]{ - } \\
    \includegraphics[width=0.525in]{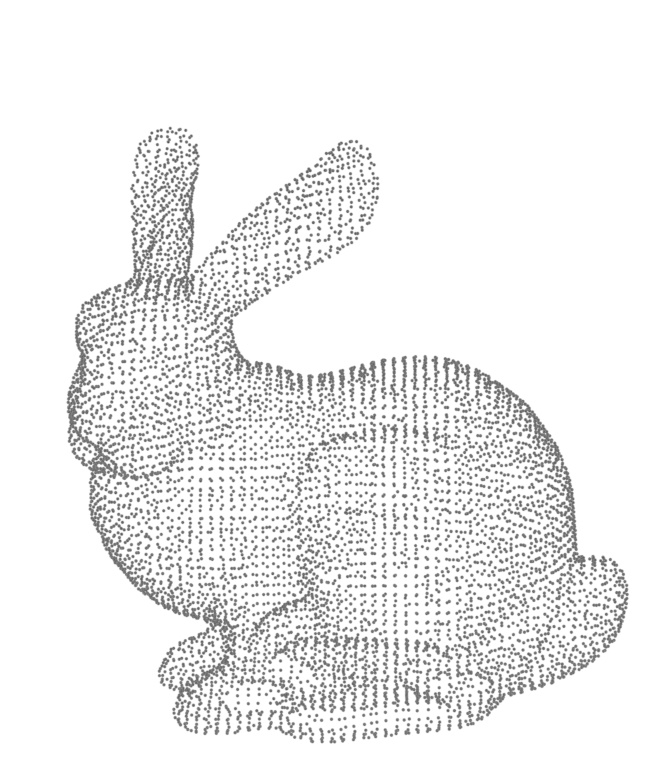}
    \includegraphics[width=0.525in]{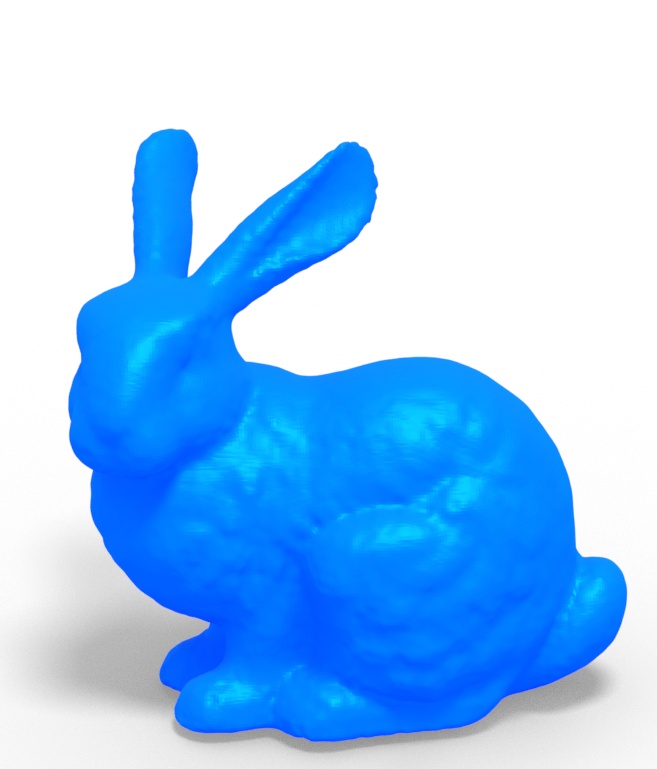}
    \includegraphics[width=0.525in]{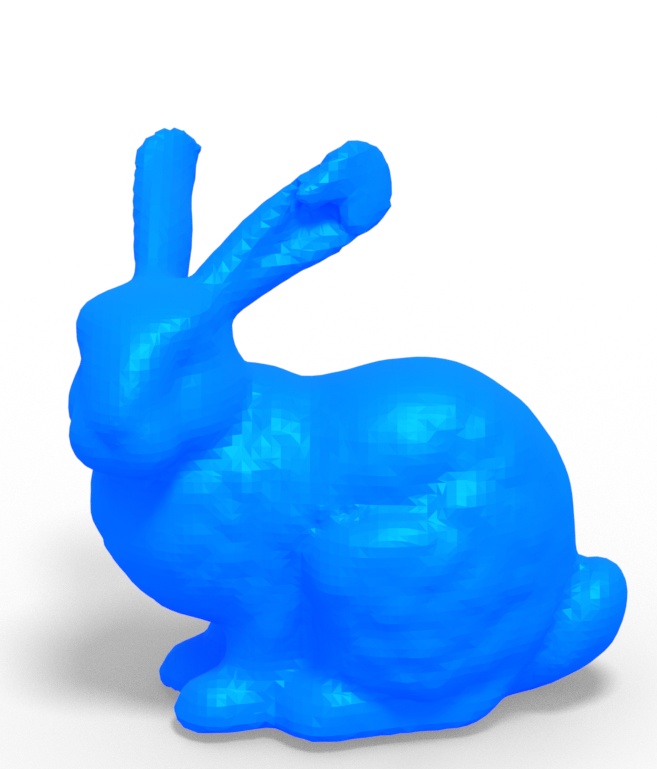}
    \includegraphics[width=0.525in]{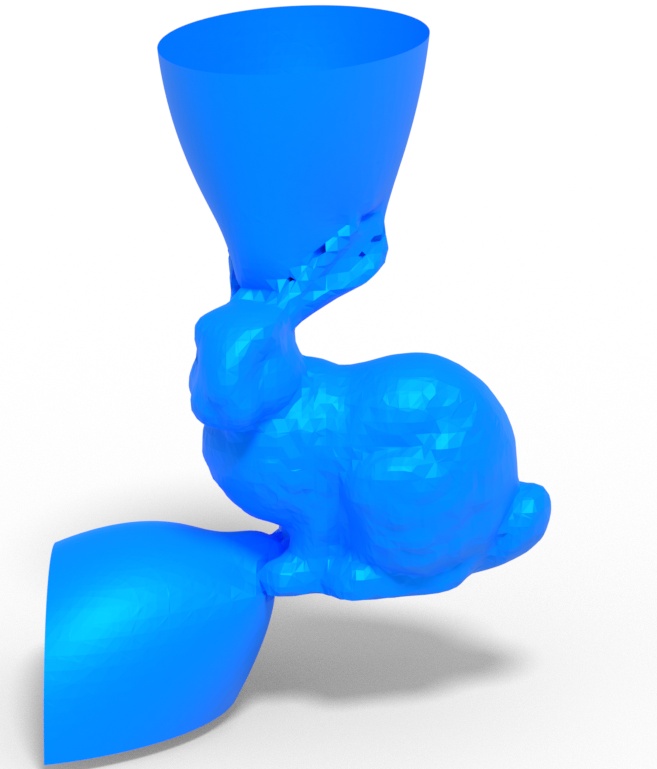}
    \includegraphics[width=0.525in]{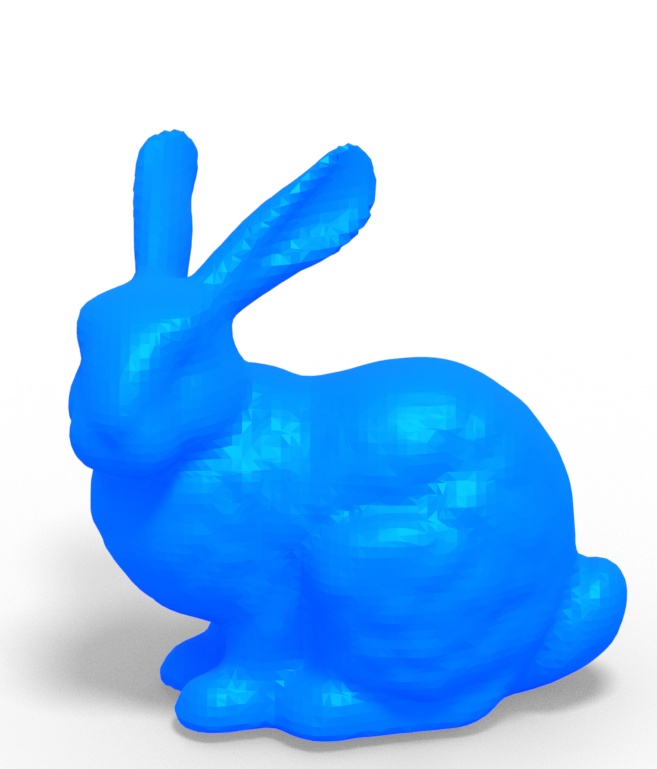}
    \includegraphics[width=0.525in]{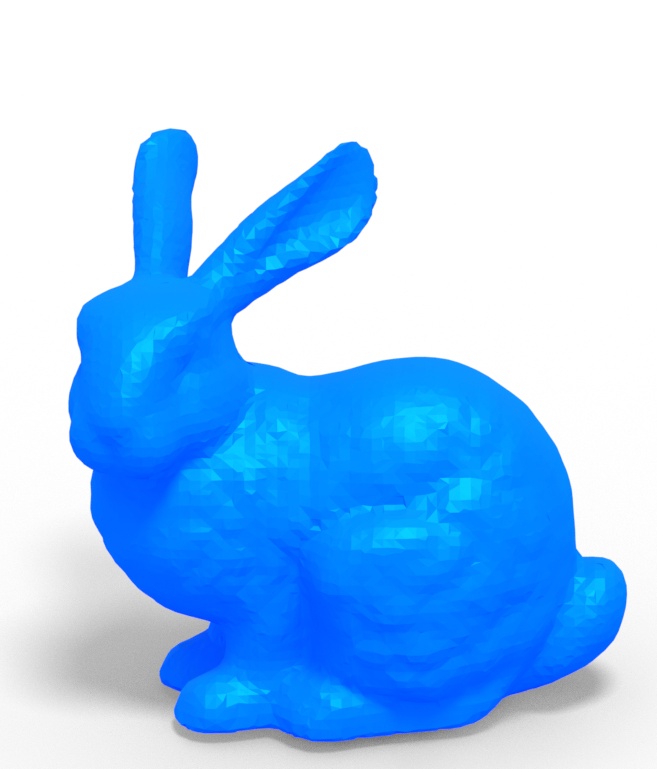}\\
    \makebox[0.525in]{8,171 pts}
    \makebox[0.525in]{(3.60\%,142.3s)}
    \makebox[0.525in]{(3.53\%,26.5s)}
    \makebox[0.525in]{(49.07\%,7271.1s)}
    \makebox[0.525in]{(3.19\%,5.5s)}
    \makebox[0.525in]{ - } \\
    \includegraphics[width=0.525in]{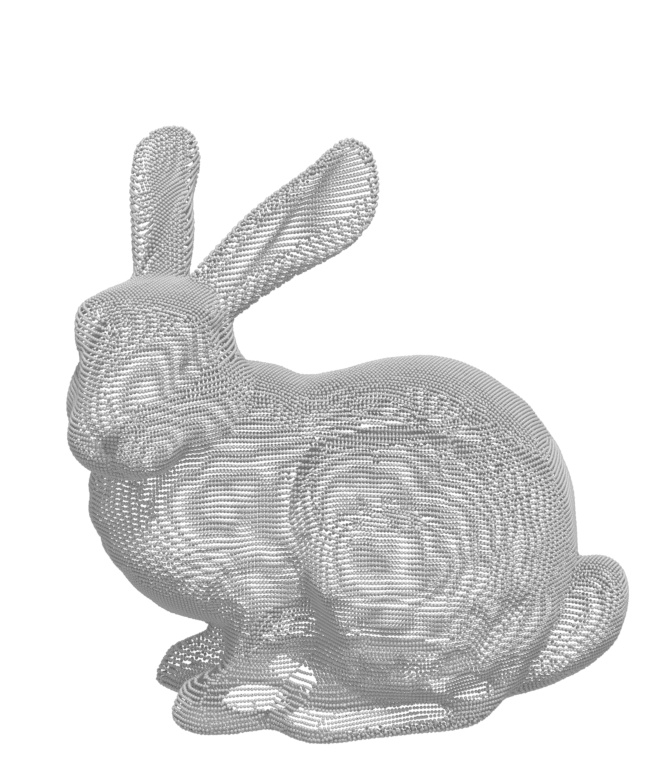}
    \includegraphics[width=0.525in]{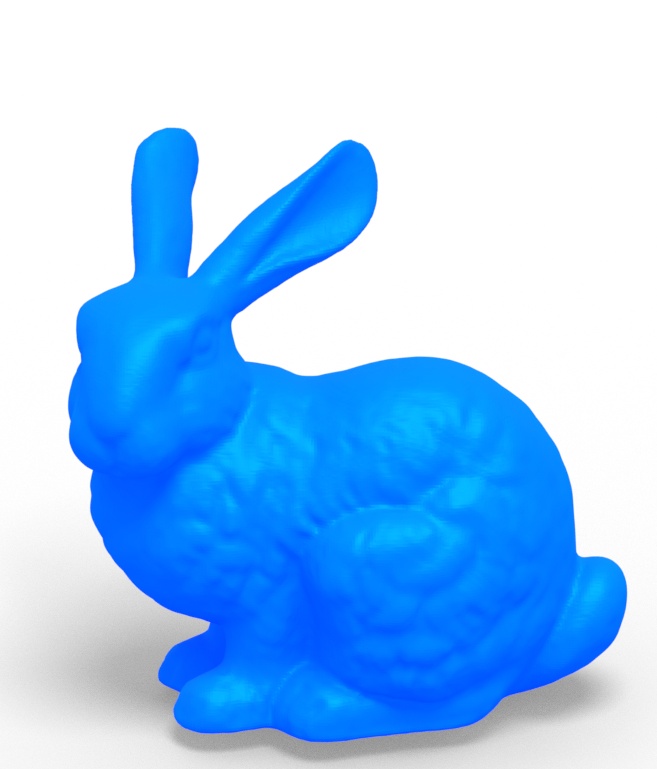}
    \includegraphics[width=0.525in]{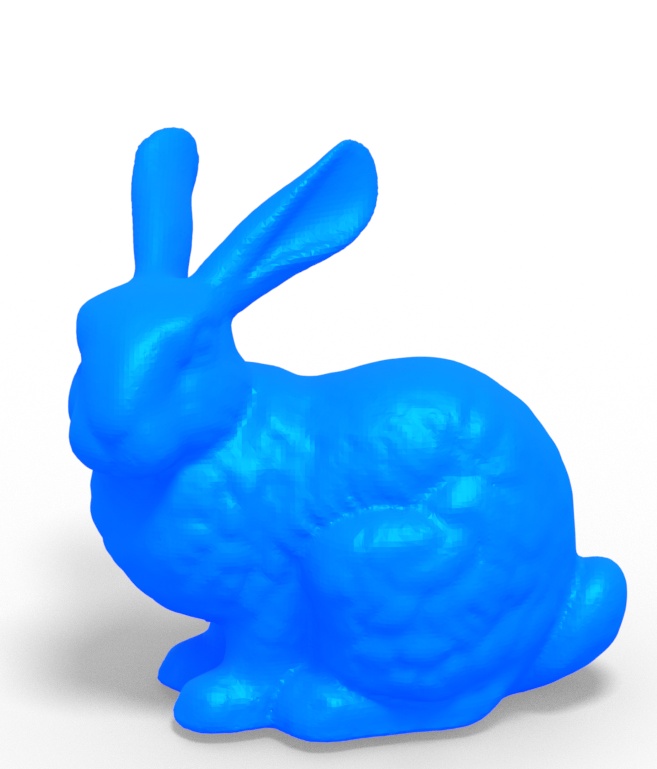}
    \includegraphics[width=0.525in]{./figures/NA.pdf}
    \includegraphics[width=0.525in]{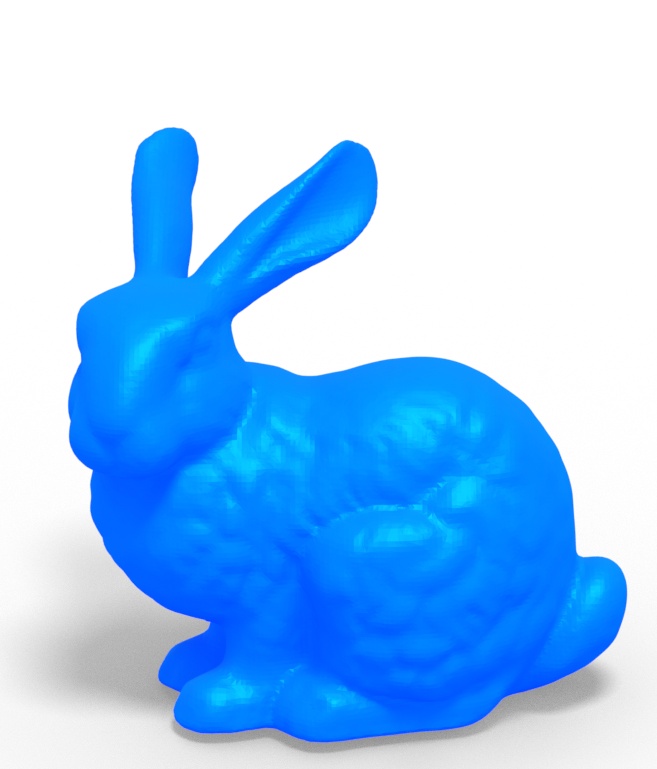}
    \includegraphics[width=0.525in]{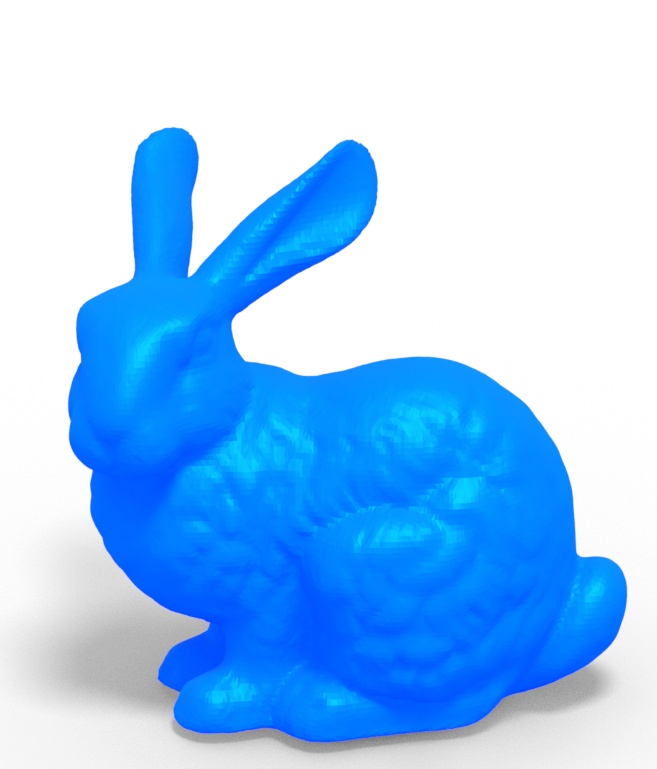}\\
    \makebox[0.525in]{35,947 pts}
    \makebox[0.525in]{(3.54\%,154.0s)}
    \makebox[0.525in]{(3.56\%,87.6s)}
    \makebox[0.525in]{(-,-)}
    \makebox[0.525in]{(3.10\%,9.7s)}
    \makebox[0.525in]{ - } \\
    \makebox[0.525in]{Input}
    \makebox[0.525in]{DPSR}
    \makebox[0.525in]{Dipole}
    \makebox[0.525in]{VIPSS}
    \makebox[0.525in]{iPSR}
    \makebox[0.525in]{GT mesh}\\
    \end{scriptsize}
    \caption{Stanford Bunny in different resolutions. The 2-tuple below each figure is the reconstruction error and running time (seconds). }
    \label{fig:bunny_ear}
\end{figure}

\begin{figure}
    \centering
    \includegraphics[width=0.50in]{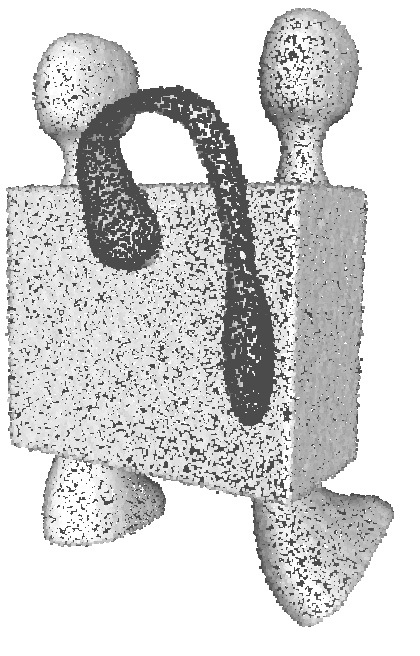}
    \includegraphics[width=0.50in]{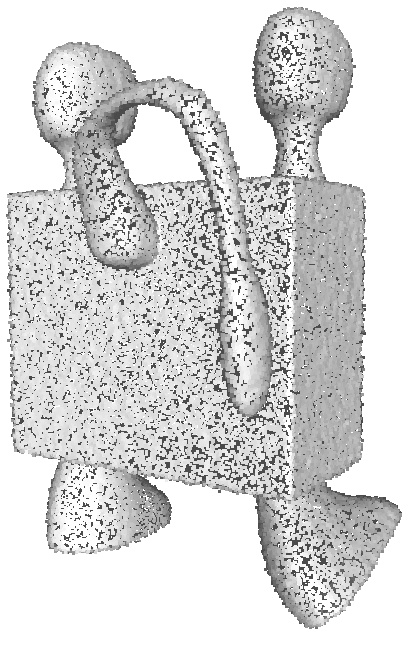}
    \includegraphics[width=1.05in]{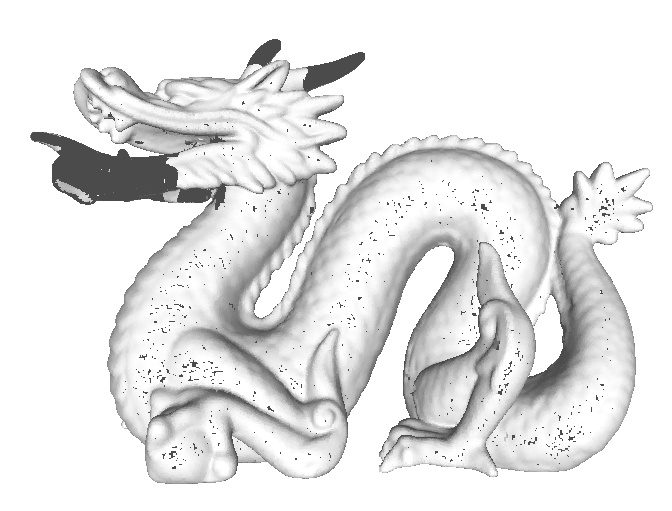}
    \includegraphics[width=1.05in]{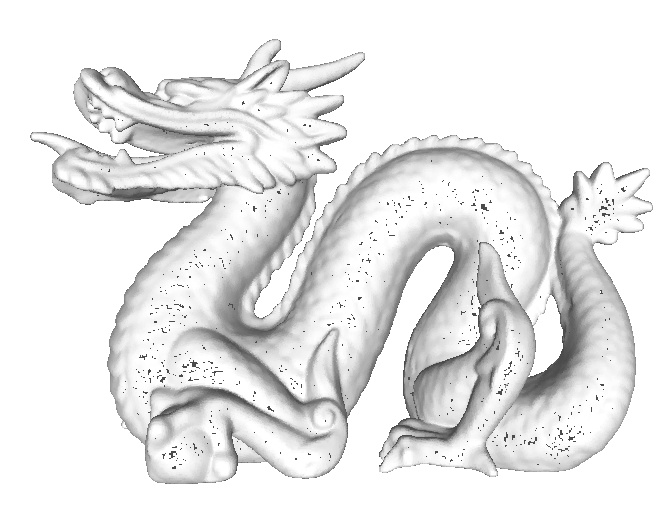}\\
    \makebox[0.50in]{Dipole}
    \makebox[0.50in]{iPSR}
    \makebox[1.05in]{Dipole}
    \makebox[1.05in]{iPSR}\\
    \caption{Comparing randomly-initialized iPSR with Dipole on orientation prediction. Points are rendered as oriented disks with grey front side and black back side. The black disks indicate the points with flipped normals.
    }
    \label{fig:compare_dipole}
\end{figure}

\begin{figure}
    \centering
    \includegraphics[width=0.76in]{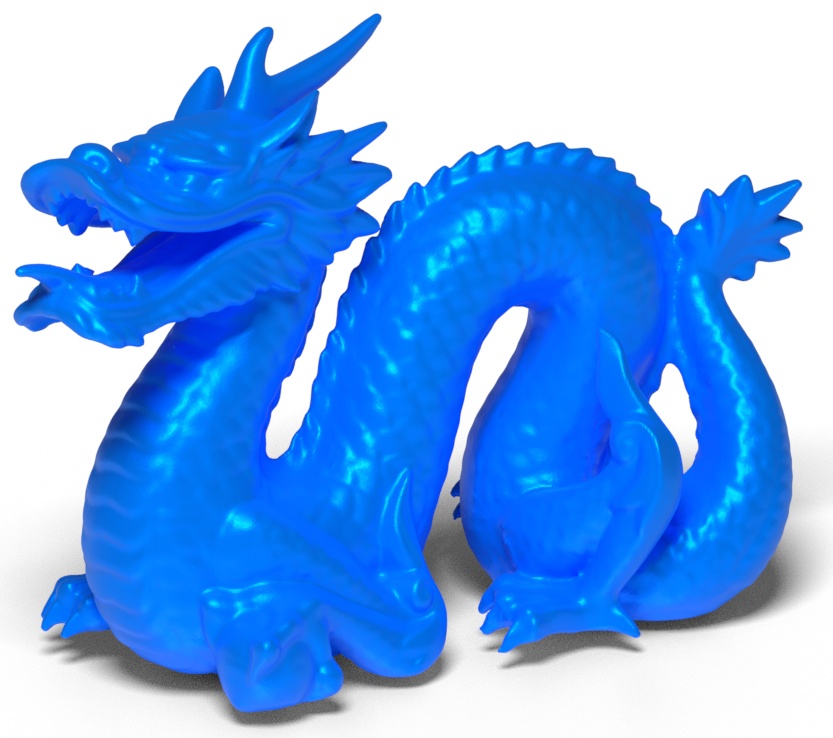}
    \includegraphics[width=0.76in]{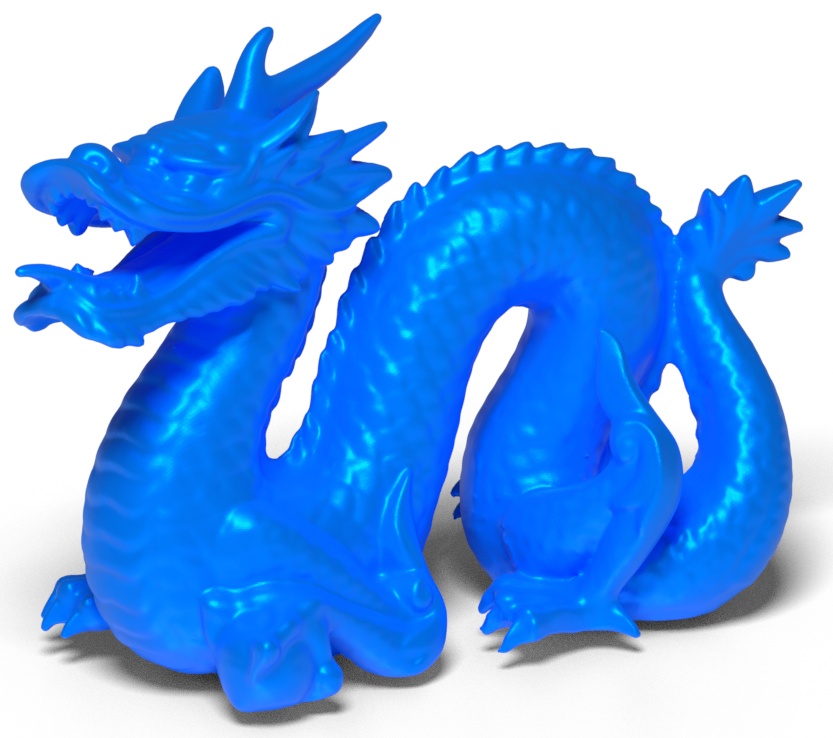}
    \includegraphics[width=0.76in]{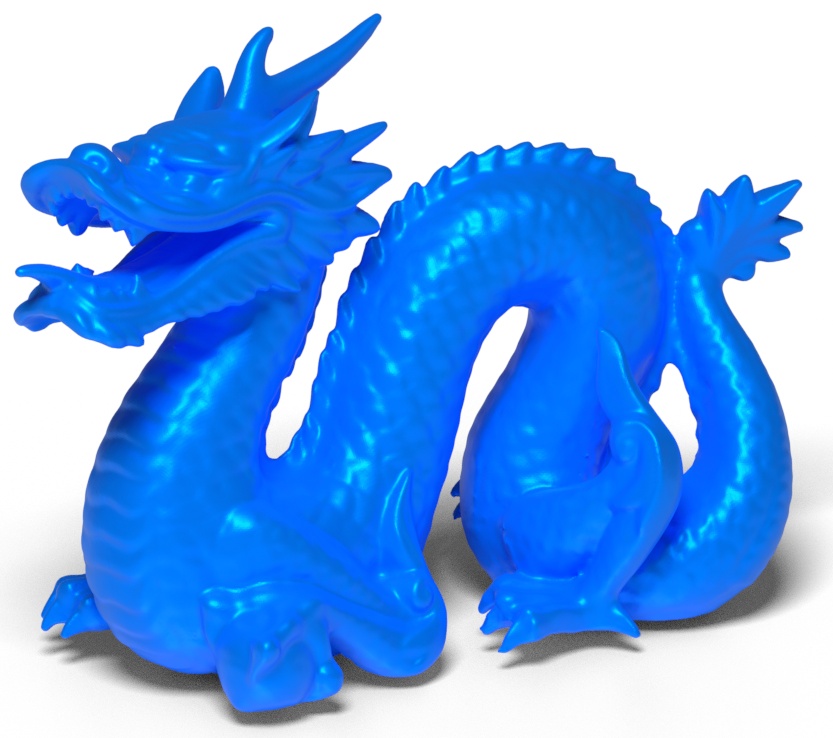}
    \includegraphics[width=0.76in]{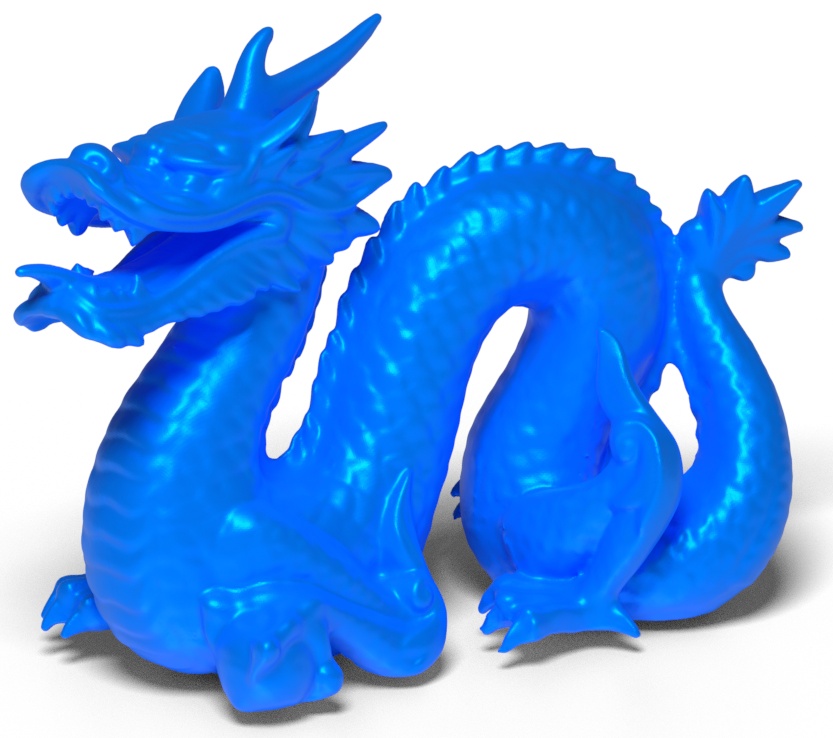}\\
    \makebox[0.76in]{GT mesh}
    \makebox[0.76in]{screened PSR}
    \makebox[0.76in]{iPSR+random}
    \makebox[0.76in]{iPSR+visibility}\\
    \makebox[0.76in]{(437K vertices)}
    \makebox[0.76in]{$\varepsilon=0.31\%$}
    \makebox[0.76in]{$\varepsilon=0.45\%$}
    \makebox[0.76in]{$\varepsilon=0.45\%$}\\
    \includegraphics[width=3.3in]{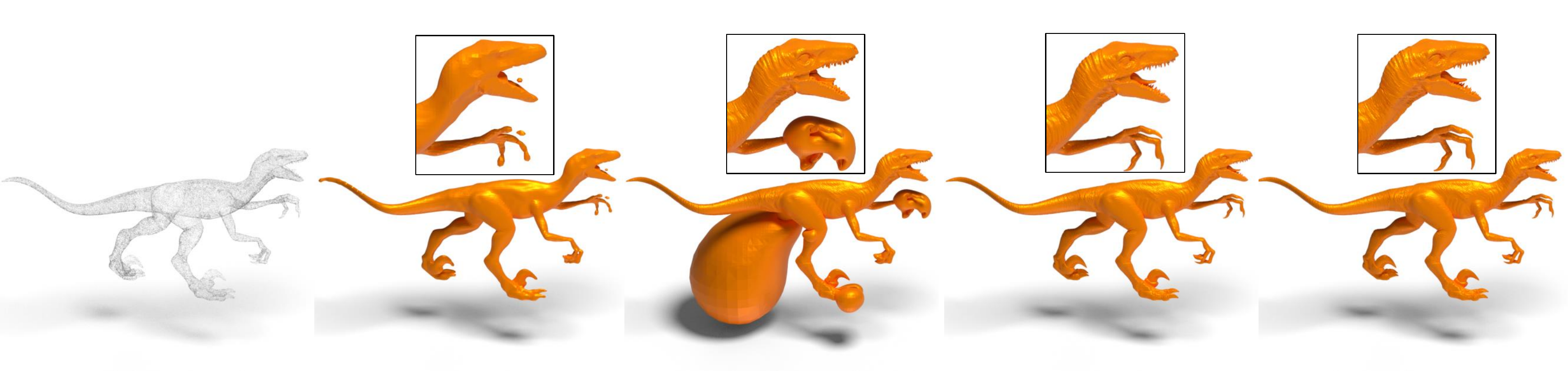}\\
    \makebox[0.63in]{Input}
    \makebox[0.63in]{DPSR}
    \makebox[0.63in]{Dipole}
    \makebox[0.63in]{iPSR+random}
    \makebox[0.63in]{GT mesh}\\
    \makebox[0.63in]{(1.5M vertices)}
    \makebox[0.63in]{$\varepsilon=2.70\%$}
    \makebox[0.63in]{$\varepsilon=28.43\%$}
    \makebox[0.63in]{$\varepsilon=0.10\%$}
    \makebox[0.63in]{}\\
\caption{Reconstruction quality. 
Row 1: iPSR is insensitive to initialization. Both random initialization (18 iterations) and visibility-based  initialization (9 iterations) yield visually identical final results, whose quality is comparable to the screened PSR which takes correct normals as input.   Row 2: Comparison with DPSR and Dipole. DPSR cannot recover fine detail due to its fixed voxelization resolution $256\times 256\times 256$. Dipole can reconstruct geometry detail well, but the flipped normals on the legs lead to large artifacts.
}
\label{fig:compare_psr}
\end{figure}

Figure~\ref{fig:robustness} shows the results of robustness test. Since iPSR inherits the robust features of screened PSR, we observed that it is resilient to noise, outliers, high genus, non-uniform, missing regions and sparse-but-structured inputs. 
In contrast, each of the other methods has one or more weaknesses.
We set the screened weight $\alpha=10$ of iPSR for all models. For the two noisy hand models, we run iPSR with $\alpha=10$ till convergence and then applied screened PSR with $\alpha=0$ to produce the final results. The final results of DPSR and Dipole on the two noisy hand models are also computed using $\alpha=0$ for a fair comparison. Figure~\ref{fig:manmade} shows results of man-made models. We observed the iPSR is more robust to thin structures and holes than DPSR and Dipole.

\begin{figure}[!htbp]
    \centering
    \begin{scriptsize}
    \includegraphics[width=0.75in]{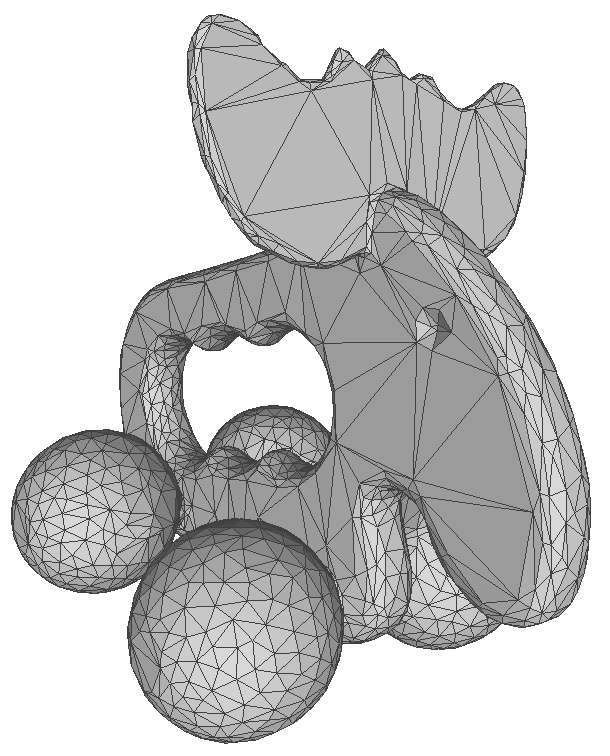}
    \includegraphics[width=0.75in]{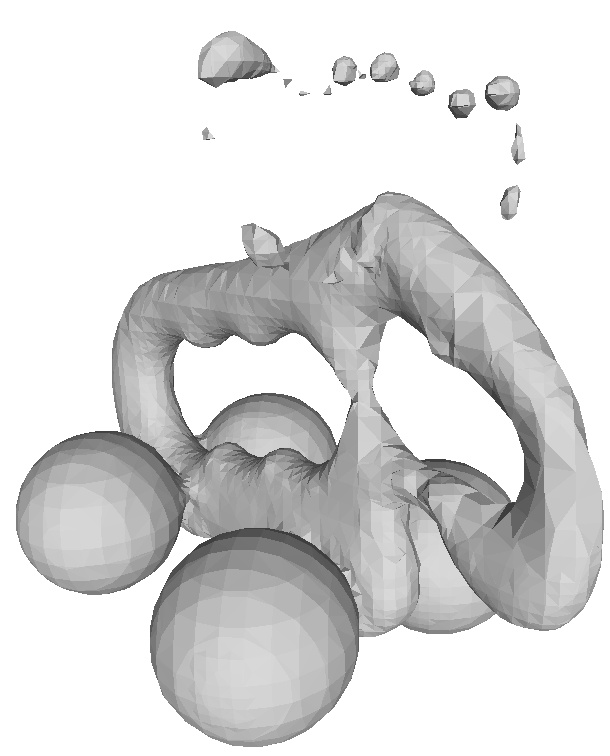}
    \includegraphics[width=0.75in]{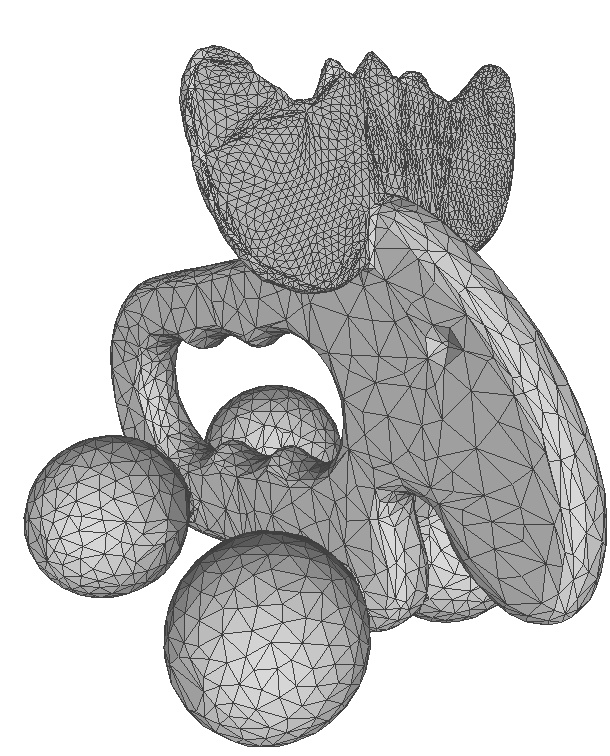}
    \includegraphics[width=0.75in]{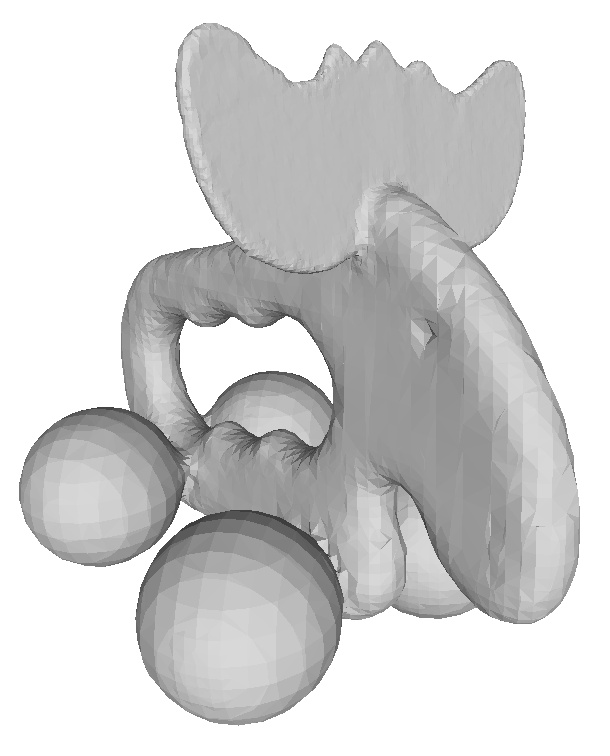}\\
    \makebox[0.75in]{$m=2,500$}
    \makebox[0.75in]{iPSR}
    \makebox[0.75in]{$m=5,786$}
    \makebox[0.75in]{iPSR}\\
    \includegraphics[width=1.05in]{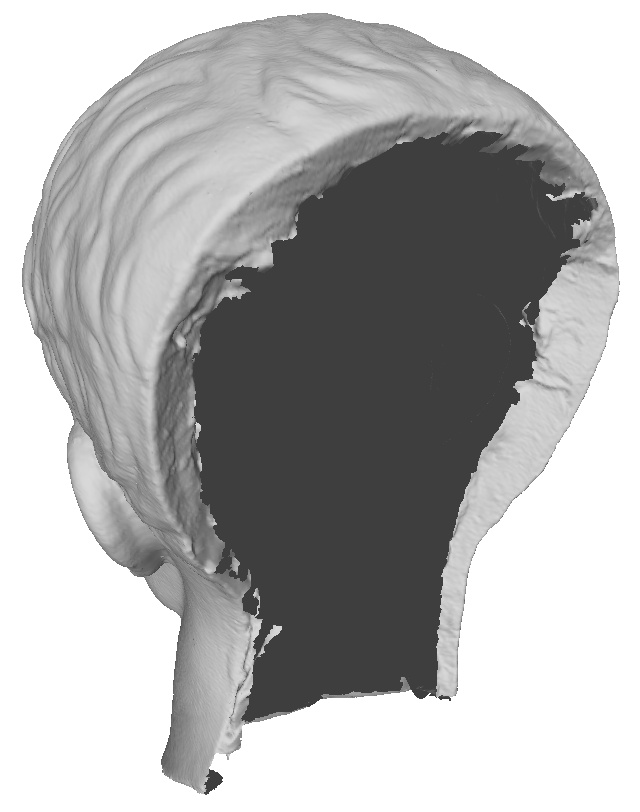}
    \includegraphics[width=1.05in]{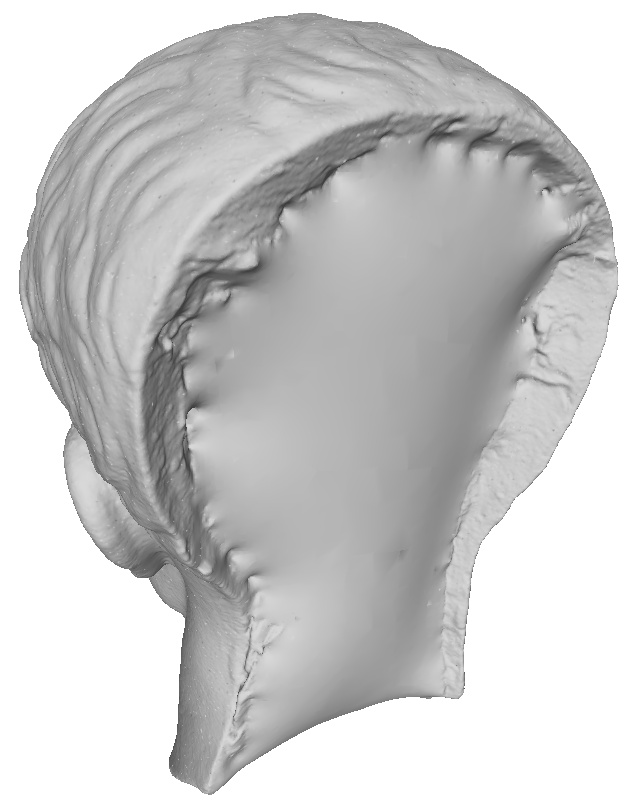}
    \includegraphics[width=1.05in]{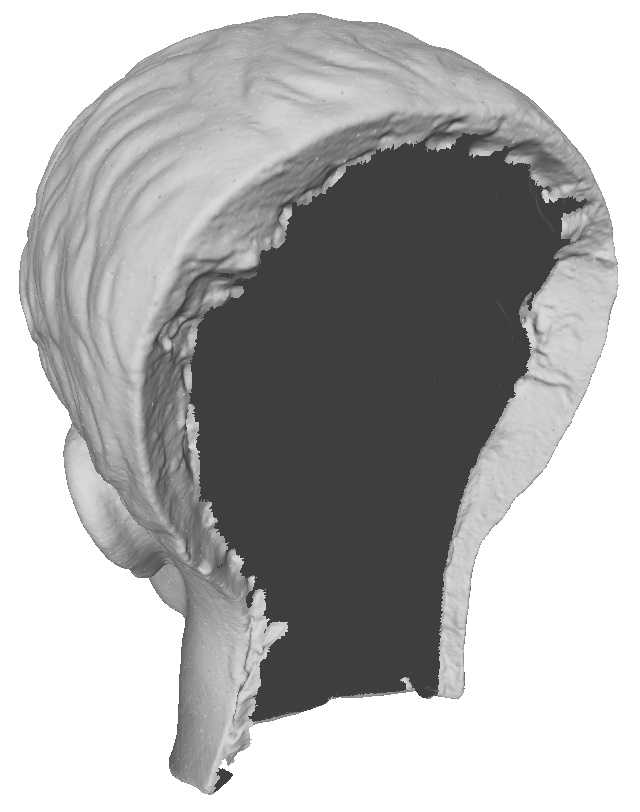}\\
    \makebox[1.05in]{GT mesh}
    \makebox[1.05in]{iPSR}
    \makebox[1.05in]{After post-processing}\\
    \includegraphics[width=1.05in]{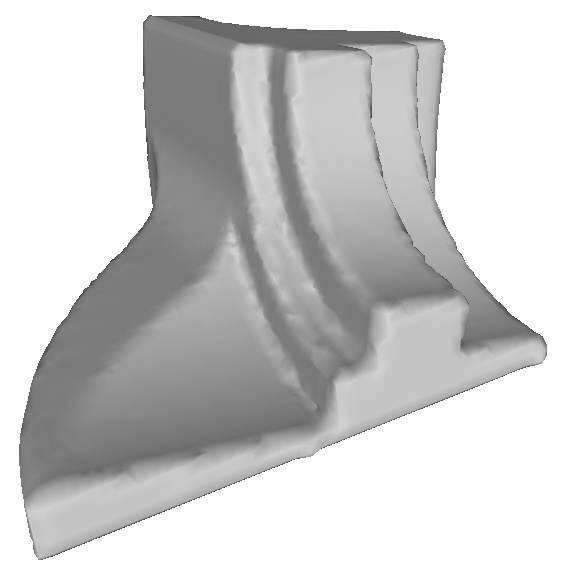}
    \includegraphics[width=1.05in]{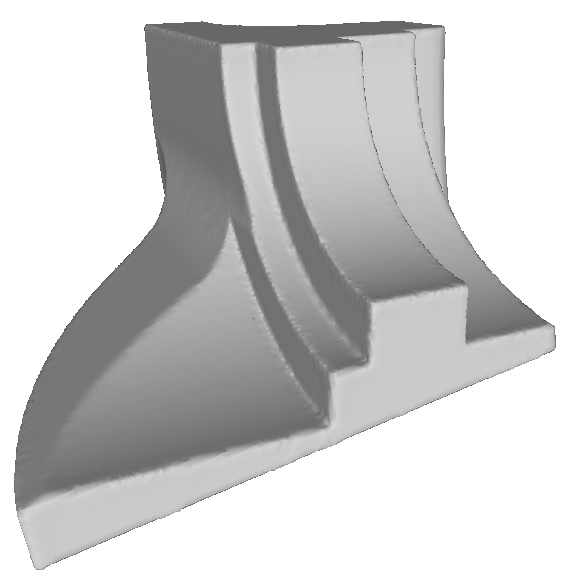}
    \includegraphics[width=1.05in]{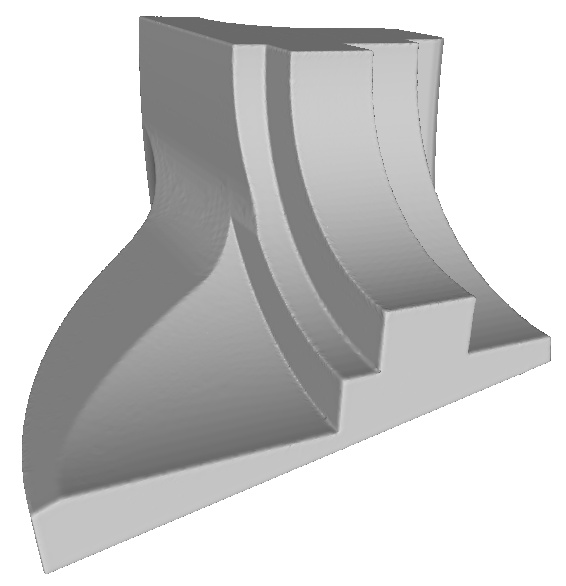}\\
    \makebox[1.05in]{$m=6,475$}
    \makebox[1.05in]{$m=38,840$}
    \makebox[1.05in]{$m=155,354$}\\
    \end{scriptsize}
    \caption{Limitations. Row 1: iPSR with random initialization generates disconnected components when sampling rate is too low. Row 2: The output is always a watertight surface. 
    Row 3: It cannot preserve sharp edges and corners.
    }
    \label{fig:limitations}
\end{figure}

Figure~\ref{fig:bunny_ear} compares DPSR, Dipole, VIPSS and our method in terms of reconstruction quality on Stanford Bunny with varying resolutions. 
Except for VIPSS, all methods can produce results for the highest resolution properly.
As the number of points decreases, defects of various degree show up on Bunny's ears due to thin structure and relatively low local sampling rate.
iPSR produces the least artifacts among all the methods.
We also observed that Dipole may produce flipped normals (see Figure~\ref{fig:compare_dipole}). Such normal inconsistency leads to incorrect local shapes, e.g., the large defects at the legs of Raptor (Figure~\ref{fig:compare_psr}). 
iPSR is insensitive to initialization and can yield visually identical results with random initialization and visibility initialization. See Figure~\ref{fig:compare_psr} (top).
With a properly chosen octree depth, iPSR can recover fine detail well. See Figure~\ref{fig:compare_psr} (bottom). In contrast, DPSR can only generate a rough model due to its fixed voxelization resolution.

\section{Discussion and Conclusion}
We extended the popular Poisson surface reconstruction method by
eliminating its requirement of point orientation. We proposed a simple
yet effective orientation strategy and showed that even with randomly
initialized point normals, our enhanced Poisson surface reconstruction
can proceed iteratively and yield visually pleasing, smooth
surfaces. Our iPSR method inherits the scalability and robustness
features of PSR, and works well for both sparse and dense raw
points. Throughout the paper, we demonstrated high-fidelity
reconstruction results on the AIM$@$SHAPE dataset and large-scale 3D scenes.

Our approach calls for possible further improvements that must be
addressed in follow-up research. First, iPSR may fail on thin
structure with very low sampling rate. Consider the toy model as an
example (refer to Figure~\ref{fig:limitations} (row 1)). With random initialization, the shape is broken into many
parts. This problem can be fixed by adaptively increasing the
sampling rate. It is also possible to initialize normals by using some simple orientation methods that can predict reasonable normals based on local information. 

Second, our method can produce only a closed manifold surface. For open models,
we have to adopt simple post-processing to remove the unnecessary
fill. Specifically, for each vertex of the reconstructed mesh, we
compute the distance to the closest sample in the input. If the
distance is greater than a user-specified threshold, we discard the
vertex and its adjacent faces (refer to Figure~\ref{fig:limitations}
(row 2)).

Third, our method cannot preserve sharp features due to the smooth
nature of the solution of Poisson's equation. Although increasing the input
sampling rate can reduce the blurring artifact (refer
to Figure~\ref{fig:limitations} (row 3)), it is more desirable
to introduce post-processing to recover sharp edges and corners. Some advanced isosurfacing methods, such as neural marching cube~\cite{Chen2021}, are also helpful to reconstruct sharp edges. 

Fourth, in our current implementation, we simply treat the screened
PSR~\cite{Kazhdan2013} in its entirety as a black box and feed it with
updated normals in each iteration. Notice that the sample positions
remain unchanged throughout the iterative procedure, implying that the
Laplacian matrix of Poisson's equation is fixed. One possible way
towards significant reduction of the running time is to pre-factorize
the Laplacian matrix (e.g., using Cholesky decomposition). Then in
each iteration, the sparse linear system can be solved using backward
substitution, which has (near-)linear time complexity. 

Fifth, although we have not seen any case that iPSR cannot converge if the input points come from smooth models, we do not have theoretical guarantee on convergence. We leave the rigorous analysis of iPSR and the sufficient and necessary conditions of convergence as an open problem.

Last but not the least, 
the odd-layered structure exhibited in the implicit functions computed by iPSR plays a critical role in normal correction. We think this type of structure is general for other implicit function based methods, thereby worth further investigation.

%\appendix


\begin{acks}
We would like to thank the anonymous reviewers for their constructive comments. Special thanks go to Reviewer 1 for the careful reviews and giving us concrete suggestions to improve exposition. This research has been partially supported by National Natural Science Foundation of China (61872347, 62072446), Special Plan for the Development of Distinguished Young Scientists of ISCAS (Y8RC535018), National Science Foundation (IIS-1715985 \& 1812606 to Qin), Singapore Ministry of Education (MOE-T2EP20220-0005 and RG20/20) and RIE2020 Industry Alignment Fund – Industry Collaboration Projects (IAF-ICP) Funding Initiative, as well as cash and in-kind contribution from the industry partner(s).
\end{acks}

\bibliographystyle{ACM-Reference-Format}
\bibliography{reference}

\end{document}